\documentclass[12pt,preprint]{aastex}
\usepackage{float,epsfig}
\usepackage{pstricks}
\usepackage{lscape}

\def\pn{\par\noindent}
\def\chandra{{\it Chandra}}
\def\xmm{XMM--{\it Newton}}
\def\cgs{erg cm$^{-2}$ s$^{-1}$} 
\def\gsimeq{\hbox{\raise0.5ex\hbox{$>\lower1.06ex\hbox{$\kern-1.07em{\sim}$}$}}} 
\def\lsimeq{\hbox{\raise0.5ex\hbox{$<\lower1.06ex\hbox{$\kern-1.07em{\sim}$}$}}} 
\def\spose#1{\hbox to 0pt{#1\hss}}
\def\simlt{\mathrel{\spose{\lower 3pt\hbox{$\mathchar"218$}}
     \raise 2.0pt\hbox{$\mathchar"13C$}}}
\def\lsimeq{\mathrel{\spose{\lower 3pt\hbox{$\mathchar"218$}}
     \raise 2.0pt\hbox{$\mathchar"13C$}}}
\def\ls{\mathrel{\spose{\lower 3pt\hbox{$\mathchar"218$}}
     \raise 2.0pt\hbox{$\mathchar"13C$}}}
\def\simgt{\mathrel{\spose{\lower 3pt\hbox{$\mathchar"218$}}
     \raise 2.0pt\hbox{$\mathchar"13E$}}}
\def\gsimeq{\mathrel{\spose{\lower 3pt\hbox{$\mathchar"218$}}
     \raise 2.0pt\hbox{$\mathchar"13E$}}}
\def\gs{\mathrel{\spose{\lower 3pt\hbox{$\mathchar"218$}}
     \raise 2.0pt\hbox{$\mathchar"13E$}}}

\slugcomment{}

\shorttitle{XMM-COSMOS paper}
\shortauthors{Brusa et al.}

\begin{document}

%% This is the end of the preamble.  Indicate the beginning of the
%% paper itself with \begin{document}.

%% LaTeX will automatically break titles if they run longer than
%% one line. However, you may use \\ to force a line break if
%% you desire.

%\twocolumn[  

%\title{Obscured and unobscured X-ray selected AGN population in COSMOS}\altaffiltext{$\star$}{
\title{The XMM-Newton Wide-Field Survey in the COSMOS field (XMM-COSMOS): 
demography and 
multiwavelength properties of obscured and unobscured luminous AGN}
\altaffiltext{$\star$}{Based on data collected at: the NASA/ESA {\em
Hubble Space Telescope}, obtained at the Space Telescope Science
Institute, which is operated by AURA Inc, under NASA contract NAS
5-26555; the Subaru Telescope, which is operated by
   the National Astronomical 
Observatory of Japan; the European Southern Observatory, Chile, under Large
   Program 175.A-0839; Kitt Peak
   National Observatory, Cerro Tololo Inter-American Observatory, and the 
National Optical Astronomy Observatory, which are operated by the Association
   of Universities for Research in Astronomy, Inc. (AURA) under cooperative
   agreement 
with the National Science Foundation; and the Canada-France-Hawaii Telescope
   operated by the National Research Council of Canada, 
the Centre National de la Recherche Scientifique de France and the University
   of Hawaii.}

%% Use \author, \affil, and the \and command to format
%% author and affiliation information.
%% Note that \email has replaced the old \authoremail command
%% from AASTeX v4.0. You can use \email to mark an email address
%% anywhere in the paper, not just in the front matter.
%% As in the title, you can use \\ to force line breaks.

\author{
M. Brusa\altaffilmark{1}, 
F. Civano\altaffilmark{2},
A. Comastri\altaffilmark{3},
T. Miyaji\altaffilmark{4,5},
M. Salvato\altaffilmark{6,7,8},
G. Zamorani \altaffilmark{3},
N. Cappelluti\altaffilmark{1,9},
F. Fiore\altaffilmark{10},
G. Hasinger\altaffilmark{6},
V. Mainieri\altaffilmark{11},
A. Merloni\altaffilmark{7,1},
%%%%%%%%%%%%%%%%%%%%%%%%%%%%%%%%%%%%%%%%%%%%%%%%%%%%%%%
A. Bongiorno\altaffilmark{1}, 
P. Capak\altaffilmark{12},
M. Elvis\altaffilmark{2},
R. Gilli\altaffilmark{3},
H. Hao\altaffilmark{2},
K. Jahnke\altaffilmark{13},
A.M. Koekemoer\altaffilmark{14},
O. Ilbert\altaffilmark{15},
E. Le Floc'h\altaffilmark{16},
E. Lusso\altaffilmark{17,3},
M. Mignoli\altaffilmark{3},
E. Schinnerer\altaffilmark{13},
J.D. Silverman\altaffilmark{18},
E. Treister\altaffilmark{19},
J.D. Trump\altaffilmark{20},
C. Vignali\altaffilmark{17,3}, 
M. Zamojski\altaffilmark{12}, %}
%%%%%%%%%%%%%%%%%%%%%%%%%%%%%%%%%%%%%%%%%%%%%%%%%%%%%%%
T. Aldcroft\altaffilmark{2}, 
H. Aussel\altaffilmark{21},  
S. Bardelli\altaffilmark{3},
M. Bolzonella\altaffilmark{3},
A. Cappi\altaffilmark{3},      
K. Caputi\altaffilmark{22},      
T. Contini\altaffilmark{23},      
A. Finoguenov\altaffilmark{1},
A. Fruscione\altaffilmark{2},
B. Garilli\altaffilmark{24},
C.D. Impey\altaffilmark{20},
A. Iovino\altaffilmark{25},
K. Iwasawa\altaffilmark{3},
P. Kampczyk\altaffilmark{26},
J. Kartaltepe\altaffilmark{19,25},
J.P. Kneib\altaffilmark{27},
C. Knobel\altaffilmark{26},
K. Kovac\altaffilmark{26},
F. Lamareille\altaffilmark{23},
J-F. Leborgne\altaffilmark{23},
V. Le Brun\altaffilmark{27},
O. Le Fevre\altaffilmark{27},
S.J. Lilly\altaffilmark{26},
C. Maier\altaffilmark{26}, 
H.J. McCracken\altaffilmark{28},
R. Pello\altaffilmark{23},
Y-J Peng\altaffilmark{26},
E. Perez-Montero\altaffilmark{26},
L. de Ravel\altaffilmark{27},
D. Sanders\altaffilmark{12},
M. Scodeggio\altaffilmark{24},
N.Z. Scoville\altaffilmark{8},
M. Tanaka\altaffilmark{11},
Y. Taniguchi\altaffilmark{29},
L. Tasca\altaffilmark{24,27},
S. de la Torre\altaffilmark{27},
L. Tresse\altaffilmark{27},
D. Vergani\altaffilmark{3},
E. Zucca\altaffilmark{3}}

\altaffiltext{1}{Max-Planck-Institut f\"ur extraterrestrische Physik, Giessenbachstrasse 1, D--85748 Garching bei M\"unchen, Germany} 
\altaffiltext{2}{Harvard-Smithsonian Center for Astrophysics, 60 Garden Street, Cambridge, MA 02138 } %elvis
\altaffiltext{3}{INAF --  Osservatorio Astronomico di Bologna, via Ranzani 1, I--40127 Bologna, Italy}
\altaffiltext{4}{Instituto de Astronom\'ia, Universidad Nacional Aut\'onoma de M\'exico, Ensenada, M\'exico (mailing address: PO Box 439027, San Ysidro,  CA, 92143-9024, USA)} 
\altaffiltext{5}{Center for Astrophysics and Space Sciences, University of California at San Diego, Code 0424, 9500 Gilman Drive, La Jolla, CA 92093, USA}
\altaffiltext{6}{IPP - Max-Planck-Institute for Plasma Physics, Boltzmannstrasse 2, D-85748, Garching, Germany}
\altaffiltext{7}{Excellence Cluster Universe,   Boltzmannstrasse 2, D-85748, Garching bei Muenchen, Germany}
\altaffiltext{8}{California Institute of Technology, 1201 East California blvd, Pasadena, 91125, CA} 
\altaffiltext{9}{University of Maryland, Baltimore County, 1000 Hilltop Circle, Baltimore, MD 21250, USA}
\altaffiltext{10}{INAF --  Osservatorio Astronomico di Roma, via Frascati 33,  Monteporzio-Catone (Roma), I-00040, Italy}
\altaffiltext{11}{European Southern Observatory, Karl-Schwarzschild-str. 2,  85748 Garching bei M\"unchen, Germany} 
\altaffiltext{12}{Spitzer Science Center, California Institute of Technology, MC 220-6, 1200 East California Boulevard, Pasadena, CA 91125}
\altaffiltext{13}{Max-Planck-Institut für Astronomie, K\"onigstuhl 17, D-69117 Heidelberg, Germany}
\altaffiltext{14}{Space Telescope Science Institute, 3700 San Martin Drive, Baltimore, MD 21218, USA}
\altaffiltext{15}{Laboratoire d'Astrophysique de Marseille, BP 8, Traverse du Siphon, 13376 Marseille Cedex 12, France}
\altaffiltext{16}{Laboratoire AIM, CEA/DSM - CNRS - Universit\'e Paris Diderot, DAPNIA/Service d'Astrophysique, B\^at. 709, CEA-Saclay, F-91191 Gif-sur- Yvette C\'edex, France}
\altaffiltext{17}{Dipartimento di Astronomia, Universit\`a di Bologna, via Ranzani 1, 40127, Bologna, Italy} 
\altaffiltext{18}{Institute for the Physics and Mathematics of the Universe (IPMU), University of Tokyo, Kashiwanoha 5-1-5, Kashiwa, Chiba 277-8568, Japan}
\altaffiltext{19}{Institute for Astronomy, University of Hawaii, 2680 Woodlawn Drive, Honolulu, HI 96822}
\altaffiltext{20}{Steward Observatory, University of Arizona, 933 North Cherry Avenue, Tucson, AZ 85 721} %Impey 
\altaffiltext{21}{Institut d'Astrophysique de Paris, UMR7095 CNRS, Universit\'e Pierre \& Marie Curie, 98 bis boulevard Arago, 75014 Paris, France} 
\altaffiltext{22}{SUPA Institute for Astronomy, The University of Edinburgh, Royal Observatory, Blackford Hill, Edinburgh EH9 3HJ, UK} 
\altaffiltext{23}{Observatoire Midi-Pyr\'en\'ees, Laboratoire d'Astrophysique (UMR 5572), 14 avenue E. Belin, F-31400 Toulouse, France} 
\altaffiltext{24}{INAF IASF - Milano, via Bassini 15, 20133 Milano, Italy} 
\altaffiltext{25}{National Optical Astronomy Observatory, 950 North Cherry Avenue, Tucson, AZ 85 721} 
\altaffiltext{26}{Department of Physics, ETH Zurich, CH-8093 Zurich, Switzerland}
\altaffiltext{27}{Laboratoire d'Astrophysique de Marseille, Traverse du Siphon, F-13376 Marseille, France} 
\altaffiltext{28}{Observatoire de Paris, LERMA, 61 Avenue de l'Observatoire,  75014 Paris, France} 
\altaffiltext{29}{Physics Department, Graduate School of Science and Engineering, Ehime University, 2-5 Bunkyo-cho, Matsuyama, Ehime 790-8577. Japan}

%% title and affiliation information. No date will appear since the author
%% does not have this information. The dates will be filled in by the
%% editorial office after submission.

\begin{abstract}

We report the final optical identifications of the medium-depth 
($\sim 60$ ksec), 
contiguous (2 deg$^{2}$) XMM-Newton survey of the COSMOS field.
XMM-{\it Newton} has detected $\sim1800$ X-ray sources down 
to limiting fluxes of $\sim 5\times10^{-16}$, $\sim 3\times10^{-15}$,
and $\sim 7\times10^{-15}$ \cgs\ in the 0.5-2 keV, 2-10 keV and 5-10 keV bands, 
respectively ($\sim 1\times10^{-15}$, $\sim 6\times10^{-15}$,
and $\sim 1\times10^{-14}$ \cgs, in the three bands, respectively, over
50\% of the area).
The work is complemented by an extensive collection of multi-wavelength 
data from 24$\mu$m to UV, available from the COSMOS survey, for each of 
the X--ray sources, including spectroscopic redshifts for $\gs50$\% 
of the sample, and high-quality photometric redshifts for the rest.
The XMM and multiwavelength  flux limits are well matched: 1760 (98\%) of the 
X-ray sources have optical counterparts, 1711 ($\sim95$\%) have IRAC
counterparts, and 1394 ($\sim78$\%) have MIPS 24 $\mu$m detections. 
Thanks to the redshift completeness (almost 100\%) we were able to 
constrain the high-luminosity tail of the X--ray luminosity
function confirming that the peak of the number density 
of logL$_{X}>44.5$ AGN is at z$\sim2$.  
Spectroscopically-identified obscured and unobscured AGN, as well as 
normal and starforming galaxies, present well-defined optical and infrared 
properties. We devised a robust method to identify a sample of $\sim150$
high redshift (z$>1$), obscured AGN candidates for which optical spectroscopy
is not available. We were able to determine that the fraction of the obscured 
AGN population at the highest (L$_{X}>10^{44}$ erg s$^{-1}$) X--ray
luminosity is $\sim 15-30$\% when selection effects are taken into account, 
providing an important observational constraint for X--ray background synthesis. 
We studied in detail the optical spectrum and the overall spectral energy
distribution of a prototypical Type 2 QSO, caught in a stage
transitioning from being starburst dominated to AGN dominated, 
which was possible to isolate only thanks to the combination of X-ray and infrared 
observations.

\end{abstract}

%% Keywords should appear after the \end{abstract} command. The uncommented
%% example has been keyed in ApJ style. See the instructions to authors
%% for the journal to which you are submitting your paper to determine
%% what keyword punctuation is appropriate.

%% From the front matter, we move on to the body of the paper.
%% In the first two sections, notice the use of the natbib \citep
%% and \citet commands to identify citations.  The citations are
%% tied to the reference list via symbolic KEYs. The KEY corresponds
%% to the KEY in the \bibitem in the reference list below. We have
%% chosen the first three characters of the first author's name plus
%% the last two numeral of the year of publication as our KEY for
%% each reference.

\section{Introduction}

First introduced to explain the properties we observe today in
{\it normal} galaxies such as their color bimodality
(Blanton et al. 2003, Bell et al. 2004,
Faber et al. 2007) and the so-called
local scaling relations (Ferrarese \& Merrit 2000, Gebahrdt et al. 2000,
G\"ultekin et al. 2009),
the existence of a ``feedback" between the accreting Super Massive Black Hole
(SMBH) and the host galaxy in which it resides ultimately challenged our
understanding of {\it Active Galactic Nuclei} (AGN).
Indeed, the widely accepted scenario of galaxy-AGN coevolution
(e.g. Silk \& Rees 1998, Fabian \& Iwasawa 1999, Granato et al. 2004, Di Matteo et al.
2005, Menci et al. 2008) points towards a physical
coupling responsible for the  self-regulated SMBH and galaxy growth.
In particular, according to the theoretical scenario emerging from extensive 
semi-analytical models and hydrodynamics simulations, 
a key event in the SMBH vs. host galaxy co-evolution is represented 
by the highly obscured AGN phase, 
when large quantities of 
gas driven to the center in the merger of two disk galaxies
were available to efficiently feed and
obscure the growing black hole and build galaxy stellar mass 
through significant episodes of star formation. 
This phase ends when strong winds and shocks from the central AGN heat
the interstellar medium, blowing away the dust and gas, thus cleaning
the line of sight and inhibiting further star-formation
(see e.g. Granato et al. 2004, Di Matteo et al. 2005; Croton et al. 2006; 
Sijacki et al. 2007; Menci et al. 2008; Hopkins et al. 2008). 

In this general framework for merger-induced AGN feedback, capable of passing 
numerous observational tests (e.g., Sanders et al. 1988,  Hopkins et al. 2006, 
Veilleux et al. 2009), the differences between ``obscured"
and ``unobscured" AGN are no longer and uniquely described under
a geometrical unification model (in which they are simply related to
orientation effects, Antonucci \& Miller 1985,
Antonucci 1993, Urry \& Padovani 1995), but can be interpreted as
due to the fact that the {\it same} objects are observed in {\it different
evolutionary} phases.
This hypothesis is consistent with the finding, mainly from
X--ray surveys, that absorption
is much more common at low luminosities (see e.g. 
Ueda et al. 2003, la Franca et al. 2005, Maiolino et al. 2007,
Hasinger 2008) and, possibly, at high 
redshift (La Franca et al. 2005; Ballantyne et al. 2006; 
Treister \& Urry 2006; Hasinger 2008).
The luminosity and redshift dependence of the obscuring fraction may  
be naturally linked to the AGN radiative power (related to the intrinsic 
X--ray luminosity, see e.g. Lawrence \& Elvis 1982)
which is able to ionize and expel gas (more common at high-z) and dust 
from the nuclear regions, nicely fitting the current 
framework of AGN formation and evolution sketched above (see, for example, 
Hopkins et al. 2006).
The complete picture is likely to be more complex, depending
on many other parameters such as the BH mass, the Eddington ratio, 
the star formation activity in the host galaxy and, in particular, 
the timescales associated with the AGN duty cycle activity. 

A correct and complete identification of unobscured, obscured and highly
obscured AGN at all redshifts (and especially in the z=1-3 interval, where
most of the feedback is expected to happen) is therefore crucial
for a comprehensive understanding of the still little explored phase of
the common growth of SMBHs and their host galaxies.
While it is relatively straightforward to select unobscured AGN from optical
multicolor surveys (e.g. Richards et al. 2002) and/or from spectroscopic samples (e.g. the VVDS survey,
Gavignaud et al. 2006, Bongiorno et al. 2007), the most efficient, reddening
independent method to select obscured, type--2 AGN is the presence of luminous X-ray emission
(L$_{[2-10keV]}>10^{42}$ erg s$^{-1}$) and hard X-ray colors
(see Brandt \& Hasinger 2005 for a review).
Indeed, hard (2-10 keV) X-ray nuclear emission is an almost unambiguous mark of 
the presence of an AGN, given the extremely small contamination from 
star-formation induced emission at these luminosities and frequencies. 
Combined with the fact that hard X-rays are not seriously 
affected 
by obscuration due to neutral gas along the line of sight
(up to N$_{\rm H}\sim10^{23}$ cm$^{-2}$), it is clear 
that hard X-ray selection is the most effective method to uncover 
unobscured to moderate obscured AGN 
and study in detail their demographics (see Comastri \& Brusa 2008; 
Brandt \& Alexander 2010 for a review).

Since the launch in 1999 of both the XMM--{\it Newton} and {\it Chandra} satellites,
a large ($>$ 30) number of surveys covering a wide fraction of the area vs. depth plane
(see Fig.~1 in Brandt \& Hasinger 2005; see also Hickox 2009) have been performed,
and our understanding of AGN properties and evolution has received a major boost.
Thanks to vigorous programs of multiwavelength follow-up campaigns,
sensitive X--ray observations turned
out to be highly efficient in unveiling  weak and/or "elusive" 
accreting black holes, in a variety of otherwise {\it non-active} galaxies
(i.e. not recognized as AGN from the optical spectra or continuum 
emission), such as (among others): 
X--ray Bright Optically Normal Galaxies (XBONG, e.g. Comastri et al. 2002,
Severgnini et al. 2003, Civano et al. 2007), Extremely Red Objects 
(e.g. Alexander et al. 2002, Brusa et al. 2005, Severgnini et al. 2005), 
Sub Millimeter Galaxies (e.g. Alexander et al. 2005, Laird et al. 2009), 
high-z starforming systems (e.g. Daddi et al. 2007, Fiore et al. 2008, 2009, 
Treister et al. 2009b), 
Lyman Break Galaxies (LBG, e.g. Brandt et al. 2002, Nandra et al. 2002, Aird 
et al. 2008).
In many of these cases, the AGN responsible for the X-ray emission is overwhelmed at longer wavelength
by the host galaxy light and/or the obscuration might be connected to processes within the host galaxy itself, 
such as the presence of dust lanes or starburst disks 
(see e.g. Martinez-Sansigre et al. 2005, Ballantyne et al. 2006, 
Ballantyne 2008, Hopkins et al. 2009). 
This suggests that the accretion activity (especially
in high-redshift sources) can be unambiguously revealed thanks to the presence of a strong X--ray
emission (see e.g. discussion in Brusa et al. 2009) and, therefore, the combination of both
X-ray and optical classifications, coupled with the multiwavelength analysis, 
can be crucial to fully assess the nature of the candidate AGN. 

The high level of completeness in redshift determination for a large number
of X--ray selected AGN (up to a few thousands) has made possible a robust determination
of the luminosity function  and evolution of unobscured and mildly obscured AGN
which turned out to be luminosity dependent:
the space density of bright QSOs ($L_X > 10^{44}$ erg s$^{-1}$) peaks at z$\sim$ 2--3,
to be compared with the z$\sim$0.7--1 peak of lower luminosity  Seyfert galaxies
(Ueda et al. 2003; Hasinger et al. 2005;  La Franca et al. 2005, Silverman et al. 2008; 
Ebrero et al. 2009, Yencho et al. 2009). 
Based on the Ueda et al. (2003) work, Marconi et al. (2004) and Merloni (2004) were 
the first to propose that SMBH undergo a ``anti--hierarchical" evolution, in the form of a differential 
growth (earlier and faster for more massive black holes).
This anti--hierarchical behavior observed in AGN evolution (similar to that observed in
normal galaxies, e.g. Cowie et al. 1996) provided an important and independent confirmation that
the formation and evolution of SMBH and their host galaxies are likely different aspects
of the same astrophysical process. It should be noted that the Luminosity 
Dependent Density Evolution (LDDE) parameterization, corresponding to a strong downsizing, 
has been recently questioned by Aird et al.(2010). Their preferred model for the evolution of the XLF is a 
luminosity and density evolution model (LADE), where the shift in the redshift peak 
of AGN space density as function of X--ray luminosity is much weaker than in LDDE models. 

A full characterization of the AGN bolometric luminosity function 
can be obtained only exploring the entire area vs. flux plane (e.g. 
combining samples from deep and large area surveys) and adopting different 
AGN selection methods (X--ray vs. optical vs. infrared).  
Nevertheless, it seems plausible that the blow-out phase  
associated with the obscured growth represents a 
relatively short, but very powerful, episode in the QSO lifetime.
For this reason, the probability to detect rare and luminous X--ray 
events is maximized by large area surveys and, in particular, by hard 
X--ray observations (to cope with obscuration effects) and associated 
deep multiwavelength coverage. 

The  XMM--{\it Newton} wide-field survey in the COSMOS field (hereinafter:
XMM-COSMOS, Hasinger et al. 2007) is an important step forward in addressing the
topics described above.
The $\sim2$ deg$^2$ area of the HST/ACS COSMOS Treasury program (Scoville et al. 2007a,b; 
Koekemoer et al. 2007) has been surveyed with \xmm\ for a total of $\sim$1.55 Ms during AO3, AO4 and
AO6 cycles of XMM observations (Cappelluti et al. 2007, 
Cappelluti et al. 2009, hereafter C09).
XMM-COSMOS provides an unprecedently large sample of point-like X-ray
sources ($\gsimeq1800$), detected over a large, contiguous area, with {\it
complete} ultraviolet to mid-infrared (including Spitzer data) and radio coverage, and
extensive spectroscopic follow--up granted through the zCOSMOS
(Lilly et al. 2007; 2009) and Magellan/IMACS (Trump et al. 2007; 2009) projects.
The excellent multiband photometry available in this area allows a robust 
photometric redshift estimate for the faint sources
not reachable by optical spectroscopy, thus allowing a virtually complete
sample of X-ray sources. 
The XMM-COSMOS project is described in Hasinger et al. (2007),
while the X--ray point source catalog and counts from the complete XMM-COSMOS survey
are presented in a companion paper (C09).
The present paper, which extends the work presented by Brusa et al. 2007
(hereiafter B07) on the optical identifications of the X--ray point sources
in the XMM-COSMOS survey, discusses the multiwavelength
properties of this large sample of X-ray selected AGN.
Several works have already appeared in the literature and are based on 
this catalog, or on previous versions of it, e.g.:
the derivation of AGN photometric redshifts (Salvato et al. 2009);
the IMACS AGN spectroscopic campaign (Trump et al. 2009);
the clustering properties of spectroscopically confirmed AGN (Gilli et al. 2009);
the space density of high-redshift QSOs (Brusa et al. 2009);
the host galaxies properties of AGN in COSMOS (Gabor et al. 2009);
the on-going and co-evolving starformation and  AGN activity at z$<0.8$ (Silverman et al. 2009a);
the environments of AGN within the galaxies density field (Silverman et al. 2009b);
the relation of the X--ray and optical emission in BL AGN (Lusso et al. 2010). 

The paper is organized as follows:
Section~2 presents the multiwavelength datasets drawn from the COSMOS survey and
used in the paper; Section 3 describes the method used to identify
the X-ray sources and its statistical reliability; the XMM-COSMOS multiwavelength
catalog is presented in Section~4, while Section~5 reports the redshift information
for the X-ray sources. The number density evolution of luminous XMM-COSMOS  sources 
is presented in
Section~6. Section~7 describes the X--ray to optical and near infrared properties
of the identified population and of the obscured AGN candidates, while in 
Section~8 we discuss the obscured AGN fraction as a function of the X--ray
luminosity. 
Section~9 presents the spectrum and SED of a prototype obscured QSO at z$\sim1.6$,
and in Section 10 we summarize the most important results.
Throughout the paper, we adopt the cosmological parameters $H_0=70$ km s$^{-1}$
Mpc$^{-1}$, $\Omega_m$=0.3 and $\Omega_{\Lambda}$=0.7 (Spergel et al. 2003).
In quoting magnitudes, the AB system will be used, unless otherwise stated. 

\section{Multiwavelength Datasets} 

\subsection{X--ray}

The catalogue used in this work includes 1848 point--like sources 
above a given threshold with a maximum likelihood detection algorithm in at
least one of the  soft (0.5--2 keV), hard (2--10 keV) or ultra-hard (5--10
keV) bands down to nominal limiting fluxes of $\sim$ 5$\times 10^{-16}$, 
$\sim$ 3$\times 10^{-15}$  and $\sim$ 7$\times 10^{-15}$ erg cm$^{-2}$ 
s$^{-1}$, respectively (i.e., the flux of the faintest source
detected in the band, see C09).
The adopted likelihood threshold corresponds to a probability $\sim
4.5\times10^{-5}$ that a catalog source is a spurious background fluctuation
(see Cappelluti et al. 2007 and C09 for more details). 
In the present analysis we used the source list created from 53 out of the 
55 XMM-COSMOS fields; for the additional 65, faint sources detected 
when the pointings obtained in AO6 are included, the identification
is not completed yet.
For this reason, the number of XMM-COSMOS sources is slightly lower than that discussed in C09. 
Twenty-six faint sources in our catalog are coincident with diffuse 
XMM sources in the catalog by Finoguenov et al. (2009). 
These sources are flagged as ``possibly extended'' in the pointlike catalog 
and are excluded from the following analysis.   
The inner part of the COSMOS field has been imaged for a total of 
1.8 Ms by {\it Chandra} (Elvis et al. 2009), with 36 pointings 
of $\sim50$ ks each in a 6x6 array. The mosaic covers an area of 
$\sim 0.92$ deg$^2$ (about half of the XMM-COSMOS field) 
down to a limiting flux of $\sim 2\times10^{-16}$ erg cm$^{-2}$ s$^{-1}$ 
in the soft band and $\sim 7.3\times10^{-16}$ erg cm$^{-2}$ s$^{-1}$ 
in the hard band, i.e. 3-4 times deeper than XMM-COSMOS.  
Of the 1822 XMM-sources, 945 (51.9\%) have been observed by \chandra\
with an exposure larger than 30 ks, and  875 of them are present in the 
C-COSMOS point-like source catalog (Elvis et al. 2009, Puccetti et al. 
2009). 
Of the 70 sources not recovered by \chandra, more than half are in regions 
with relatively low exposure (between 30 and 50 ks) and are detected at
faint XMM fluxes, mostly in the hard band ($\ls 6\times10^{-15}$ \cgs). 
The remainder are either sources with only hard XMM detections (14) or, 
after a visual inspection, they can be associated with spurious 
sources (15), 
consistent with the expected fraction of spurious sources in the XMM-COSMOS 
field (see Elvis et al. 2009). 
Twenty-five of the 875 XMM-COSMOS sources with \chandra\ detection (2.8\%) 
are {\it resolved} in two different \chandra\ sources, 
lying at distances between 2 and 10 arcsec from each other and 
therefore being likely blurred in the
XMM large PSF (see discussion in C09\footnote{The twenty-five XMM-COSMOS 
sources with two \chandra\ counterparts are listed here for completeness. 
XID numbers refer to column 1 in the Cappelluti et al. 2009 catalog table: 
\#35, \#82, \#131, \#208, 
\#215, \#307, \#336, \#354, 
\#365, \#380, \#384, \#419, 
\#2591, \#2618, \#5141, \#5208, 
\#5355, \#5556, \#10764, \#31163, 
\#53328, \#54468, \#60133, \#60275, \# 5210}). 
We will further discard these 25 sources in the following analysis. 
The XMM-COSMOS sample presented in this catalog consists 
therefore of 1797 X-ray sources, 850 with \chandra\ detection (47.3\%). 
In the following, we use the subarcsec accurate
\chandra\ positions to control-check the optical/NIR identifications proposed in
Section 3, and to assess the reliability of the proposed identifications. 

\subsection{Optical, near infrared and {\it Spitzer} photometry} 

As we will describe in the next Section, 
the XMM-COSMOS catalog has been cross-correlated with an updated version of 
the optical multiband catalog of Capak et al.~2007
(``optical catalog'' hereafter), the CFHT/K band catalog 
(Mc Cracken et al. 2010, ``K-band catalog'' hereafter), the IRAC  catalog 
(Sanders et al. 2007, Ilbert et al. 2009, 
``IRAC catalog'' hereafter) and the 24 $\mu$m 
MIPS catalog (Le Floc'h et al. 2009, ``MIPS catalog'' hereafter). 
The optical catalog\footnote{publicly 
available at: \\
 http://irsa.ipac.caltech.edu/data/COSMOS/tables/ib. \\ 
The detection image, I band mask, and SExtractor settings for the optical
catalog were significantly modified from the version presented in Capak 
at al~2007, which is now superseeded.
All point-like sources brighter than 17th magnitude were modeled and 
subtracted from the detection image. This allowed the de-blending 
contrast to be set at a higher value, thus significantly decreasing 
the number of spurious sources without missing objects around bright 
sources and significantly reducing the number of close pairs blendings.
In addition to lowering the contrast parameter, several of the detection
parameters were adjusted to decrease the number of noise peaks at 
the faint end of the catalog.
The combination of the change in background, de-blending, and detection
results in small changes in the centroid and flux for some sources.} 
 contains about 1.5 million objects detected in at least 
one of the Subaru bands (b,v,g,r,i,z) down to an AB magnitude limit 
of $\sim27$.
The K-band catalog contains about 5$\times10^5$ galaxies detected
at a S/N$>$5 down to K(AB)=23.5 (see details in Mc Cracken et al. 2010). 
The IRAC catalog contains about 4$\times10^5$
objects detected in the 3.6 micron (IRAC channel 1) band and it is 90\%
complete at $>1 \mu$Jy (AB=23.9). For each source in the IRAC catalog,
the photometry from all the other IRAC channels is also reported. 
The MIPS catalog, obtained in Cycles 2 and 3, has very accurate photometry
(Sanders et al. 2007). The catalog has been cleaned of spurious sources (mostly
asteroids, Le Floc'h et al. 2009) and contains $\sim 50000$ sources. 
 In the XMM-COSMOS area there are $\sim36000$ sources detected with a signal to noise
$>5$ (implying  a 24$\mu$m flux limit of 80$\mu$Jy). 
We also matched the X-ray counterparts with the 70 $\mu$m MIPS catalog 
(Frayer et al. 2009, Kartaltepe et al. 2010) which contains $\sim1500$ sources down to 
$\sim4$ mJy, and with the VLA COSMOS catalog ($\sim 2400$ sources at a limiting
flux of $\sim 50\mu$Jy/beam , Schinnerer et al. 2007, Bondi et al. 2008).

\section{Identification of X-ray sources}

We have matched a counterpart to each of the 1797 X-ray sources
as follows. 
\begin{itemize}
\item[$\bullet$] First, we associated the X-ray positions to the optical
ones (I-band) using the statistical method described in B07 
(the ``likelihood ratio technique'', Sutherlands \& Saunders 1992, Ciliegi et al. 2003) 
in order to isolate the most obvious associations, and, at the 
same time, to pick up problematic cases (i.e. sources with 2 or more
different counterparts with comparable likelihood or sources too faint to be
identified). 
The method calculates the probability 
that a source is the correct association by weighting the information
on the X--ray to optical distance, the surface density of (possible)
false coincidence background objects and the brightness of the 
chosen counterpart (see B07 for other details). 
We divided the sources in three different classes:
``reliable id'', ``ambiguous id'' (comprising mostly 2 relatively
bright sources with similar probability of being the correct optical counterpart) 
and ``not identified'' (comprising mostly faint or undetected sources for
which the association is not statistically significant). Roughly, 
the 1797 sources were split into the three classes with the following
percentages: 80\%, 10\%, 10\%, respectively, in agreement with the analysis
reported in B07 that was limited to a subsample ($\sim 700$ sources, 40\%) 
of the complete XMM-COSMOS sample. 

\item[$\bullet$] Then, we cross-correlated the optical positions with the K-band and IRAC catalogs. 
We created I-band (using ACS data in the inner 1.7 deg$^2$ 
area, the Subaru I-band data elsewhere), K-band and IRAC (3.6$\mu$m and
8.0$\mu$m)  20$''$x20$''$ cutouts around each of the 1797 X-ray positions and
visually checked the correctness of the optical/IR matches. 
Most of the ``not identified'' sources turned out to be associated
to bright, isolated K-band or IRAC counterparts and therefore were moved
into the ``reliable id'' sample. However, the diversity of the SED of 
objects in the sky led a large number of sources showing up in the
IR, adding ambiguity to some of the proposed ``reliable" associations. 
The distribution of the sources in the three classes after the
correlation with the K-band and IRAC catalogs is: 1458 sources (81.3\%)
in the ``reliable ID'' class, 319 sources (17.7\%) in
the ``ambiguous ID'' class and 20 sources (0.9\%) in the ``not identified''
class.   
\end{itemize}

\subsection{Chandra-XMM matches and position} 

{\it Chandra} subarcsecond accurate X-ray positions are available for 850 of the 1797
XMM-COSMOS sources (see Sect 2). Of these 850 objects, 712 were in the
``reliable ID'' class (83.7\%), 135  (15.9\%) were in the 
``ambiguous ID'' class, and 
the remaining 3 were ``not identified'' (0.4\%). The lower
  percentage of not identified sources in the \chandra\ detected subsample 
  can be due to the fact that many of the objects in this class might be
  spurious XMM sources and therefore not even detected by \chandra\
  (see also discussion in Elvis et al. 2009). Indeed, many of the sources
without an optical id have a faint XMM flux, close to the detection limit.

Of the 712 objects in the class of ``reliable ID'', \chandra\ pointed
to a counterpart different from the one proposed in 12 cases (1.7\%,
Figure 1, upper left panel).
This fraction is lower for the brightest (flux limited, see below) sample 
(1.3\%) and rises up to$\sim6$\% for the faintest sources. 
Among the 135 sources in the ``ambiguous ID'' class, in 114/135
cases (84.4\%) \chandra\ points to one of the two proposed associations 
(Figure 1, upper right panel), choosing the primary identifications in
50\% of the cases and the secondary in the remaining half. 
For the remaining 21 objects (15.6\%) the \chandra\ information
was not good enough to solve the ambiguity 
in the identification (Figure 1,
lower left panel). 
Finally, all of the 3 ``not identified'' XMM sources were associated with 
a \chandra\ source at a large distance ($>5''$) from the XMM position
and coincident with a relatively bright optical counterpart (Figure 1, 
lower right panel). 
%%%%%%%%%%%%%%%%%%%%%%%%%%%%%%%%%%%%%%%%%%%%%%%%%
\begin{figure*}[!t]
\includegraphics[width=8cm]{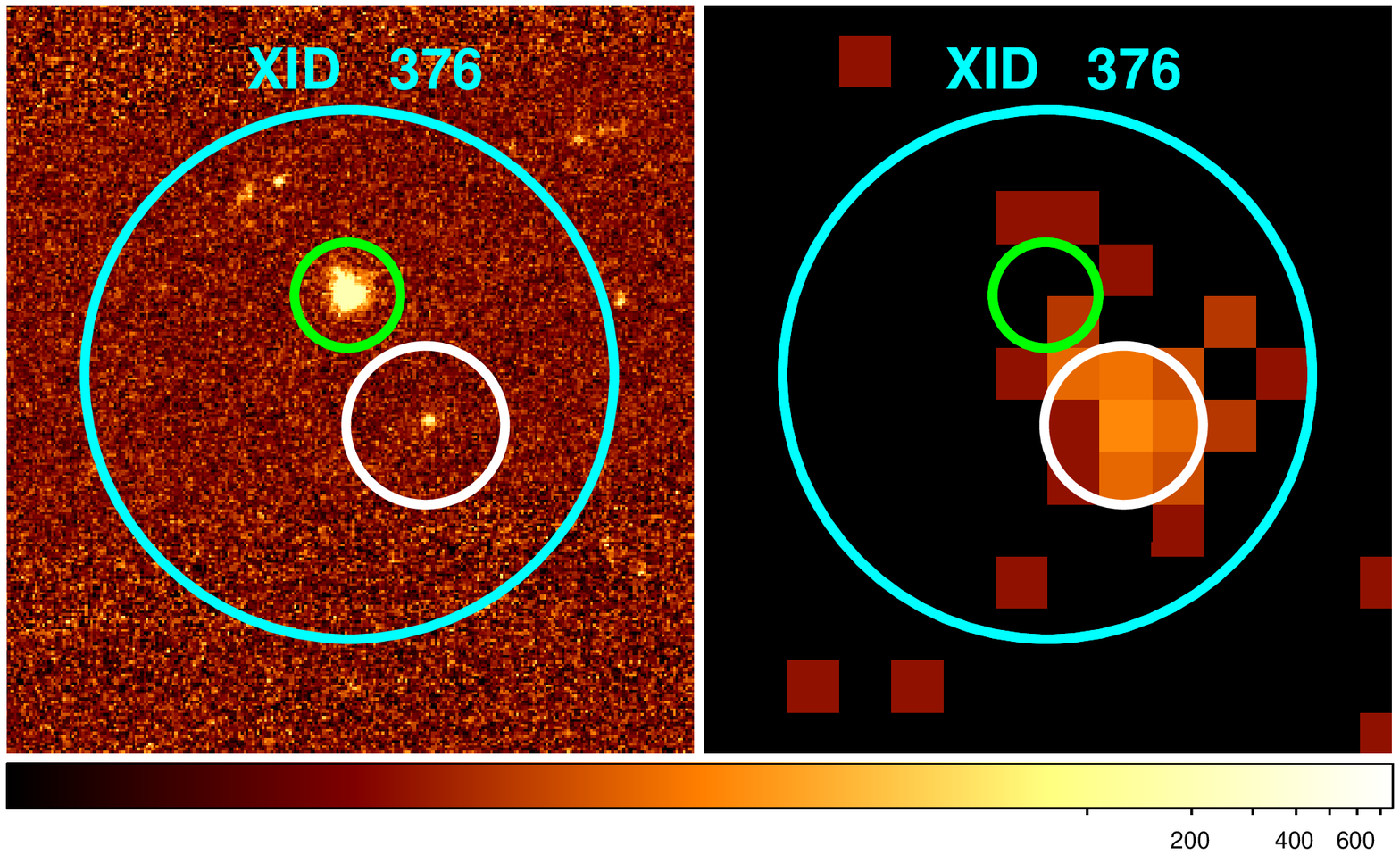} 
\includegraphics[width=8.cm]{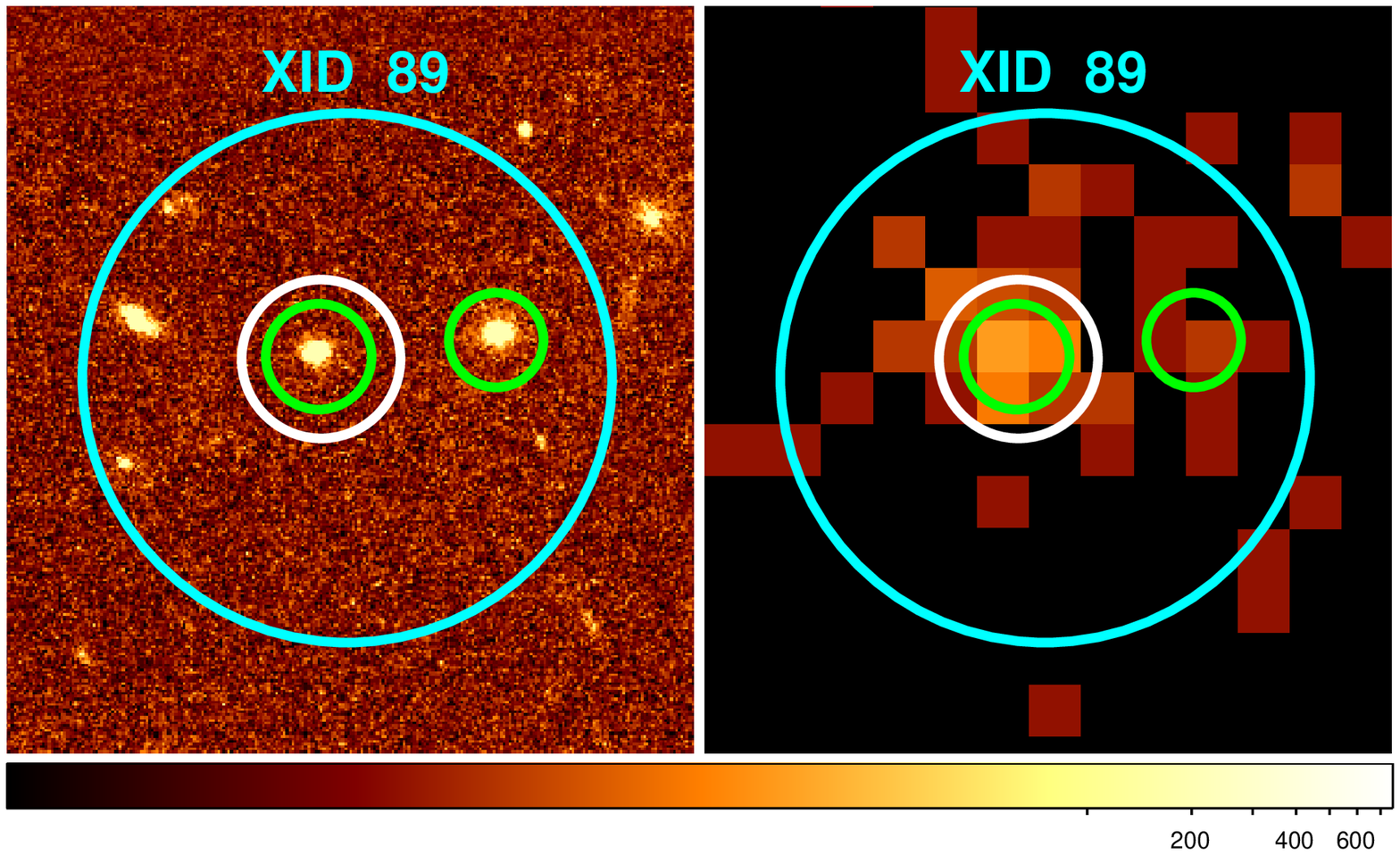} 
\includegraphics[width=8cm]{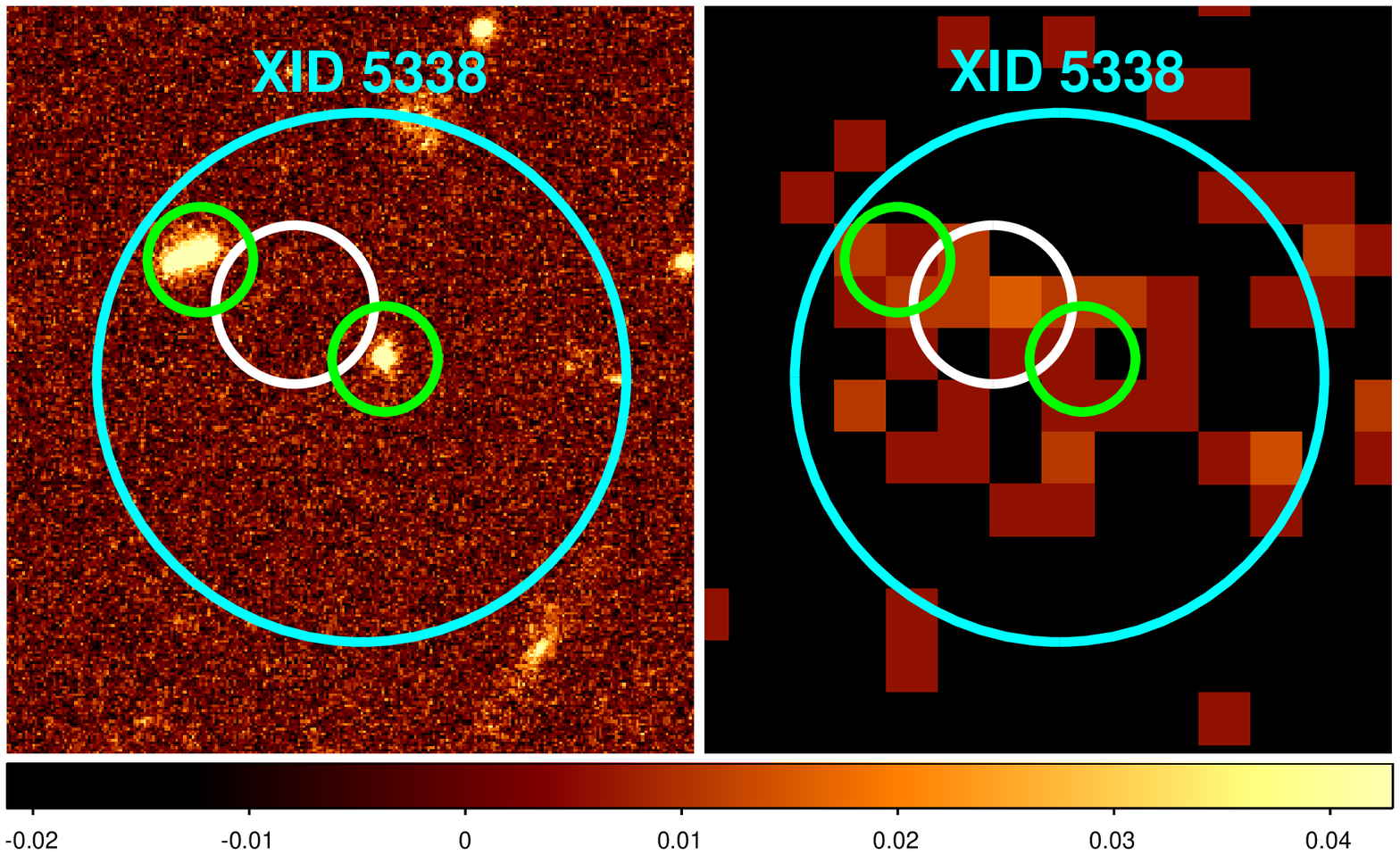}
\hspace{0.2cm}
\includegraphics[width=8cm]{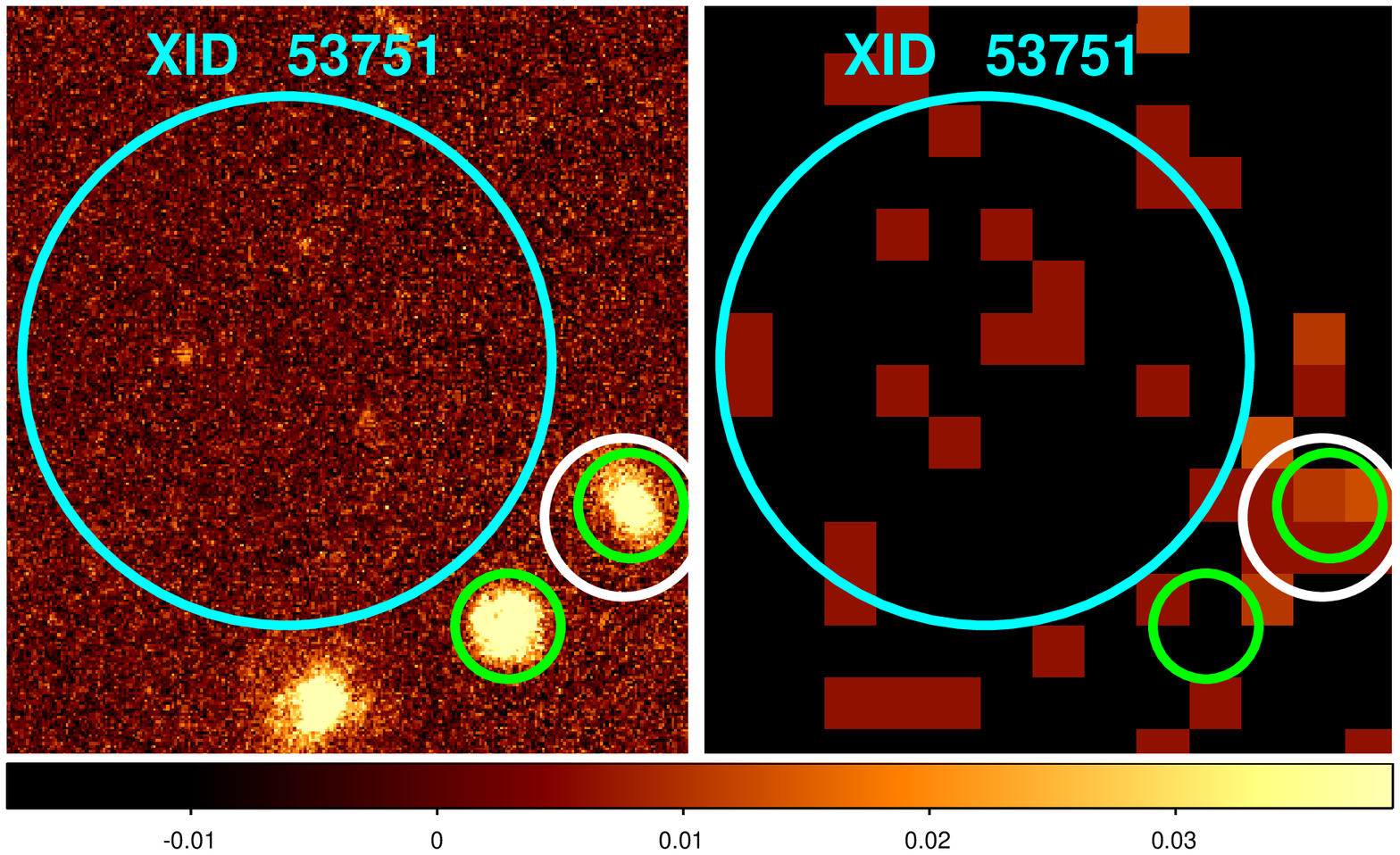}\\
\caption{HST/ACS cutouts showing four examples of the use of C-COSMOS 
data in the process of identifying 
XMM-COSMOS sources (see Section 3.1). 
({\it upper left}): the C-COSMOS position pointed to a counterpart different 
from the proposed (``reliable") one; ({\it upper right}): the C-COSMOS position 
pointed to one of the possible counterparts (``ambiguous" class); ({\it lower
left}): the C-COSMOS position doesn't help in choosing a counterpart
in the ambiguous class; ({\it lower right}): the C-COSMOS position  
helped in choosing the correct counterpart for sources in the
``not identified" class. In all four examples shown, the left image is 
the ACS cutout and the right image is the \chandra\ cutout. The size 
of the cutouts is $\sim 12"$. In all the 
cutouts, the XMM position is marked by the large, 5" radius cyan circle;
the \chandra\ position by a smaller, 1.5" radius white circle; 
and the candidate counterpart(s) from the XMM alone analysis 
with a green, 1" radius, circle. {The scale bars below the upper cutouts 
represent a logaritmic scale of the X--ray counts, while, as a comparison, 
the scale bars below the bottom panels represent the scale of the HST/ACS 
emission.}}
 \label{id}
 \end{figure*}
%%%%%%%%%%%%%%%%%%%%%%%%%%%%%%%%%%%%%%%%%%%%%%%%
We incorporated the \chandra\ information in our ID process
and, at the end, we were able to reliably identify 1577 XMM sources
(87.7\%); we still classify as ``ambiguous ID'' 203 sources (11.3\%) and we 
could not identify 17 sources (1.0\%). 
If we use the 1.7\% of wrong identifications obtained above as an estimate 
of the failure rate of the likelihood ratio technique applied to the XMM-COSMOS
data, we estimate that  $\sim$12-13 sources out of the 747
sources in the ``reliable ID'' class without \chandra\ coverage
can still be associated with a wrong counterpart.  
 
To maximize the completeness over a well defined large area and, at the same
time, keep selection effects under control, we further defined 
a subsample of sources detected 
above the limiting fluxes corresponding to a sky coverage of more than 1 deg$^{2}$
in at least one  of the three X--ray energy ranges considered, namely:
1$\times 10^{-15}$ \cgs, 6$\times 10^{-15}$
\cgs, and 1$\times 10^{-14}$ \cgs, in the 0.5--2 keV, 2--10 keV or 5--10 keV
bands, respectively (see also Table 2 in C09). 
This subsample includes 1651 X--ray sources, that will be used in the following
to investigate the multiwavelength properties of XMM counterparts (see Section 7).  
The combination of area and depth is similar to the one of the sample 
studied by Silverman et al. (2008, see their Fig. 8) in the Chandra Multiwavelength 
Project survey (ChaMP, Kim et al. 2007); the main difference is given by the considerably higher redshift
completeness obtained in COSMOS thanks to the much deeper
coverage in the optical and near IR bands and the systematic use of
photometric redshifts (see below). This drastically limits 
the need for substantial corrections  for incompleteness (see discussion in Silverman et al. 2008). 
When imposing the limiting flux thresholds, the breakdown of the objects in the three
classes is as follows: 1465 ``reliable ID'' (88.7\%), 175 ``ambiguous'' 
(10.6\%) and 11 (0.7\%) ``not identified''. 
A summary of  the identification process and the breakdown in the
classes of ``reliable ID'', ``ambiguous ID'' and ``not ID'', before and after
the \chandra\ control-check, and before and after the inclusion of thresholds in X-ray fluxes is
reported in Table 1.   

%%%%%%%%%%%%%%% TABLE 1 %%%%%%%%%%%%%%%%%%%%%%%%%%%%%%%%%%%%%%
%
\begin{deluxetable}{rccccc}
\tablecaption{Summary of optical identification of point-like XMM sources}
\tablehead{
\colhead{sample} & \colhead{Total sources} & \colhead{Reliable$^a$ (\%)} &
\colhead{Ambiguous$^b$ (\%)} & \colhead{unidentified (\%)}}
\startdata
\chandra\ area & 850 &  712 (83.8\%) & 135 (15.9\%) & 3 (0.3\%) \\
(before \chandra\ check) & & & & \\
\chandra\ area & 850 &  829 (97.5\%) & 21 (2.5\%) & 0 (0.\%) \\
(after \chandra\ check) & & & & \\
\hline 
XMM-COSMOS area & 1797 & 1458 (81\%) & 319 (18\%) & 20 (1.0\%)\\ 
(before \chandra\ check) & & & & \\
\hline
XMM-COSMOS area & 1797 & 1577 (87.7\%) & 203 (11.3\%) & 17 (1.0\%)\\ 
(after \chandra\ check) & & & & \\
\hline
XMM-COSMOS area & 1651 & 1465 (88.7\%) & 175 (10.6\%) & 11 (0.7\%)\\ 
(after flux thresholds) & & & & \\
\hline
\enddata
\tablenotetext{a}{We classified as ``reliable'' those sources for which the multiwavelength
analysis consistenly point to the same counterpart, expected to be correct in
98\% of the cases (see Section~3 for details).}  
\tablenotetext{b}{We classified as ``ambiguous'' those sources for which 
the multiwavelength analysis didn't allow us to reliably associate a unique counterpart
to the X--ray sources; for these objects, two sources with comparable likelihood 
are present in the X--ray error-box. See Section~3 for details.}  

\end{deluxetable}
%
%%%%%%%%%%%%%%% TABLE 1 %%%%%%%%%%%%%%%%%%%%%%%%%%%%%%%%%%%%%%

\section{The XMM-COSMOS multiwavelength catalog} 

The photometry at different wavelengths (except MIPS) has been cross-correlated
with the positions of the XMM-COSMOS counterparts (see Section 3) 
using a 0.5$\arcsec$ radius, and all the matches have been visually
inspected, in order to remove false associations or include
obvious matches at larger distances ($>0.5$"). A significant number 
of sources ($\sim 10$\%) of the XMM-COSMOS counterparts turned out to
be associated with blended objects in the IRAC catalog (e.g. the IRAC source
is a clear blend of two different objects). Therefore, the photometry 
in the four different IRAC bands has been re-extracted at the position of the 
XMM-COSMOS counterparts, using a PSF fitting routine, allowing us a  
better determination of the fluxes for these blended IRAC sources. 
The MIPS catalog has been cross-correlated with the IRAC positions
of the XMM-COSMOS counterparts using a radius of 2.5$\arcsec$ and 
all the matches have been validated through a visual inspection. 
At the end, 1760 (98\%) sources in the XMM-COSMOS sample 
have associated entries from the ``optical catalog", 1745 (97\%)
have photometry from the ``K-band catalog", 1711 (95\%) sources have IRAC 3.6$\mu$m 
photometry  retrieved from the ``IRAC catalog" or re-extracted at the position of
the optical counterparts, and 1394 ($\sim 78$\%) have 24$\mu$m MIPS counterparts. 
Only two sources (XID 5120 and XID 556) have no associated photometry
in any of the optical or infrared catalogs used. Both of them are
close to bright stars (I$<16$). 

Table 2 lists the basic properties of the complete XMM-COSMOS counterpart sample:
XMM-COSMOS IAU designation (column 1); 
the XMM-COSMOS identifier number (from C09, column 2); 
the coordinates of the optical/IR counterpart (J2000, column 3 and 4); 
the X--ray fluxes in the soft, hard and ultra-hard bands (from C09; columns 5,6, and 7, respectively); 
a flag identifying the sources included in the flux-limited sample (column 8); 
the X--ray hardness ratio, HR, defined as HR=(H-S)/(H+S) where H are the hard band counts and S the soft band counts, respectively (column 9); 
the Chandra-COSMOS identifier number (from Elvis et al. 2009, column 10); 
the flag for the optical identification, according to the classes described in Table 1 (column 11); 
the identifier number from Capak et al. (2007) catalog\footnote{The Capak et al. (2007) photometric catalog is available at IRSA, at the link: http://irsa.ipac.caltech.edu/data/COSMOS/tables/photometry/cosmos\_phot\_20060103.tbl.gz} (column 12); 
the identifier number as reported in the optical catalog and in Ilbert et al. (2009) (column 13); 
the r-band and I-band magnitudes (from Capak et al. 2007; column 14 and 15);
the K-band magnitude (from Mc Cracken et al. 2010; column 16); 
the magnitudes in the four IRAC channels (from Ilbert et al. 2009; columns 17-20); 
the MIPS 24$\mu$m magnitude (from Le Floc'h et al. 2009; column 21); 
the spectroscopic redshift (see Section 5; column 22); 
the spectroscopic classification according to the classes outlined in Section 5.1 (column 23); 
the origin of the spectroscopic redshift, with relevant notes (column 24); 
the photometric redshift (from Salvato et al. 2009; column 25).
In case of ambiguous sources, all the  possible counterparts of the XMM-COSMOS sources 
are listed. 
Table 2 is also available in ASCII format at the link:
http://www.mpe.mpg.de/XMMCosmos/xmm53\_release/.

The distribution of the X-ray to optical distances 
and of the I-band magnitudes for the XMM-COSMOS counterpart sample,
derived considering the first counterpart listed  in Table 2 for the 
203 sources with an ``ambiguous" counterpart, are very similar to those 
reported in B07. More specifically, about 90\% of the reliable optical counterparts have an 
X--ray to optical separation ($\Delta(X-O)$) smaller than $3''$ which is 
better than the $\sim 80$\% found 
in XMM--{\it Newton} data of comparable depth (see 
e.g. Fiore et al. 2003, hereinafter F03, Della Ceca et al. 2005, Loaring et al. 2005). 
The improvement is likely given by
the combination of the accurate astrometry of the positions of the X-ray
sources (tested with simulations in Cappelluti et al. 2007), the full
exploitation of the multiwavelength catalog (optical and infrared) for the
identification process, 
and the availability of {\it Chandra} accurate positions. 
The majority of the XMM-COSMOS counterparts ($\sim$68\%) have optical 
(I$_{\rm  AB}$) magnitudes in the range $20<I_{\rm AB}<$24, with 
a median magnitude of I$_{\rm AB}=21.98$ (with dispersion of 1.35), and two symmetric 
tails at fainter and brighter magnitudes. 
The median magnitude of optical counterparts is  slightly fainter than, 
but still consistent with, the median value presented by B07. 
%%%%%%%%%%%%%%%%%%%%%%%%%%%%%%%%%%%%%%%%%%%%%%%%%
\begin{figure*}[!t]
\begin{center}
\includegraphics[width=5.cm]{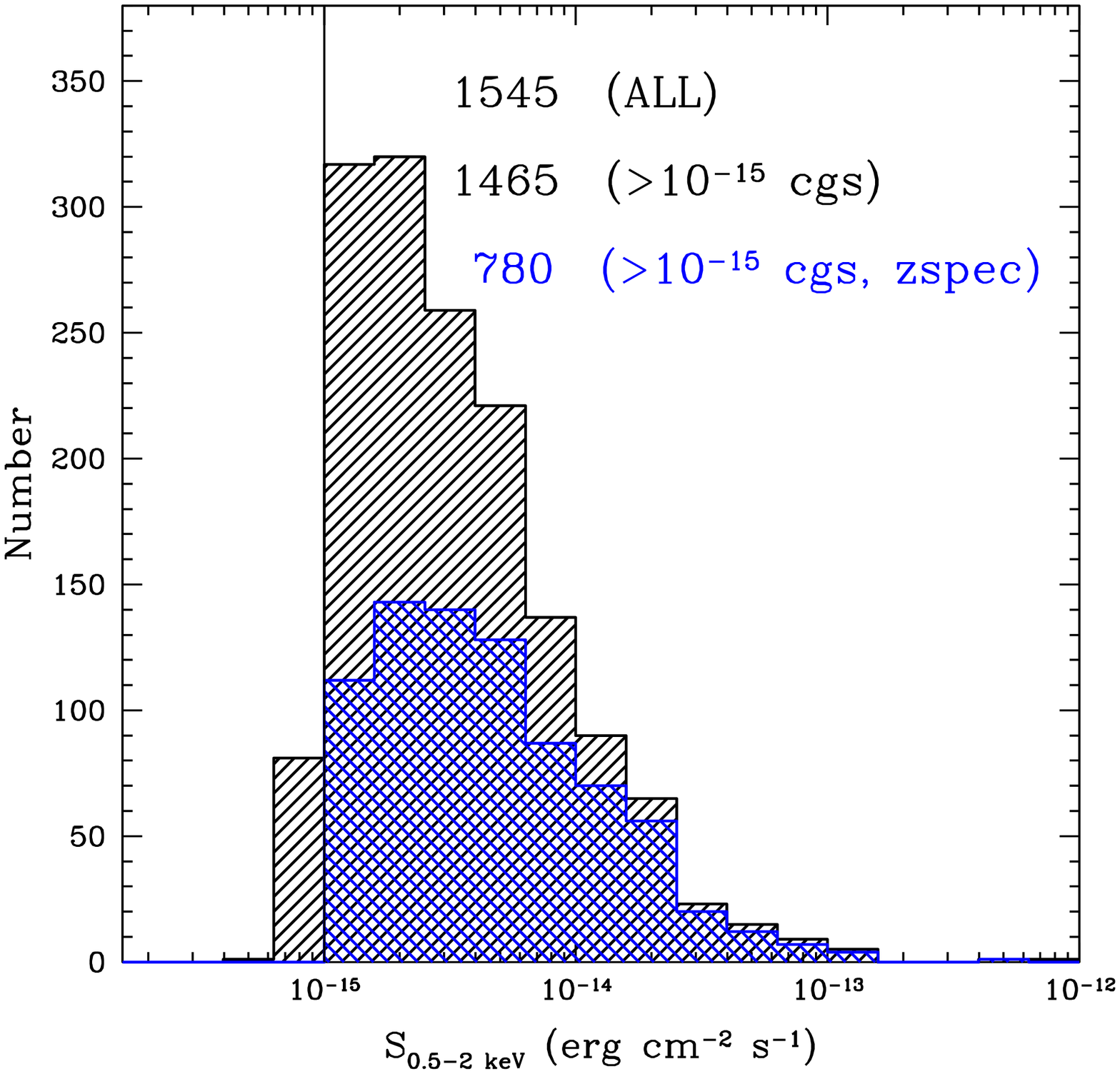} 
\includegraphics[width=5.cm]{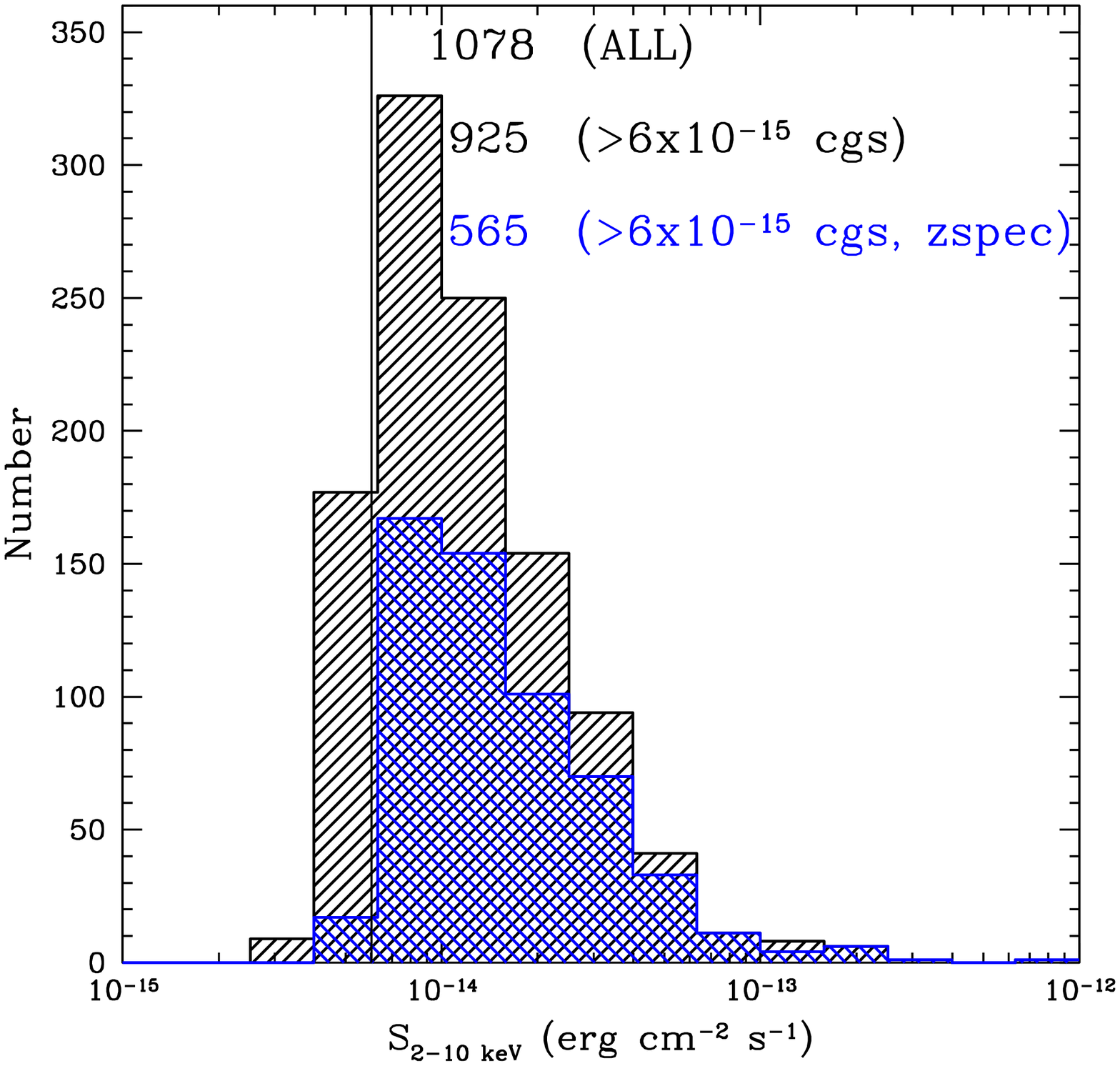} 
\includegraphics[width=5.cm]{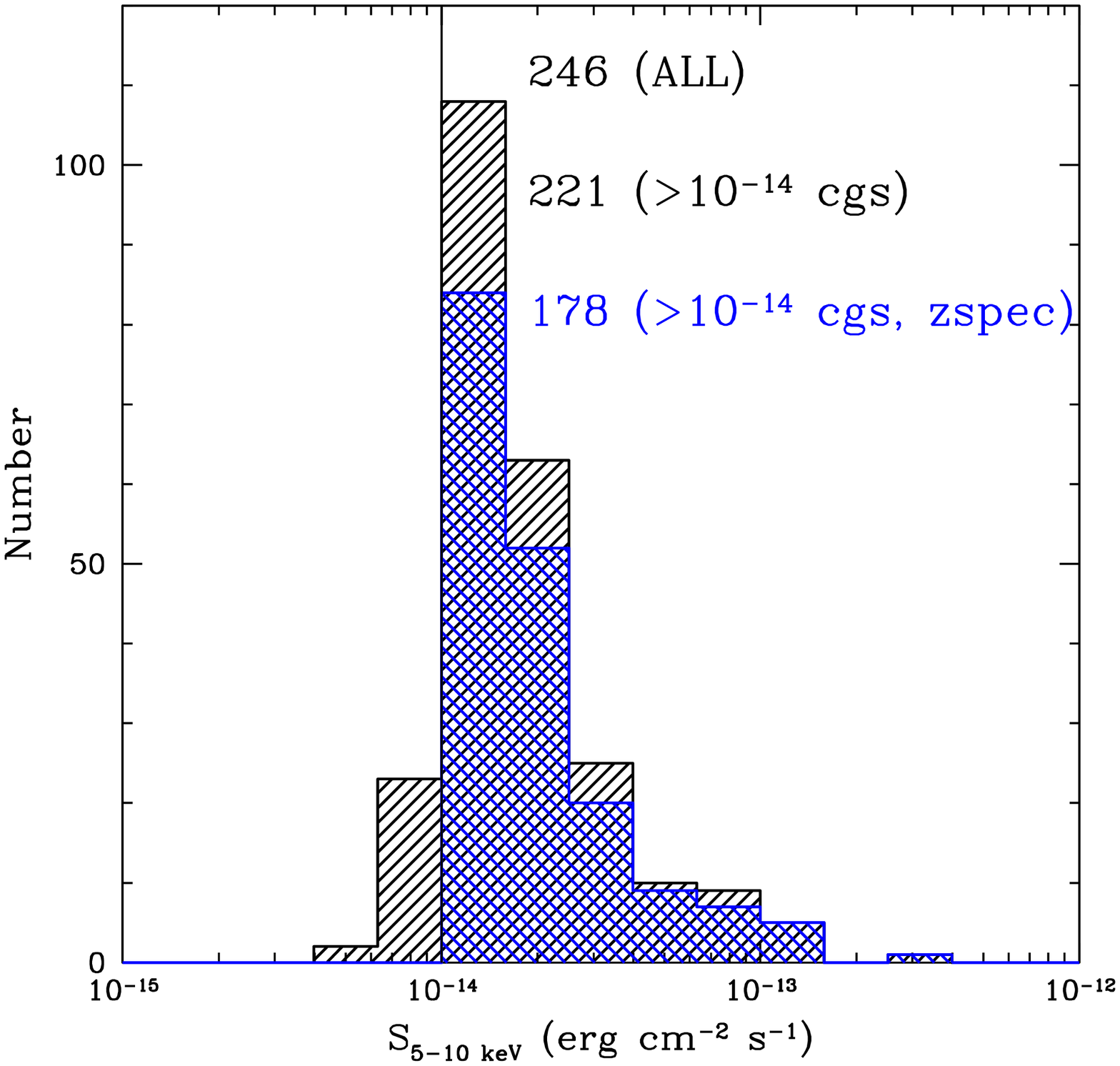}
\caption{Histograms of the flux distributions for the soft (0.5-2 keV, 1545 sources; left), 
  hard (2-10 keV, 1078 sources; middle) and ultra-hard (5-10 keV, 246 sources; right) samples.
  The vertical line shows the flux at which the sky coverage is 1 deg${^2}$
   (with 1465, 925 and 221 soft, hard, and ultra-hard sources
   respectively). 
   The blue histograms show the flux distributions for the subsamples  of
   sources with spectroscopic redshift, including 780, 565 and
   178 sources in the soft, hard and ultra-hard band, at fluxes higher than the
   adopted thresholds, respectively. }
 \label{fluxhisto}
\end{center}
 \end{figure*}
%%%%%%%%%%%%%%%%%%%%%%%%%%%%%%%%%%%%%%%%%%%%%%%%
The above mentioned distributions do not change if we make a different 
choice for the optical counterpart  of the 203 ``ambiguous''  sources, i.e. 
the second most likely optical  counterpart is considered  instead. 
As an example, the median optical  magnitude would be I$_{\rm AB}=21.96$ (with 
dispersion 1.36).   The same applies to other quantities (e.g. K-band magnitudes,
photometric redshifts etc.). 
Therefore, from a statistical  point of view, the multiwavelength properties
discussed in this paper can be considered representative of the overall X-ray
point source population, at the sampled X--ray fluxes. 

\section{Spectroscopic and photometric redshifts and source breakdown} 

Good quality spectroscopic redshifts for the proposed counterparts are available from 
Magellan/IMACS and MMT observation campaigns ($\sim 530$ objects, Trump et al. 2007, 2009), 
from the VIMOS/zCOSMOS bright project ($\sim 500$ objects, Lilly et al. 2009), 
from the VIMOS/zCOSMOS faint project ($\sim 80$ objects, Lilly et al. 2007) 
or were already present either in the SDSS survey catalog 
($\sim 100$ objects, Adelman-McCarthy et al. 2005, Kauffman et al. 2003\footnote{These 
  sources have been retrieved from the NED, NASA Extragalactic Database and
  from the SDSS archive.})
or in the literature ($\sim 95$ objects, Prescott et al. 2006).
Additional $\sim$ 40 objects have redshifts from ongoing spectroscopic
campaigns designed to target high-redshift, faint objects with the 
DEIMOS instrument at KeckII telescope (PI: P. Capak, M. Salvato, N. Scoville).
In summary, a total of 890 unique, good-quality spectroscopic redshifts are
available for XMM-COSMOS sources, 852 when the XMM-COSMOS flux limited sample  
is considered, corresponding to a substantial fraction ($\sim 52$\%, 
852/1651) of the sample investigated in this paper.

Figure~\ref{fluxhisto} shows the flux distributions in the soft, hard and 
ultra-hard bands, respectively, for the entire X--ray source population
(black histograms) and for the subsample with spectroscopic identifications 
(blue histograms). The vertical line shows the flux at which the sky coverage 
is 1deg${^2}$ (with 1465, 925 and 221 soft, hard, and ultra-hard sources
respectively). The spectroscopic completeness increases with the energy
band considered: it is $\sim53$\% (780/1465) in the soft band, $\sim61$\% 
(565/925) in the hard band, while the highest percentage
is reached in the ultra-hard band (178/221, $\sim80$\%).  
The high spectroscopic completeness in the ultra-hard sample reflects 
the fact that the sources are detected at bright fluxes and, therefore, 
associated  on average to bright optical counterparts 
(median I$_{\rm AB}\sim20.7$, to be compared to the median magnitude of 
the entire sample, I$_{\rm AB}\sim22$, see Section 4) therefore favoring 
the spectroscopic follow-up.

Photometric redshifts for all the XMM--COSMOS sources have been obtained 
exploiting the COSMOS multiwavelength database and are presented in
Salvato et al. (2009). 
Since the large majority of the XMM--COSMOS sources are AGN, in addition to the standard photometric
redshift treatments for normal galaxies, a new set
of SED templates has been adopted, together with a correction for long--term
variability and  luminosity priors for point-like sources (see Salvato et
al. 2009 for further details). 
Thanks also to the availability of the intermediate band {\it Subaru} filters
(Taniguchi et al. 2007), crucial in picking up emission lines 
(see also Wolf et al. 2004), we were able, for the first time for an 
AGN sample, to obtain photometric redshift  accuracy comparable to that 
achieved for inactive galaxies ($\sigma_{\Delta z/(1+z)} \sim 0.015$ and 
$\sim 5$\% outliers) at I$\lsimeq$22.5. 
At fainter magnitudes (22.5 $<I<24.5$) the dispersion
increases to $\sigma_{\Delta z/(1+z)} \simeq 0.023$ with $\sim$ 10\% 
outliers, still remarkably good for an AGN sample. 
A photometric redshift is available for all but 31 (24) objects out of 
1797 (1651) objects in the complete (flux-limited) XMM-COSMOS counterpart sample.
About half of them do not have wide multiband photometry, being detected only
in the IRAC and K bands. The remaining half are affected by severe
blending problems in the IR  making the photo-z estimate unreliable. 
%%%%%%%%%%%%%%%%%%%%%%%%%%%%%%%%%%%%%%%%%%%%%%%%
\begin{figure*}[!t]
\begin{center}
\includegraphics[width=8.cm]{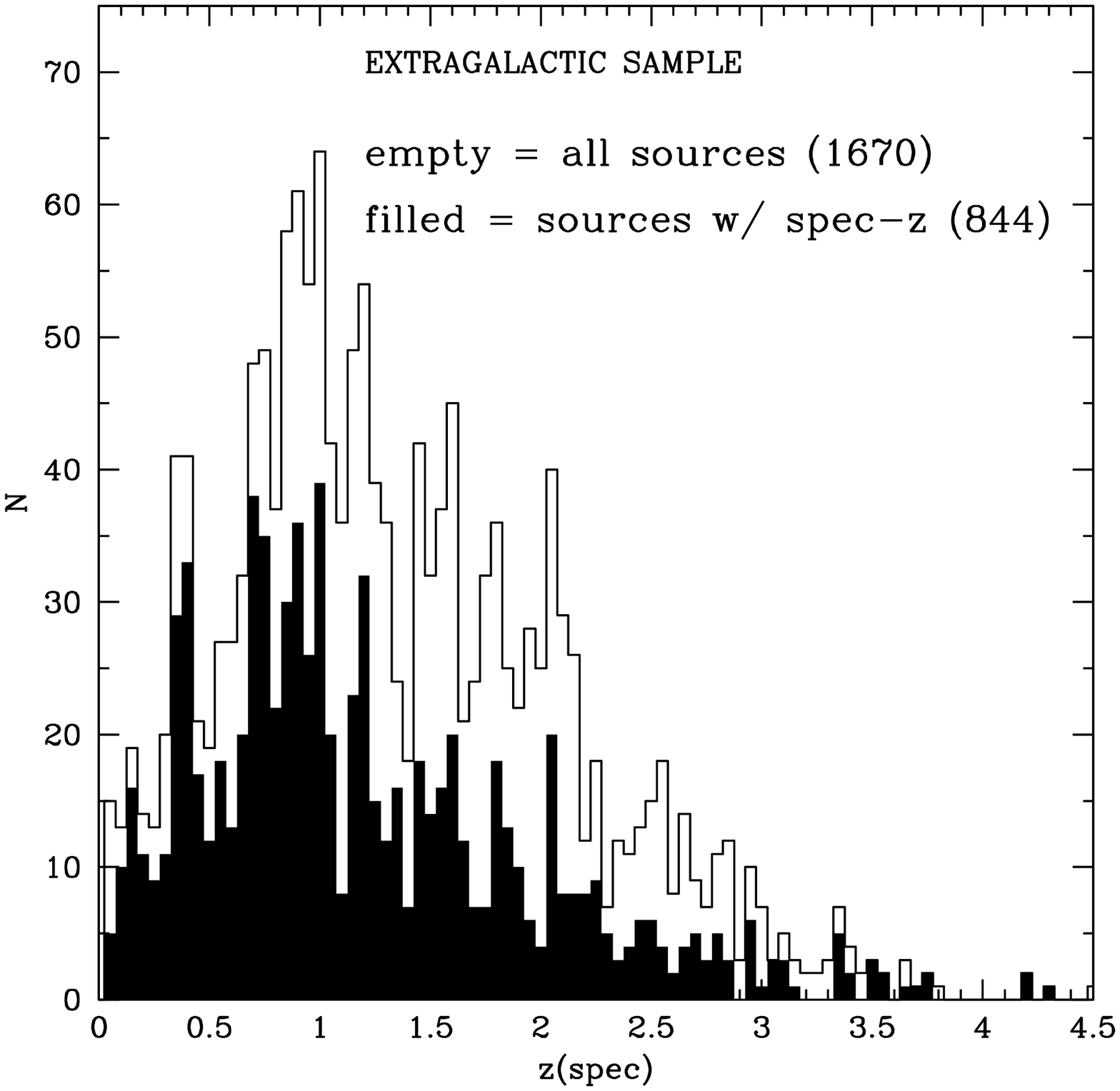}
\includegraphics[width=8.cm]{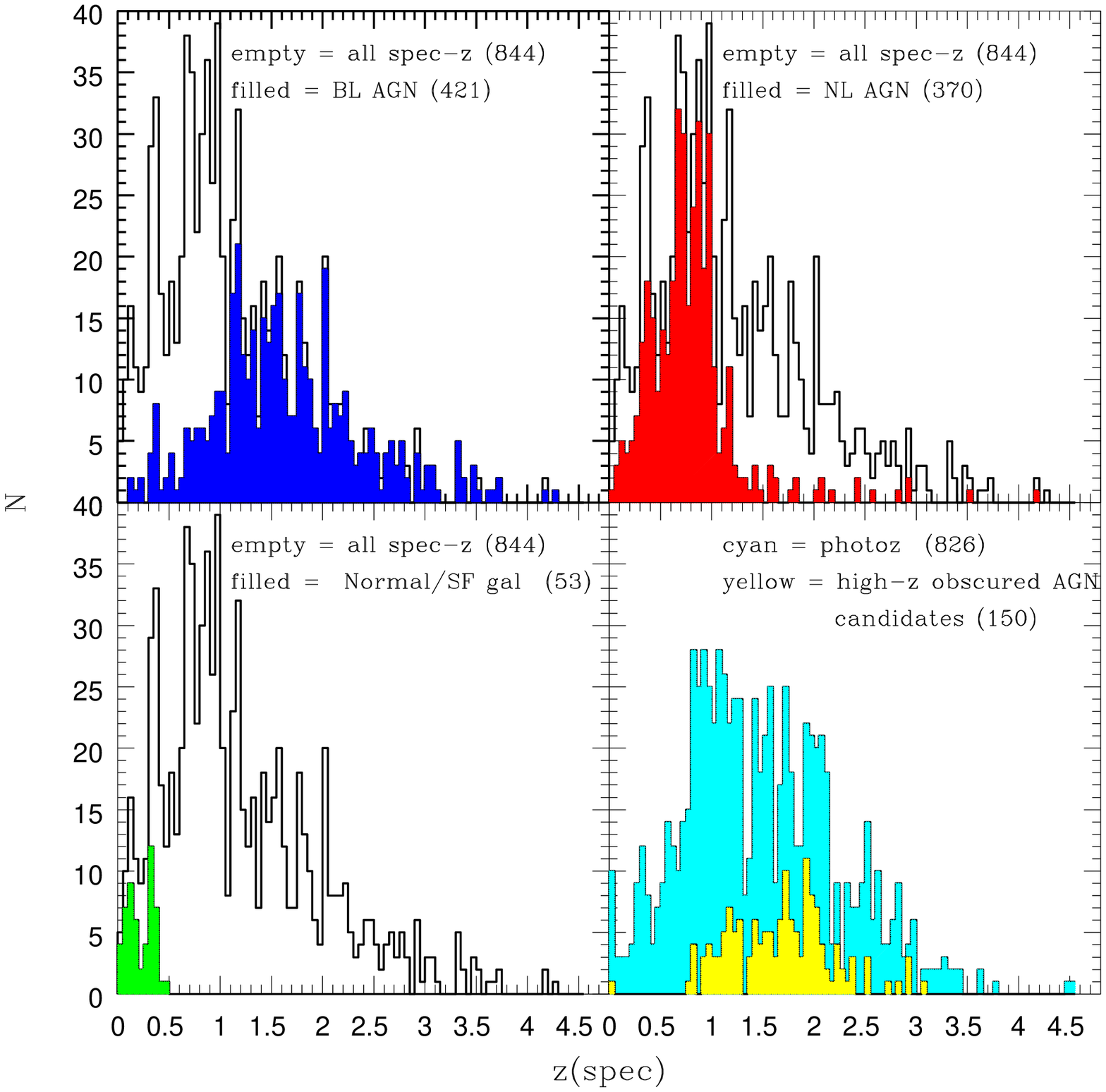}
\end{center}
\caption{({\it Left panel}): Redshift distribution of the extragalactic XMM-COSMOS counterparts
  in $\Delta$z=0.05.  
  Open histogram = all sources for which a spectroscopic or photometric redshift is available (1670); 
  filled histogram = sources with measured spectroscopic redshift (844). ({\it Right panel}): 
  Source breakdown of the spectroscopic sample (open histogram) in 3 different classes (solid histograms) 
  based on  the combined optical 
  and X--ray classification (upper-left: BL AGN; upper-right: NL AGN; lower-left: non-AGN; see 
  text for details). The lower-right panel shows the distribution of the sources with photometric\
  redshifts only (cyan histogram) and, as a comparison, the redshift distribution for the
  high-z obscured AGN candidates discussed in Section 7. The redshift bin in all the panels 
is $\Delta$z=0.05.}
 \label{zdist}
 \end{figure*}
%%%%%%%%%%%%%%%%%%%%%%%%%%%%%%%%%%%%%%%%%%%%%%%%

Considering the spectroscopic and photometric sample, we have an almost 100\%
completeness in redshift for the XMM-COSMOS sources: only 28 sources 
do not have a photometric or spectroscopic measured redshift (21, i.e. 
less than 1.3\%, in the flux limited sample). 

The left panel of Figure~\ref{zdist} shows, as an open histogram, the redshift 
distribution of the objects in the XMM-COSMOS counterpart sample. Only extragalactic
sources are considered here. The black solid histogram shows the distribution of 
the objects with available spectroscopic redshifts. The redshift distribution 
shows prominent peaks at various redshifts (z$\sim 0.12$, z$\sim 0.36$, z$\sim 0.95$, 
z$\sim 1.2$, z$\sim 2.1$), both in the spectroscopic and photometric subsamples. 
In particular, the structure at z$\sim0.36$ is the one with the highest overdensity 
with respect to a smooth distribution, being 
present over many wavelengths in the COSMOS field (see Lilly et al. 2009),
and is likely responsible for the excess signal observed in the clustering amplitude of
X--ray sources around $r_p\sim5-15 h^{-1}$ Mpc described in details in Gilli 
et al. (2009). 

The XMM-COSMOS counterpart sample has 46 spectroscopically confirmed
galactic stars. Additional 53 objects are classified as stars on 
the basis of the inspection of the optical finding charts and/or 
the observed spectral energy distribution, for which the best-fit 
solution is obtained with a stellar template (see Salvato et al. 2009). 
Overall, 99 objects are classified as stars, making up the 5.5\%
of the entire XMM-COSMOS counterpart sample (6.5\% when the soft sample 
is considered), in agreement with results from other large area
X--ray surveys at similar limiting fluxes (e.g., Covey et al. 2009
from the Champ Extended Stellar Survey, Chess). We refer to a paper
in preparation (Stelzer et al. 2010) for a detailed analysis of the
stars contents in XMM-COSMOS.

%%%%%%%%%%%%%%%%%%%%%%%%%%%%%%%%%%%%%%%%%%%%%%%%
\begin{figure*}[!t]
\begin{center}
\includegraphics[width=12cm]{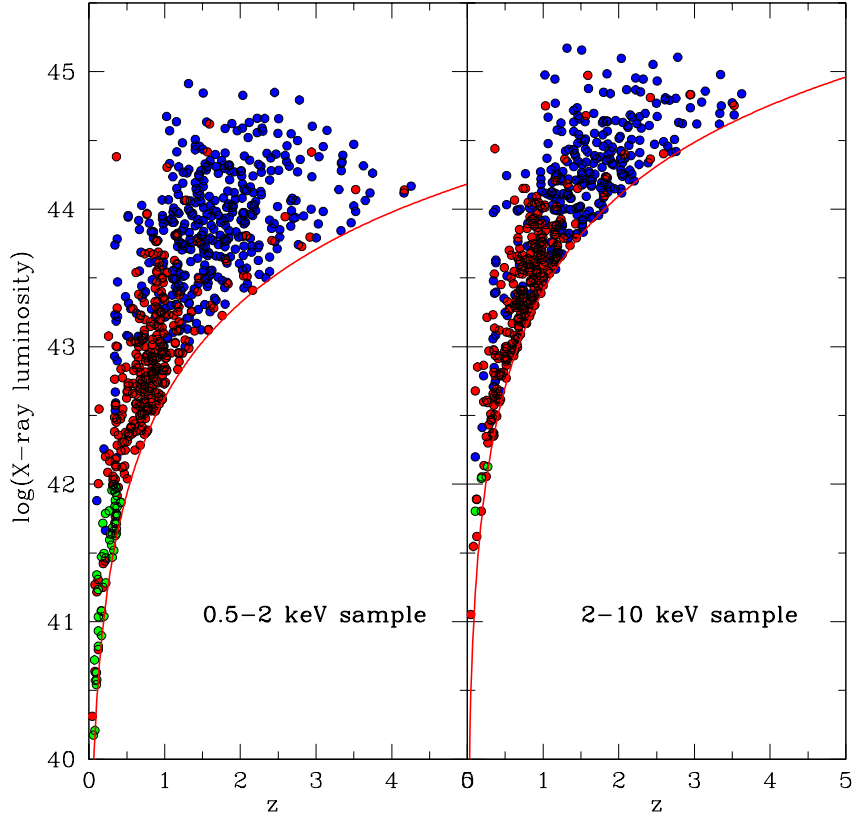}
\end{center}
\caption{({\it Left panel}): Luminosity-redshift plane for the sources with
 spectroscopic redshift detected in the soft band. Blue circles are AGN 1,
  red circles are AGN 2, green circles are ``normal" galaxies (see Section 5.1
  for the definition). The continuous line represents the completeness limit
  in the sample. 
  ({\it Right panel}): The same as left
  panel, for the hard band sample. }
 \label{lz}
 \end{figure*}
%%%%%%%%%%%%%%%%%%%%%%%%%%%%%%%%%%%%%%%%%%%%%%%%

\subsection{Spectroscopic breakdown} 

For the purposes of the present paper we divided the
extragalactic sources with available optical spectra in the XMM-COSMOS counterpart sample 
into three classes, on the basis of a combined X--ray and optical classification 
(see also Szokoly et al. 2004). 
A more detailed analysis of the optical spectra will be presented in
future papers, based also on more complete data that are rapidly being 
accumulated from on-going projects.
 
\begin{itemize}
\item {\bf Broad line AGN} (BL AGN or AGN 1 hereafter): all the objects having at least one broad (FWHM$>2000$ km s$^{-1}$) 
optical emission line in the available spectrum (421 sources, 416 in the flux limited sample);
\item {\bf Non-broad line AGN} (NL AGN or AGN 2 hereafter): 
all the objects with unresolved,  high-ionization  
emission lines,  exhibiting line ratios indicating AGN activity, 
and, when high-ionization lines are not detected, or the observed spectral
range does not allow to construct line diagnostics, objects without broad-line
in the optical spectra and with rest--frame hard X--ray luminosity
in excess than 2$\times10^{42}$ erg s$^{-1}$ typical of AGN 
(370 sources, 344 in the flux limited sample; about 1/3 from line diagnostics;
see below for details);
\item {\bf ``Normal" galaxies}: all the objects with unresolved emission 
lines consistent with spectra of starforming galaxies or with a typical galaxy 
spectrum showing only absorption lines, and with rest--frame hard X--ray luminosity 
smaller than 2$\times10^{42}$ erg s$^{-1}$, or undetected in the hard band 
(53 sources, 49 in the flux limited sample); 
\end{itemize} 

The choice of using a combination of X-ray and optical criteria in the classification
was motivated by the fact that both obscured and unobscured AGN can be misclassified in spectroscopic studies, 
given that the host galaxy light may over-shine the nuclear emission (as in the
cases of the XBONG; Moran et al. 2002, Comastri et al. 2002, Severgnini et al. 2003,
Civano et al. 2007). 
Moreover, at high redshift (z$>0.5$), the observed wavelength range 
does not allow one to construct 
the standard line ratio diagnostics (BPT Diagrams, Baldwin, Phillips and Terlevich, 1981, 
Kewley et al. 2001)
used to distinguish between starforming and Type 2 AGN at lower redshifts (z$<0.5$),  
where a more accurate classification is usually possible (see, e.g., discussion in Bongiorno 
et al. 2010 for the zCOSMOS Type 2 AGN sample). 
We chose a luminosity threshold
of 2$\times10^{42}$ erg s$^{-1}$ in the hard band\footnote{When a source is detected 
only in the soft band the corresponding luminosity threshold used is $10^{42}$ erg s$^{-1}$.}  
to discriminate between star formation processes and accretion processes as responsible
for the X--ray emission, following two arguments:
i) the most powerful local, starburst galaxy, NGC 3256 has a 2-10 keV luminosity 
lower than this value ($\sim5\times10^{41}$ erg s$^{-1}$, see Moran et al. 1988);  
ii) following Ranalli, Comastri \& Setti (2003), this luminosity level
can be due to stellar processes only in objects with SFR larger than 
$\sim$400 M$_\odot$ yr$^{-1}$ ($>2000$ M$_\odot$ yr$^{-1}$ for luminosities 
larger than 10$^{43}$ erg s$^{-1}$, typical of the large majority
of our objects). 
Furthermore, given the flux sensitivity of the XMM-COSMOS survey 
(solid line in Figure 4), objects less luminous than 2$\times10^{42}$ 
erg s$^{-1}$ cannot be detected at redshift z$\gs0.5$ (see also next section), 
where optical spectroscopic classification starts to become problematic. 

The right panel of Figure~\ref{zdist} shows the spectroscopic breakdown of 
the extragalactic sample in the three classes reported above (upper-left, upper-right,
and lower-left quadrants, respectively). We also show
the redshift distribution of the sources for which only photometric redshifts 
are available (lower-right quadrant). 
Among the spectroscopically identified sample, BL AGN make up 
$\gs50$\% in both the soft and hard subsamples, and more than 60\% in the
ultra-hard sample. The higher fraction of BL AGN in the ultra-hard
sample is mainly due to the shallower sensitivity of XMM at energies $>5$ keV,
which limits the detection to the brigthest flux / luminosities where the
fraction of obscured AGN is lower (see also Section 8).
The contribution of non-AGN sources (normal galaxies 
and stars) is maximum in the soft sample ($\sim 12$\%, 97 objects are 
classified in these two classes), decreases to $\sim3$\% in the hard band
(16 objects classified as stars or non-AGN) and only two sources are 
classified in these two classes in the ultra-hard band. 

Figure~\ref{lz} shows the luminosity-redshift plane for the sources with
spectroscopic redshift detected in the soft (left panel) and
in the hard band (right panel), respectively. Different colors refer
to different AGN or galaxy classes as detailed above. 
Among the sources in the XMM-COSMOS sample 
with spectroscopic redshifts, only 64 (12) of the objects classified as 
NL AGN are located at z$>1$ (z$>$2), to be compared with 350 (124) in the 
BL AGN sample. This is mostly due to the fact that high-redshift NL AGN 
are optically faint (typically I$\sim 23-24$) and have not been targeted
yet with dedicated spectroscopic campaigns (see also 
Eckart et al. 2006, Caccianiga et al. 2008, Stalin et al. 2010).

%%%%%%%%%%%%%%%%%%%%%%%%%%%%%%%%%%%%%%%%%%%%%%%%
\begin{figure*}[!t]
\begin{center}
\includegraphics[width=13cm]{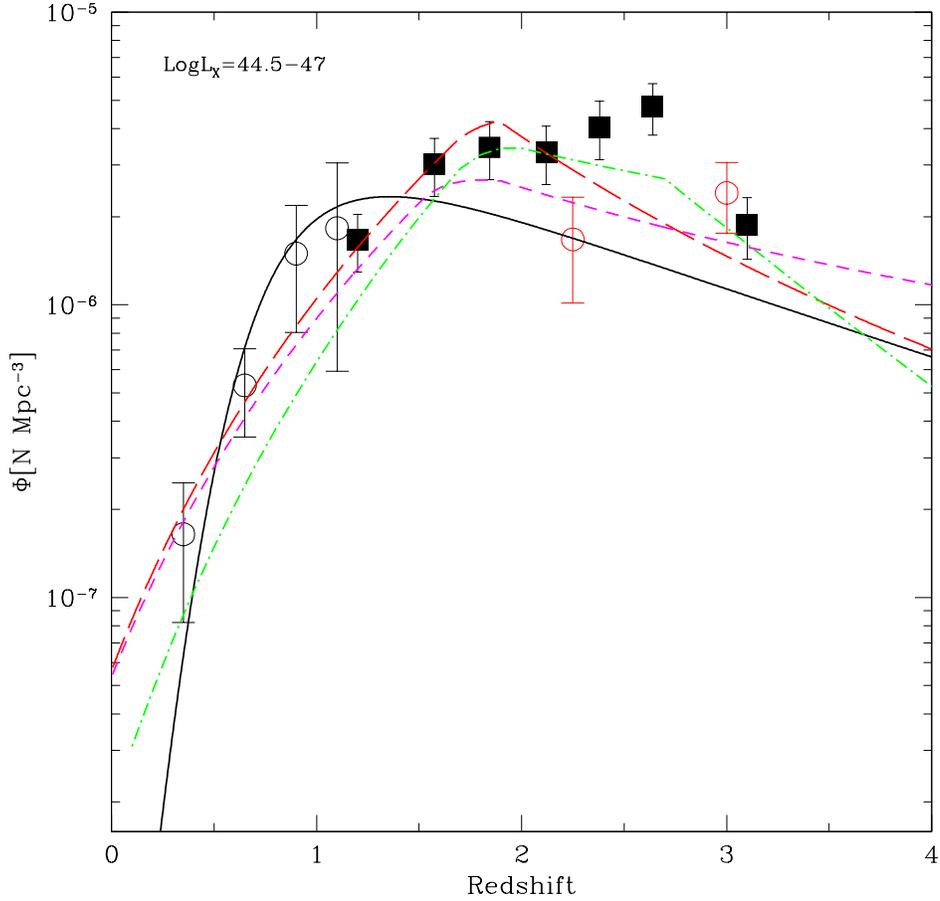}
\end{center}
\caption{The number density evolution from the XMM-COSMOS AGN 
in the luminosity range logL$_{\rm X}$=44.5-47 and in the redshift 
range z=1-3 (black squares), compared with recent results from Aird et al. (2010; 
open circles/black solid lines),
Ebrero et al. (2009; pink dot-dashed curves), Silverman et al. (2008; red dotted curves). 
The green dot-long-dashed curve represents the expectation of XRB synthesis models from 
Gilli et al. (2007).}
\label{fig:xlf_comp}
 \end{figure*}
%%%%%%%%%%%%%%%%%%%%%%%%%%%%%%

\section{Number Density Evolution}
\label{sec:acf}

The high completeness in optical identifications of the XMM-COSMOS sample 
and the availability of spectroscopic and photometric redshifts 
allows us to estimate the number densities of AGN as a function of luminosity and redshift. 
A discussion of the X-ray luminosity function is beyond the scope
of this paper; a comprehensive analysis using XMM-COSMOS and C-COSMOS data, combined with archival data, 
will be the subject of a future paper (Miyaji et al. in prep.). 
In this section, we compare our first estimates of the number density evolution of
the most luminous AGN population, for which XMM-COSMOS can be considered 
reasonably complete, with other recent measurements. 
In this analysis, we use the traditional 
$\sum 1/V_{\rm a}$ estimator (Avni \& Bahcall 1980). 

In calculating the binned number density evolution 
we considered only the subsample of sources detected in the hard (2-10 keV) 
band and included in the flux limited sample (925 sources at $S_{\rm x}=6\times 10^{-15}$ erg 
s$^{-1}$ cm$^{-2}$; see Section 5 and Figure~\ref{fluxhisto}).
We limited the analysis to sources that have ID-class ``reliable'' or ``ambiguous'' (Table 1). 
As discussed in Section~4, the statistical properties of the ``primary'' counterparts of those 
in the ``ambiguous'' class are very similar to those of the secondary objects in the same class. 
Thus, we use the primary identifications for the ``ambiguous'' class objects.  
In calculating the available volume $V_{\rm a}$, we have used  
the curve of the hard-band survey solid angle versus flux limit from C09, 
calibrated with extensive simulations and reproducing the observed 
logN-logS distribution (Cappelluti et al. 2007; C09). 

Using the hardness ratio (HR) between the soft and hard band count rates and the XMM-EPIC energy 
response functions, we have estimated the absorption column density ($N_{\rm H}$) of each source. 
If the AGN was not detected in the soft band, the upper limit was used instead. In calculating 
$N_{\rm H}$, we have assumed a $\Gamma=1.8$ power-law for the unabsorbed spectrum. 
If the HR were softer than the $\Gamma=1.8$ intrinsic power-law, we have set 
$N_{\rm H}=0$ and used the $\Gamma$ calculated from the HR. The 
derived $N_{\rm H}$ values 
are in good agreement with those of the full X-ray spectroscopic analysis of bright XMM-Newton 
sources in the sample (Mainieri et al. 2007, Mainieri et al. in preparation).
The $N_{\rm H}$ estimates have been used to derive de-absorbed X--ray 
luminosities. 

The available volumes $V_{\rm a}$ have been calculated with K-corrections assuming the above spectrum, folded
with the hard band response curve of the XMM-Newton EPIC detectors (pn + 2 MOS).
Binning in luminosity has been made using the de-absorbed luminosities in the rest-frame 2-10 keV. 
The sizes of the redshift bins have been determined adaptively such that there are 20 
objects in each bin. If the 20 objects 
bin is smaller than $\Delta z$=0.2 or larger than 0.9, these minimum and maximum sizes are used. 
The last two redshift bins have been adjusted manually such that they contain approximately 
the same number of objects.

The results are presented in Figure~\ref{fig:xlf_comp}, where the number density of XMM-COSMOS
sources with logL$_{\rm X}=44.5-47$ erg s$^{-1}$ is plotted versus the redshift (black squares).  
In the calculation of the space density we limited the redshift interval to z=3, where the 
L$_{\rm X}$ corresponds to a 2-10 keV flux above our limit (see right panel of 
Figure~\ref{lz}). 
In the same figure we plot the data points obtained from the AEGIS survey by Aird et al. (2010; 
black circles at z$<1.2$ red circles at z$>2$), along with their best 
fit luminosity function (solid black curve).
We also plot the number density evolution for the XLF models
from Ebrero et al. (2009; pink dot-dashed curve), and Silverman et al. (2008; red dotted curve), 
and the expectations from XRB synthesis models (Gilli et al. 2007; green dot-long-dashed curve). 
The XMM-COSMOS and Aird et al. (2010) datapoints at z$>2$ 
are consistent within $\sim2\sigma$ and the difference between the
models can be ascribed to the way in which the models are fit to 
the data.
The XMM-COSMOS data points seem to favor a higher
redshift (z$\sim2$) peak for the space density of luminous quasars, 
more consistent with an LDDE parameterization (represented by the Ebrero 
et al. 2009 and Silverman et al. 2008 curves, though different in the details;
see also Yencho et al. 2009), than with the 
lower redshift peak expected from the LADE model recently proposed by Aird 
et al. (2010), which is a factor of 2.5 lower than our points at 
$z\sim2.5$. In particular, the number density evolution reproduced by 
XRB synthesis models (Gilli et al. 2007) accounts for both the COSMOS and 
the Aird et al. (2010) points to within the error bars.

\section{Luminous Obscured AGN}

In this section we discuss the strategy adopted to build a sample of luminous
(logL$_{\rm X}>44$ erg s$^{-1}$) obscured AGN at moderate 
to high redshifts (1$\lsimeq z \lsimeq 2.5$), fully
exploiting the photometric and spectroscopic information presented in the previous sections.
In particular, we make use of the high spectroscopic redshift
completeness at relatively bright fluxes, and the availability of spectral classifications,
to assess the reliability/robustness of color-color
diagrams and flux ratios as diagnostics of the presence of an obscured AGN,
i.e. objects for which the optical nuclear emission is blocked 
(no broad lines in the optical spectra) and/or characterized by substantial 
(logN$_{\rm H}\gs21.5$) X--ray obscuration. 
We will then apply these diagnostics to select and study the fainter sources
for which only photometric redshifts are available.   
The devised method  may complement systematic studies based on SED fitting,
allowing one to efficiently isolate luminous, obscured AGN at high-z for more
detailed follow-up analyses.
All the statistical properties, correlation analyses
and figures will be presented for the flux limited sample of 1651 sources,
since it is less affected by incompleteness, and we expect a lower fraction
(1.3\%) of wrong identifications (see Section 3.1). We further exclude
from the analysis the 11 sources in this sample for which we are  not able
to provide a reliable identification from our multiwavelength analysis 
(see Section 3 and 3.1). 
The multiwavelength properties of fainter X--ray detected AGN 
(F$_{0.5-2 \rm{KeV}}\ls10^{-15}$ \cgs) will be investigated in the
framework of the C-COSMOS survey (Civano et al. in preparation). 

\subsection{Selection and sample construction} 
A proper source classification (AGN vs. starburst vs.
passive galaxy) should ideally be obtained via a complete analysis
of its emission over the entire electromagnetic spectrum, using both
spectroscopic (e.g. emission line widths and ratios, see e.g. Bongiorno
et al. 2010) and photometric
(SED, Lusso et al. 2010, Elvis et al. 2010) observables.
However, a complete source characterization is
difficult to obtain even in fields where the best and deepest imaging
and spectroscopic campaigns have been obtained, such as the COSMOS
field. 
A reliable information (at least in a statistical sense)
can be obtained through the analysis of the emission in bands where differences between
nuclear and star formation emission are emphasized. 
In particular, the combination of observed-frame mid-infrared, near infrared and 
X--ray to optical flux ratios has been exploited recently in the literature to isolate obscured AGN
(see, e.g., F03, Martinez-Sansigre et al. 2005, 2006,
Fiore et al. 2008, Donley et al. 2008, Dey et al. 2008).  
We will apply in the following several multiwavelength diagnostics
in order to study the properties of the most obscured sources in the XMM-COSMOS
sample. 

\subsubsection{X--ray to optical and optical to near infrared colors (X/O and $R-K$)}

%%%%%%%%%%%%%%%%%%%%%%%%%%%%%%%%%%%%%%%%%%%%%%%%
\begin{figure*}[!t]
\begin{center}
\includegraphics[width=13cm]{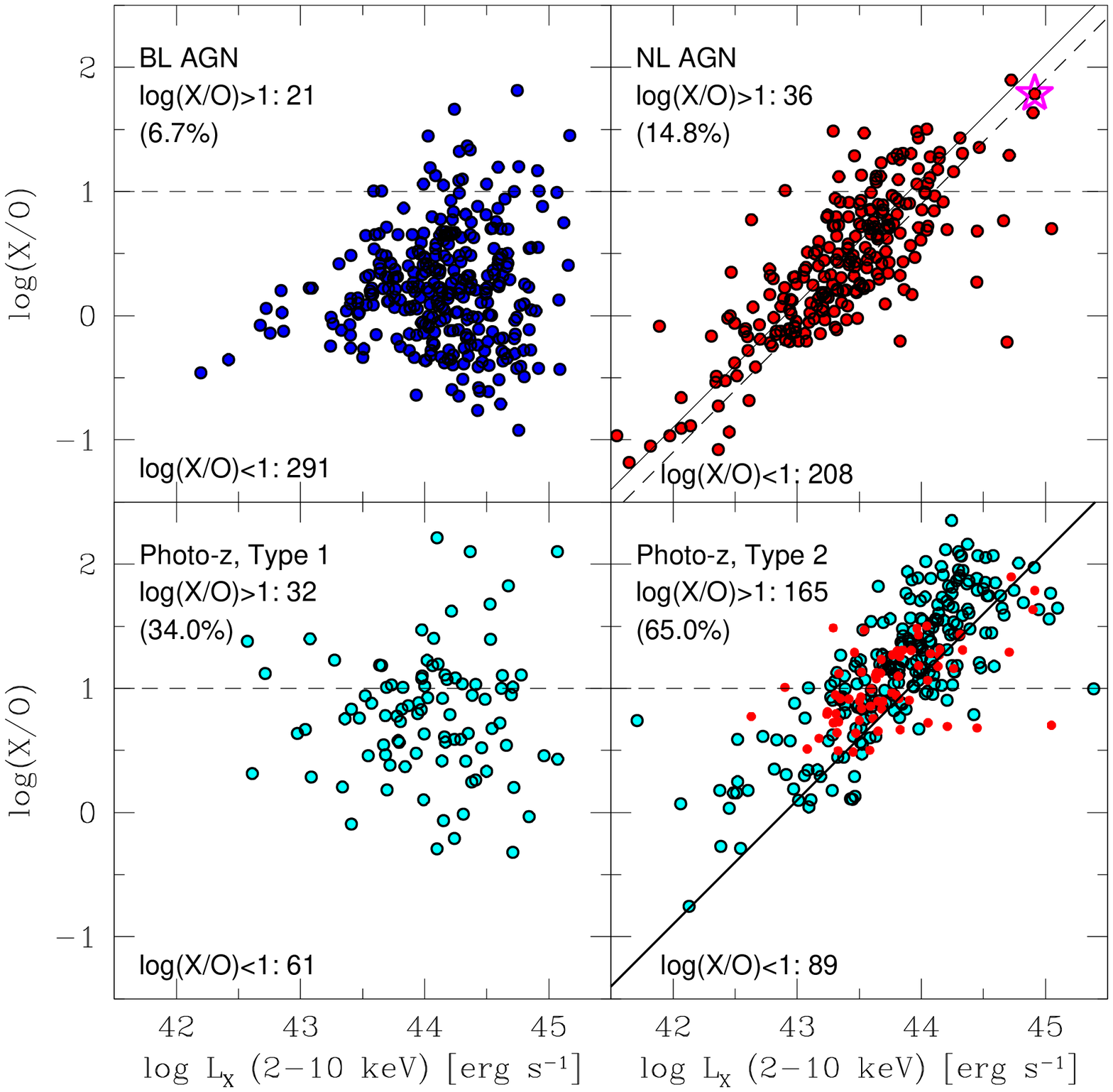}
\end{center}
\caption{{\it Upper panels}: 
  X/O vs. the deabsorbed rest-frame hard X-ray luminosity for the sources classified
  as BL AGN (left, blue circles) and for the sources classified as AGN 2 
  (right, red circles).
  The dashed diagonal line marks the F03 relation between the X/O and the 
  2-10 keV luminosity for obscured AGN. 
  The solid line marks the best fit relation obtained for the XMM-COSMOS 
  sample (see text for details). The magenta star marks the position
  in this diagram of XID 2028 (see Section 8). 
  {\it Lower panels}: 
  X/O vs. the rest-frame hard X-ray luminosity for the sources
  with photometric redshifts and classified from the 
  analysis of the spectral energy distribution as AGN dominated (left) 
  and as galaxies dominated (right). As red circles, we also plot 
  the spectroscopically identified AGN 2 (taken from the upper panel)
  with faint optical magnitudes ($R>23$). 
  The solid line marks the best fit relation obtained for
  the XMM-COSMOS spectroscopic sample of NL AGN (see upper panel)
  The numbers in the top (bottom) left  of each quadrant give the number of 
  objects in the BL and AGN 2 samples above (below) the dashed lines; 
  for the objects above the dashed lines also the percentages have 
  been reported. }
 \label{xolx}
 \end{figure*}
%%%%%%%%%%%%%%%%%%%%%%%%%%%%%%%%%%%%%%%%%%%%%%%%

Since the first source identification campaigns of hard X--rays surveys, 
a class of sources with high ($>10$) X-ray-to-optical flux ratio 
(X/O\footnote{We define X/O=f$_{2-10 keV}$/f$_{r-band}$, where the 
R--band flux is computed by converting r-band magnitudes into 
monochromatic  fluxes and then multiplying  them by the width of the 
r-band filter (Zombeck 1990)}; Hornschemeier et al. 2001, Alexander et al. 2001;
Giacconi et al. 2001; F03; Koekemoer et al. 2004; Eckart et al. 2006) 
has been found. This value of X/O is significantly higher than the average value observed 
for optically and soft X--ray selected AGN (Maccacaro et al. 1988). 
Later studies (e.g. Perola et al. 2004, Civano, Comastri \& Brusa 2005) found that the high-X/O sources tend also 
to be obscured in the X-rays, with column densities of the order of $10^{22}-10^{23}$ cm$^{-2}$,
and that sources with X/O$>10$ selected at bright fluxes (F$_{2-10}\gs10^{-14}$
\cgs) are candidate high-luminosity obscured quasars. 
Also, these sources are detected preferentially at z$>1$ and characterized on average by red optical to
near infrared colors ($R-K\gsimeq5$, in the Vega system) pointing towards a strong link
between X--ray detected Extremely Red Objects and Type 2 Quasars (Alexander et al. 2002; 
Mignoli et al. 2004, Brusa et al. 2005,  Severgnini et al. 2005, Georgantopoulos, 
Georgakakis \& Akylas 2006). 

The upper panels in Figure~\ref{xolx} show the X/O value as a function 
of the 2-10 keV X--ray luminosity, separately for the XMM-COSMOS sources 
spectroscopically identified as BL AGN (312 objects, left) and NL AGN 
(244 objects, right), detected in the 2-10 keV flux limited sample 
(925 sources). 
The numbers in the top (bottom) left of each panel give the number of 
objects in the different samples above (below) the dashed lines which
correspond to the thresholds of X/O$>10$.  
For the objects above the dashed lines also the percentages to the entire
subsamples have been reported.

%%%%%%%%%%%%%%%%%%%%%%%%%%%%%%%%%%%%%%%%%%%%%%%%
\begin{figure*}[!t]
\begin{center}
\includegraphics[width=13cm]{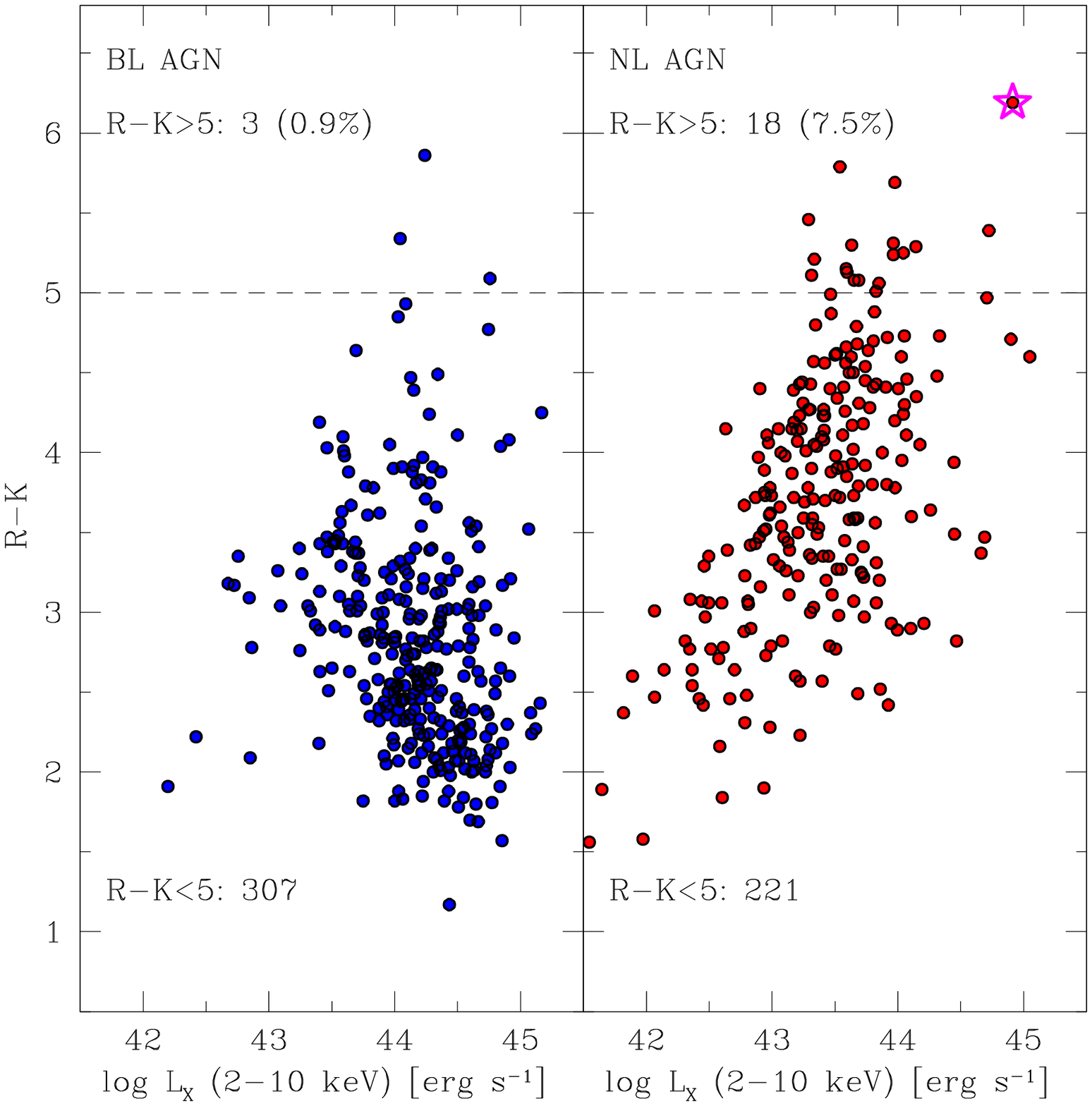}
\end{center}
\caption{{\it Left panel}: 
  $R-K$ vs. the rest-frame hard X-ray luminosity for the sources classified
  as BL AGN (left, blue circles) and for the sources classified as AGN 2 
  (right, red circles).
  The magenta star marks the position
  in this diagram of XID 2028 (see Section 8). 
  The numbers in the top (bottom) left  of each panel give the number of 
  objects in the BL and AGN 2 samples above (below) the dashed lines; 
  for the objects above the dashed lines also the percentages have been 
  reported.}
 \label{rklx}
 \end{figure*}
%%%%%%%%%%%%%%%%%%%%%%%%%%%%%%%%%%%%%%%%%%%%%%%%

%%%%%%%%%%%%%%%%%%%%%%%%%%%%%%%%%%%%%%%%%%%%%%%%
\begin{center}
\begin{figure*}[!t]
\begin{center}
\includegraphics[width=8.5cm]{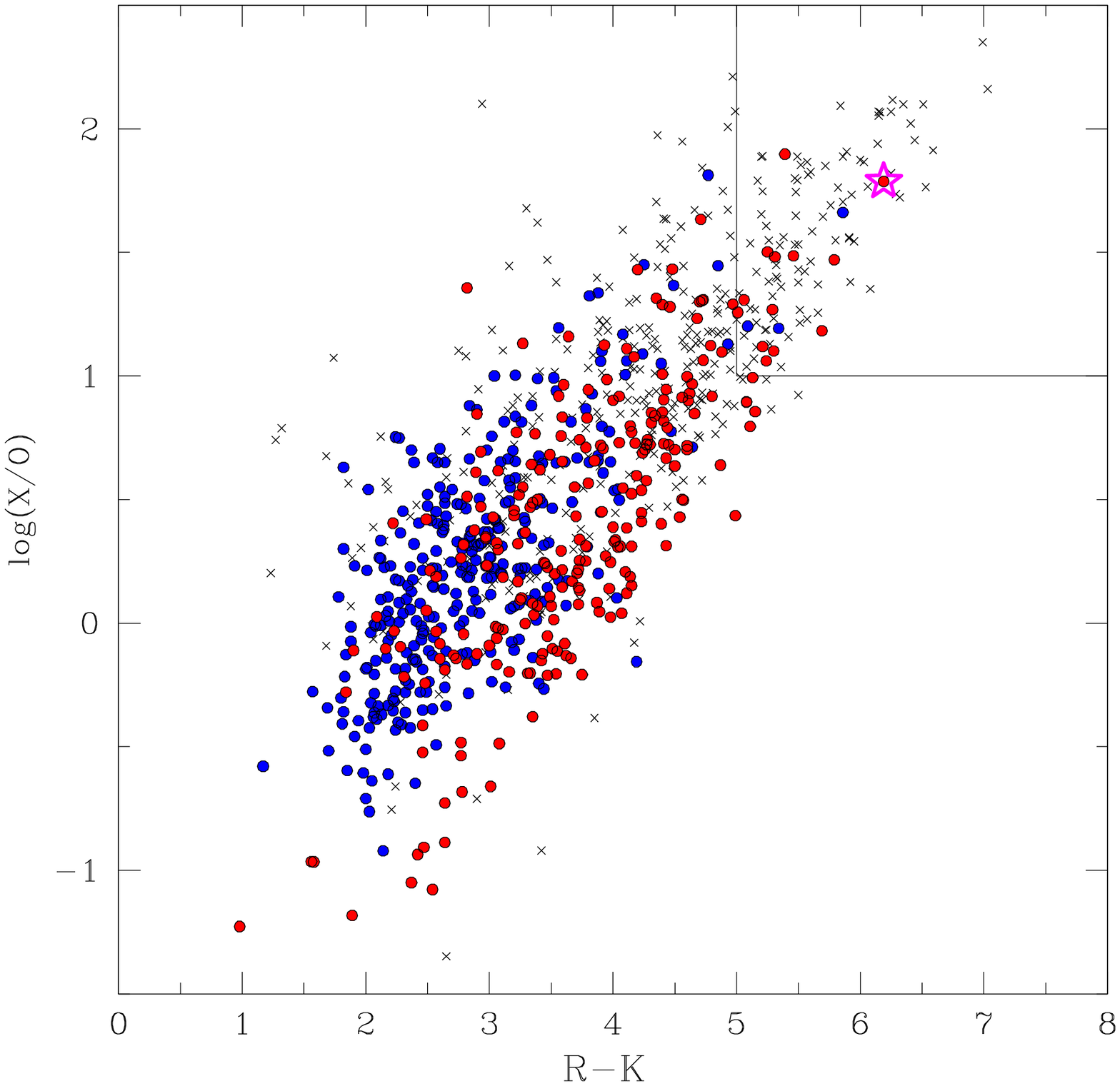}
\includegraphics[width=8.5cm]{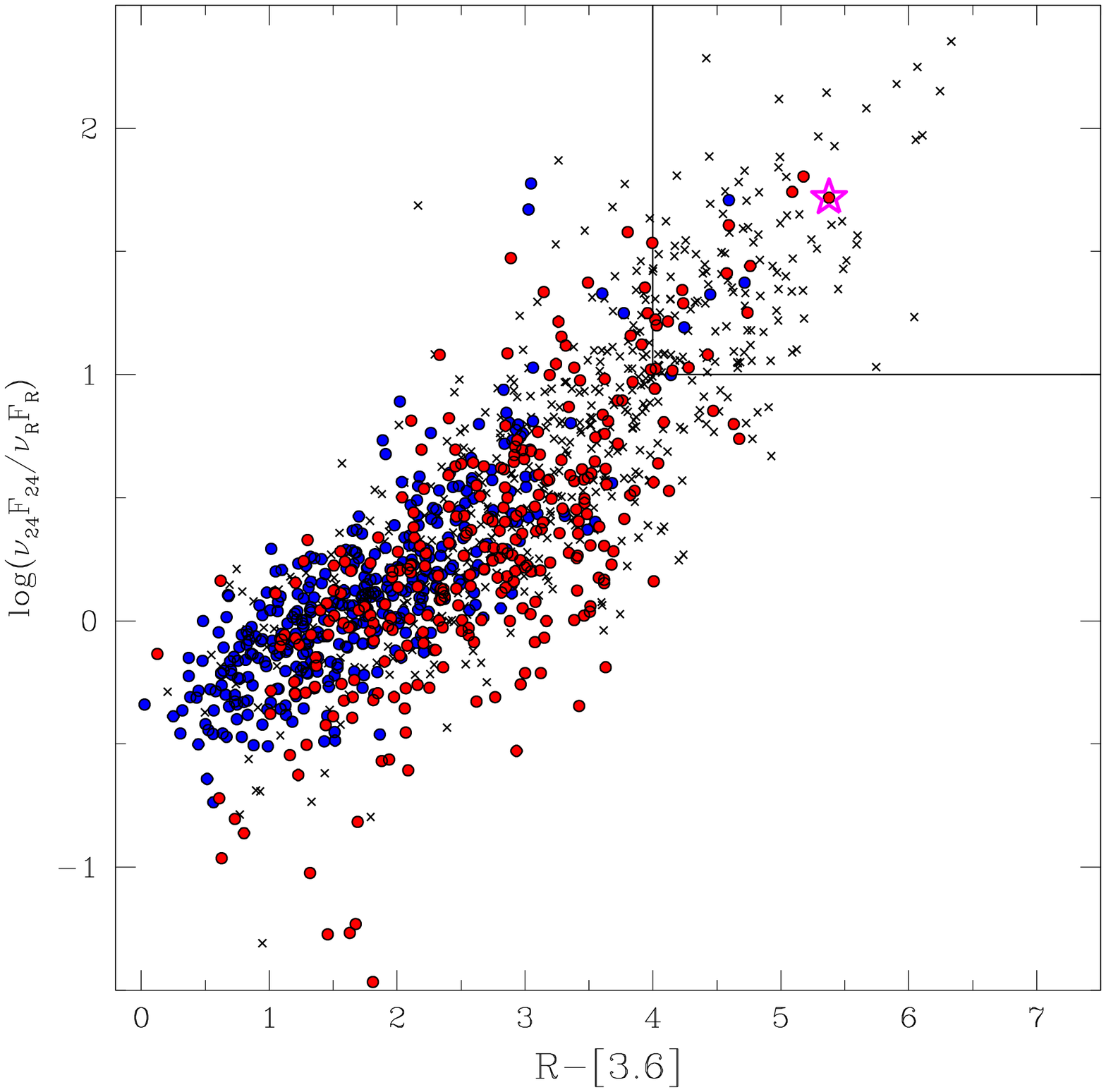}
\end{center}
\caption{{\it Left panel}: the X--ray to optical flux ratio (X/O) plotted against
the $R-K$ color (Vega magnitude, in order to be immediately comparable with other
$R-K$ selection criteria used in the literature) for the BL AGN sample (blue circles); 
the AGN 2 sample (red circles); and the sources with only photometric redshifts (crosses). 
The solid lines mark the selection for high-z obscured AGN candidate used in this work ($R-K>=5$ 
and X/O$>10$). {\it Right panel}: the ratio of the 24$\mu$m flux over the R-band flux versus the 
R-3.6$\mu$m color (AB magnitudes). The solid lines mark the selection criteria for
high-z obscured AGN candidates used in this work (R-3.6$>=4$ and $\nu_{24}F_{24}/\nu_{R}F_{R}>10$). 
In both panels, the magenta star marks the position of XID 2028 (see Section 8 for details).}
 \label{xork}
 \end{figure*}
  \end{center}
%%%%%%%%%%%%%%%%%%%%%%%%%%%%%%%%%%%%%%%%%%%%%%%%

A strong correlation between the X/O ratio and the observed X--ray 
luminosity for AGN 2 has been discovered by the first massive
identifications campaigns in Chandra and XMM-Newton surveys (e.g., F03,
Eckart et al. 2006), and it is related to the fact that the
X--ray luminosity is less affected by absorption than the optical
luminosity. 
The same correlation is also present 
in the XMM-COSMOS sample (upper right panel  of Fig.~\ref{xolx}). 
The dashed line is the original F03 relation, the solid line is the relation 
which describes our data: the slope is the same, while the zeropoint is 
different. This is mostly due to: i) the optical bands used in the X/O 
parameter, R-band, having slightly different filter curves, and ii) 
most important, the average magnitude of the spectroscopically
identified sample being fainter for the XMM-COSMOS sample with respect 
to the sample used in F03. 
On the contrary, objects in the AGN 1 sample (upper-left quadrant) do not
present such a trend. 

The lower panels of Figure~\ref{xolx} show the X/O vs. the 2-10 keV 
X--ray luminosity for the sources with photometric redshifts, separately 
for the objects with a best-fit SED template of unobscured quasars
(model numbers SED from 19 to 30 in Salvato et al. 2009; left panel;
see also Lusso et al. 2010) and for sources best-fit with a galaxy-dominated template 
(model numbers SED $<19$ in Salvato et al. 2009;  right panel).
The sources in the lower right panel of Fig.~\ref{xolx} show a 
correlation of the X/O vs. the X-ray luminosity similar to that
present for the AGN 2 sample, albeit with a somewhat larger scatter and 
higher normalization. 
The fact that sources at fainter optical magnitudes (as it is the case for the
sources in the photo-z sample) have a higher normalization of
the X/O vs. L$_{\rm X}$ relation is obviously expected and 
was already pointed out by Barger et al. (2005). 
The large and deep COSMOS spectroscopic
database allows us to test this variation using a suitably
large ($\sim 70$) sample of faint ($R>23$) sources. 
When the sources with spectroscopic redshifts fainter than $R=23$ 
are considered (plotted for comparison with red symbols in the lower right 
panel of Fig.~\ref{xolx}), they show the same higher normalization of the 
X/O vs. L$_{\rm X}$ relation. 
The observed correlation between X/O and X--ray luminosity is due to the fact that, 
while the nuclear AGN X-ray luminosity can span several decades, the host galaxy 
R-band luminosity (which dominates the optical emission in obscured sources) has a 
much smaller scatter, less than one decade (see also Treister et al. 2005).
At fainter magnitudes, lower luminosity host galaxies 
are starting to be detected, moving the points towards higher X/O
values.
In addition to these selection effects, 
it is important to note that moving the SED of 
a moderately obscured or Compton Thick AGN (see, e.g., Franceschini et al. 2000) 
to progressively higher redshifts the K--corrections in the optical and 
X--ray band work in opposite directions.
The shape of the hard X--ray spectrum 
is responsible for a strong K--correction which   
``boosts'' the X--ray flux and favors the detection of high redshift sources.
Conversely, the weak rest--frame optical--UV emission is shifted in the
R band explaining the extremely faint optical magnitudes.  
As a consequence, the optical to X--ray flux ratio changes in a non--linear 
way (see also discussion in Comastri, Brusa, Mignoli 2003) and this
can explain the higher normalization observed for the fainter sources. 

Figure~\ref{rklx} shows the $R-K$ color (Vega system)
as a function of the 2-10 keV X--ray luminosity, separately 
for the XMM-COSMOS sources spectroscopically identified as BL AGN 
and NL AGN detected in the 2-10 keV flux limited sample. 
The numbers in the top (bottom) left  of each panel give the number of 
objects in the different samples above (below) the dashed lines which
correspond to the thresholds of $R-K>5$.  
The fraction of AGN 2 with high X/O and/or red $R-K$ colors is
considerably higher than the corresponding fraction for BL AGN
(e.g. about 7.5\% of AGN 2 have $R-K>5$ while only $\sim0.9$\% of the BL AGN
have the same red color). 
When the dependence of the $R-K$ color on the X--ray luminosity for the NL AGN
sample is considered, 
there is evidence for a trend where high-luminosity sources have redder
colors than low-luminosity sources (right panel 
of Fig.~\ref{rklx}). The median $R-K$ color for the high luminosity sample 
(L$_{\rm X}>10^{43.5}$ erg s$^{-1}$) is $\langle R-K\rangle=4.17$ (0.53 dispersion), while it is 
$\langle R-K\rangle=3.47$ (0.62 dispersion) for the low-luminosity sample (L$_{\rm X}<10^{43.5}$ 
erg s$^{-1}$). This correlation is mostly driven by redshift effects,
the median redshifts of the two subsamples being z=0.93 and z=0.57, respectively. 
On the other hand, lower luminosity (lower-redshift) BL AGN 
are slightly redder, as it is  clear from the left panel of Figure~\ref{rklx}. 
In particular, the median $R-K$ color for the high luminosity sample (L$_{\rm X}>10^{44}$ erg
s$^{-1}$) is $\angle R-K\rangle$=2.58 (0.68 dispersion; similar to the average value of $R-K$ colors
obtained from optically selected, bright quasars, Barkhouse \& Hall 2001), while it is 
$\langle R-K\rangle$=3.04 (0.55 dispersion) for the low-luminosity sample (L$_{\rm X}<10^{44}$ erg s$^{-1}$).  
The probability that the two distributions are drawn from the same parent population is 
5.5$\times10^{-7}$ according to a Kolmogornov-Smirnov test. 
Given that the main difference
of the two samples is the average redshift (the low-luminosity sample being
at considerably lower redshift, z$<1$, than the high-luminosity sample, 
z$\sim2$), the most likely explanation of the observed diversity is
that, at low redshift, the host galaxy dominates the rest-frame optical emission
(as sampled by the R and K bands).
These objects also present on average an extended morphology (as derived from ACS analysis,
see Gabor et al. 2009) and are similar to the ``extended BL AGN" 
presented in B07.   

%%%%%%%%%%%%%%%%%%%%%%%%%%%%%%%%%%%%%%%%%%%%%%%%
\begin{center}
\begin{figure*}[!t]
\begin{center}
\includegraphics[width=15cm]{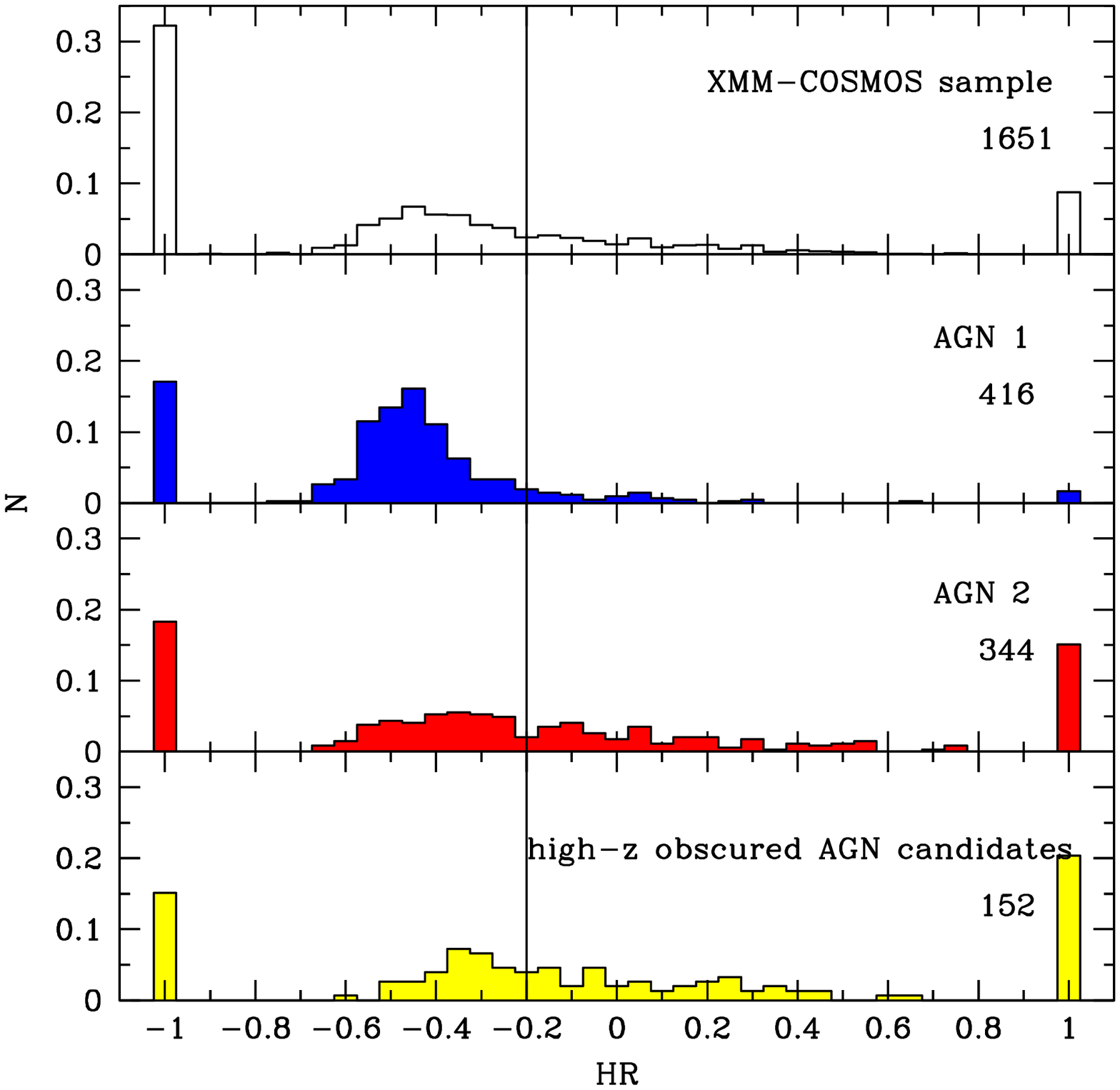}
\end{center}
\caption{   HR distribution normalized to the total number of sources in each sample for
  a) the whole XMM-COSMOS flux limited sample (1640 objects); b) the AGN 1 sample 
  (414 objects); c) the AGN 2 sample (337 objects); and d) the high-z obscured AGN candidate 
  as isolated from the previous figures (152 objects).
  The line at HR$=-$0.2 marks the threshold used in this paper to divide obscured and 
  unobscured AGN on a purely X--ray based calssification. 
  Sources detected only in the soft (0.5-2 keV) or hard (2-10 keV) band are plotted
  at values of HR$=-$1 and HR=1, respectively. 
 }
 \label{hr}
 \end{figure*}
  \end{center}
%%%%%%%%%%%%%%%%%%%%%%%%%%%%%%%%%%%%%%%%%%%%%%%%

The left panel of Figure~\ref{xork} shows the two obscured AGN indicators discussed above plotted 
one against the other, e.g., the X/O ratio versus the $R-K$ color (see also
Brusa et al. 2005). 
In addition to the spectroscopically identified population (plotted
as colored symbols) we also report as crosses the sources without spectroscopic
redshifts for the whole XMM-COSMOS catalog. In the region with $R-K>5$ and X/O$>10$ 
we have 105 sources, 16 with spectroscopic redshifts and 
13 of them classified as AGN 2. The high rest-frame, 
2-10 keV X--ray luminosities (logL$_{\rm X}\gsimeq43.5$ erg s$^{-1}$), further
classifies these sources as quasars. For the remaining 89 sources,
only photometric redshifts are available. 

\subsubsection{Mid-infrared to optical flux ratio} 
Recently, the combination of high MIPS 24$\mu$m to optical flux ratio 
and red near infrared to optical colors has been proven to be very efficient 
in selecting high-redshift starforming galaxies (Yan et al. 2005, Houck et al. 2005) 
and heavily absorbed, possibly  Compton Thick AGN (see Fiore et al. 2008, 2009, Dey et al. 
2008, Daddi et al. 2007, Sacchi et al. 2009). Given the high X--ray column density,
most of these sources may remain undetected even in the deepest X--ray exposures
and their AGN nature can be unvealed only through stacking analysis (Daddi et al. 2007, 
Fiore et al. 2008, 2009, Lanzuisi et al. 2009). However, the X--ray detected population may
constitute the most-luminous and slightly less absorbed tail of this
heavily absorbed AGN population (see also discussion in Fiore et al. 2008). 

%%%%%%%%%%%%%%%%%%%%%%%%%%%%%%%%%%%%%%%%%%%%%%%%
\begin{center}
\begin{figure*}[!t]
\begin{center}
\includegraphics[width=16cm]{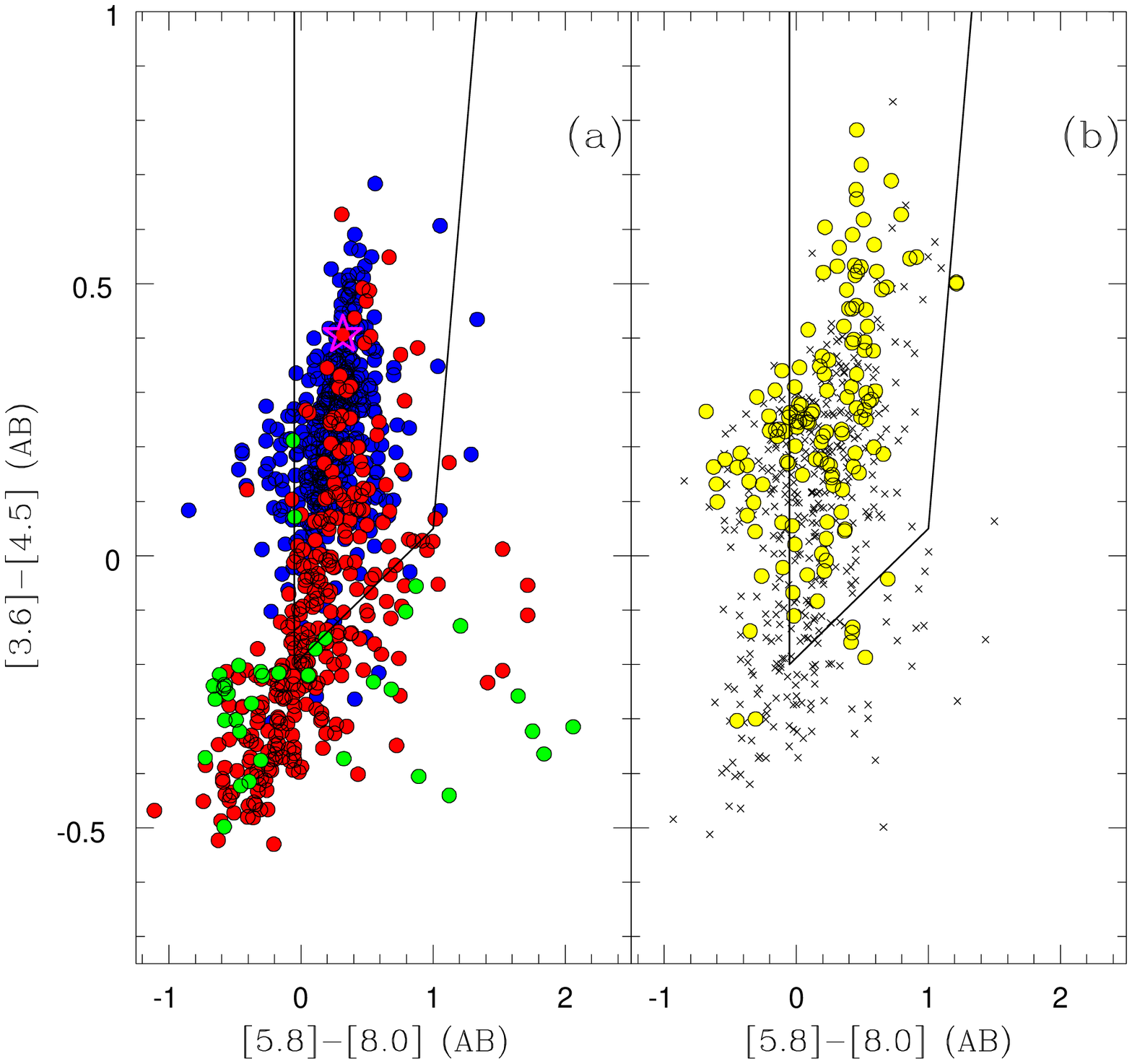}
\end{center}
\caption{{\it Panel (a)}: Stern color-color diagram for the spectroscopically
  identified sample. Symbols and colors as in previous figures.  
{\it Panel (b)}: Stern color-color diagram for the sources with only 
photometric redshifts ($\sim 600$ sources, crosses) and for the high-z 
obscured AGN candidates 
($\sim 150$ sources, yellow points)
}
 \label{sternflux}
 \end{figure*}
  \end{center}

The right panel of Figure~\ref{xork} shows the $\nu_{24}F_{24}/\nu_{R}F_{R}$ value 
versus the R-3.6 color for all the sources with MIPS detection in the XMM-COSMOS 
counterpart sample (the symbols have the same meaning as before). 
There are 137 sources with  $\nu_{24}F_{24}/\nu_{R}F_{R}>10$ and R-3.6$\geq4$ (a similar
criterion has been introduced by Yan et al. 2007); 64 of them are in common 
with the sample isolated on the basis of their combined high X/O and $R-K$ colors
(Section 7.1.1). 
Among the remaining 73 objects, ten sources have spectroscopic redshifts and classifications available: 
2 of them are classified as AGN 1 and the remaining 8 are classified as AGN 2 from our combined optical and X-ray classification. For 63 sources no spectroscopic redshift information is available.

In the following, we will refer to the sample of the 152 sources (89+63) without 
spectroscopic redshifts selected on the basis of their high X/O and/or 
$\nu_{24}F_{24}/\nu_{R}F_{R}$ 
ratio, and red optical to near infrared colors as the ``high-z obscured AGN 
candidates". 
When the spectroscopically identified population only is considered, these 
criteria appear to be robust in selecting obscured AGN:   
only about 20\% (5/25) of the sources satisfying these selection criteria 
are classified as AGN1, while in the 
XMM-COSMOS flux limited sample the fraction of AGN 1 over the total AGN1+AGN2 makes up 
$\sim55$\%. 
Photometric redshifts are available for 150/152 sources. 
The redshift distribution
for these high-z obscured AGN candidates is plotted in the bottom right of right panel of 
Fig.~\ref{zdist} (yellow histogram). This distribution does not change if we consider
only the 64 sources in common between the two selections presented in 7.1.1 
and 7.1.2. The vast majority of these sources lie at z=1-3. 

%%%%%%%%%%%%%%%%%%%%%%%%%%%%%%%%%%%%%%%%%%%%%%%%
\begin{center}
\begin{figure*}[!t]
\begin{center}
\includegraphics[width=14cm]{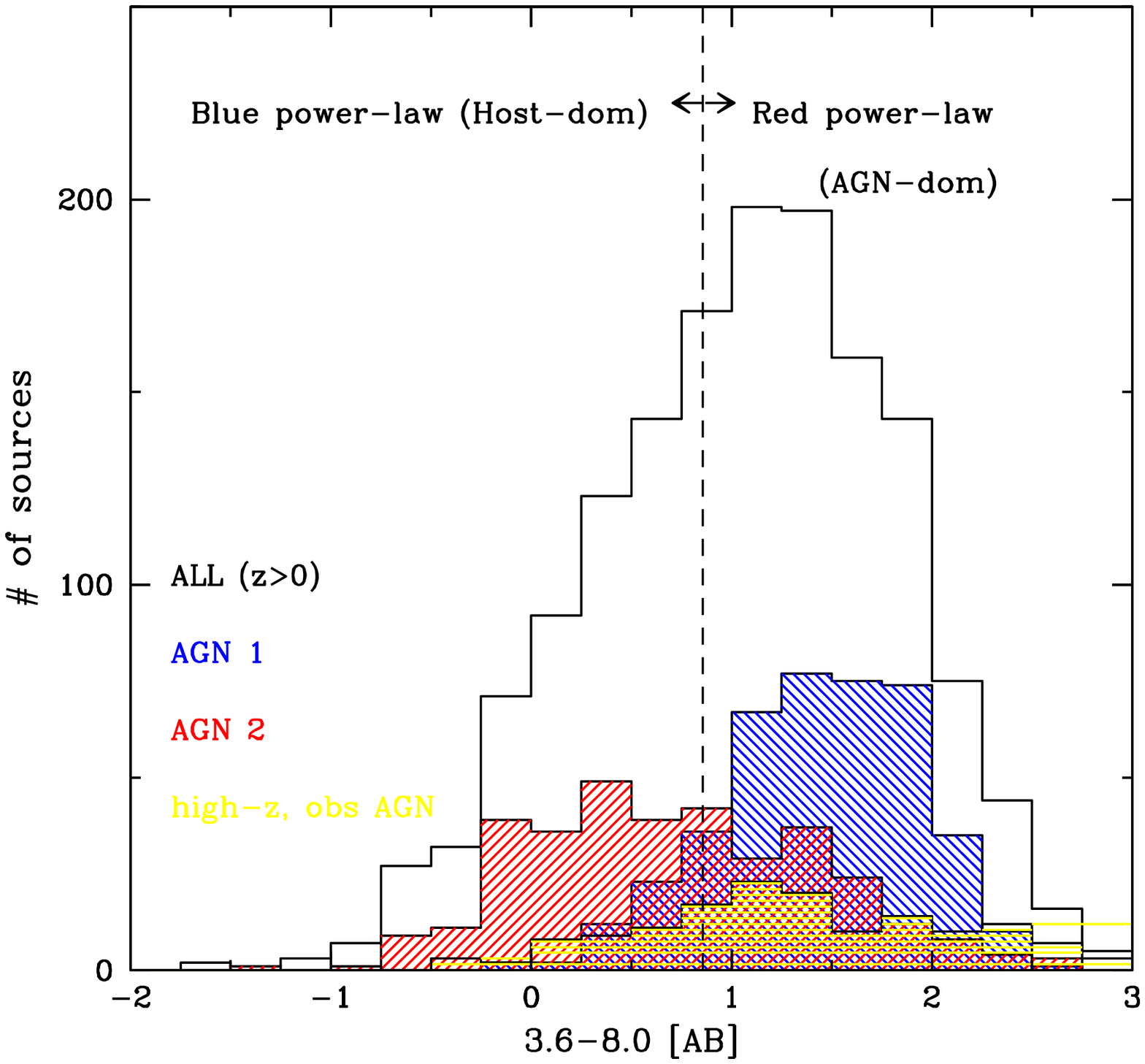}
\end{center}
\caption{
The distribution of the  [3.6]-[8.0] color for the XMM-COSMOS
counterparts, shown separately for the entire extragalactic sample (black), the
AGN 1 sample (blue), the AGN 2 sample (red) and the high-z obscured AGN candidates sample (yellow). 
The dashed line at  [3.6]-[8.0]=0.856 marks the division between ``blue" and 
``red" power-law sources.}
 \label{iraccm}
 \end{figure*}
  \end{center}
%%%%%%%%%%%%%%%%%%%%%%%%%%%%%%%%%%%%%%%%%%%%%%%%

\subsection{Hardness ratio and X--ray obscuration}  
We explore in the following the X--ray properties of the high-z candidate obscured
AGN and compare them with those of the entire XMM-COSMOS sample. 
When the number of counts in a source is inadequate to perform a spectral fit, a widely 
used tool to study the spectral properties of an X-ray source is the hardness 
ratio HR (see the description of Table 2 in Section 4 for the definition
and, e.g., Della Ceca et al. 2004, Mainieri et al. 2007).
Mainieri et al. (2007) showed that the HR and the column 
density of the sources derived from spectral analysis are quite well correlated, and that 
90\% of the sources with column densities larger than 10$^{22}$ cm$^{-2}$  have
HR$>-0.3$. 
The results in Mainieri et al. (2007) have been obtained  
for the subsample of the XMM-COSMOS counterparts described in B07;
given that the redshift distribution does not significantly differ from the one of 
the present sample (see Fig.~5 in B07 and Fig.~\ref{zdist} in this paper), 
we can rely on the HR$-$N$_{\rm H}$ calibration as described in Manieri 
et al. (2007). 
On the other hand, however, only about 50\% of the sources with HR$>-0.3$ 
have N$_{\rm H}>10^{22}$ cm$^{-2}$. In order to create a cleaner sample of 
obscured AGN from an HR selection, and to compare with previous results 
(e.g. Hasinger 2008), we use in the following a more conservative limit 
of HR$=-0.2$ to distinguish between obscured and unobscured objects. 
Above this threshold, $\sim 80$\% of the sources have N$_{\rm H}>10^{22}$
cm$^{-2}$.

Figure~\ref{hr} shows the HR distribution, normalized to the total number 
of sources, for the entire XMM-COSMOS counterparts in the flux limited sample (1640), 
the AGN 1 sample (416), the AGN 2 sample (344), 
and the high-z obscured AGN candidates, as isolated in Section~7.1. 
As it is clear from the normalized distributions, the relative fraction of hard sources 
increases from top to bottom: in particular, the fraction of sources with HR$>-0.2$ 
in the AGN 1 sample is only $\sim10$\%, while it rises to 45\% in the
AGN 2 sample and up to 53\% in the high--z obscured AGN candidates sample (80/152). 
The median values of the HR in the three different classes are HR$=-0.49$ 
for the BL AGN, HR$=-0.29$ 
for the NL AGN, and HR$=-0.17$ %(0.37) 
for the high--z obscured AGN candidates. 

The average HR values of the two obscured AGN samples (NL AGN and high-z obscured AGN candidates)
are not directly comparable, given that the average redshifts of the two samples are 
different (with median values of $\langle z \rangle =0.73$ and $\langle z\rangle=1.72$, 
respectively) and the N$_{\rm H}$-HR relation is redshift dependent. 
In addition, it is  worth to note that, if the NL AGN sample is significantly contaminated 
by misclassified low-luminosity BL AGN for which the broad line emission is diluted in 
the host galaxy light (see section 7.1 and next section 7.4), the average HR value may 
be contaminated too, resulting in an average softer value than the intrinsic one.    
However, it is also worth to point out that the redshift effect works against
hard HR at high-z and, therefore, the fact that high-z obscured AGN candidates 
are harder than NL AGN confirms the efficiency of the proposed selection
in picking up the most obscured sources.

\subsection{IRAC color-color diagrams and color indices } 

Since the advent of the Spitzer satellite, several 
color-color diagrams based on the combination of flux ratios from the
four different IRAC channels have been widely used  
to classify infrared sources and isolate obscured AGN missed in optical 
and near infrared surveys (Lacy et al. 2004, Stern et al. 2005, Hatziminaouglou
et al. 2005, Barmby et al. 2006). 

Figure~\ref{sternflux} shows the IRAC color-color diagram (as 
initially proposed by Stern et al.~2005) for the sources in the XMM-COSMOS 
counterpart sample. For this analysis only sources with IRAC detection 
in all the four bands and with good determination of the photometry  
(i.e. not in masked region, with error $<$ 25\% of the IRAC flux) 
are considered, for a total of 1326 objects. 
 Panel (a) shows the distribution in this diagram of the spectroscopically identified
population, for the three different categories described in Section 5.1: 
380 AGN 1 (blue circles), 296 AGN 2 (red circles), 37 normal galaxies (green
circles). 
The locus at [5.8]-[8.0]$>1$ and [3.6]-[4.5]$<0$
mainly selects X--ray sources which are expected to be low redshift
normal and star forming galaxies (see also discussion in Feruglio et al. 
2008). 
The dashed lines isolate the wedge for the selection of luminous quasars, 
originally proposed in Stern et al. (2005). 
While 90\% of the BL AGN are successfully recovered from the
IR selection, and reassuringly almost all (95\%) of the normal galaxies 
detected in the X--rays lie indeed outside the wedges, only 48\% of the X--ray 
selected non-broad line AGN lie within the Stern's wedge.
Similar results can be obtained using IRAC-based color-selection wedges
different from the specific one proposed by Stern et al. (2005), e.g.
those proposed by Lacy et al. (2004) or Hatziminaoglou et al. (2005). 
This analysis confirms that purely IRAC based criteria are not optimally designed to 
select AGN in which the host galaxy dominates the near-infrared energy (e.g. 
lower-luminosity AGN), and/or AGN obscured in the IRAC bands, and therefore
may miss a significant fraction of the low-luminosity, obscured AGN population. 
Most of non-broad line AGN outside the wedge are, indeed, moderate luminosity 
objects at intermediate redshifts (z$<1$, see also Cardamone 
et al. 2008, Georgantopoulos et al. 2008, Brusa et al. 2009), 
in which the host galaxy light dominates the IR emission. 

Panel (b) shows the same diagram for the population with photometric redshifts only 
(572 sources). Superimposed as yellow circles are the high-z obscured AGN candidates isolated 
through the exploitation of the multiwavelength properties described in the
previous sections. 
The vast majority (80\%) of the high-z obscured AGN candidates would have been
isolated also through the IRAC selection, confirming that these sources are luminous AGN, 
with the NIR/MIR emission  dominated by the central engine. However, at the limiting fluxes 
of the IRAC COSMOS survey the contamination from high-redshift (z$>1$) normal and starforming 
galaxies in the Stern et al. (2005) wedge starts to become important (see discussion in Sajina 
et al. 2005) and the selection criterion would become efficient only when combined 
with an additional criterion (in this case, the X--ray emission). 

%%%%%%%%%%%%%%%%%%%%%%%%%%%%%%%%%%%%%%%%%%%%%%%%
\begin{center}
\begin{figure*}[!t]
\begin{center}
\includegraphics[width=14cm]{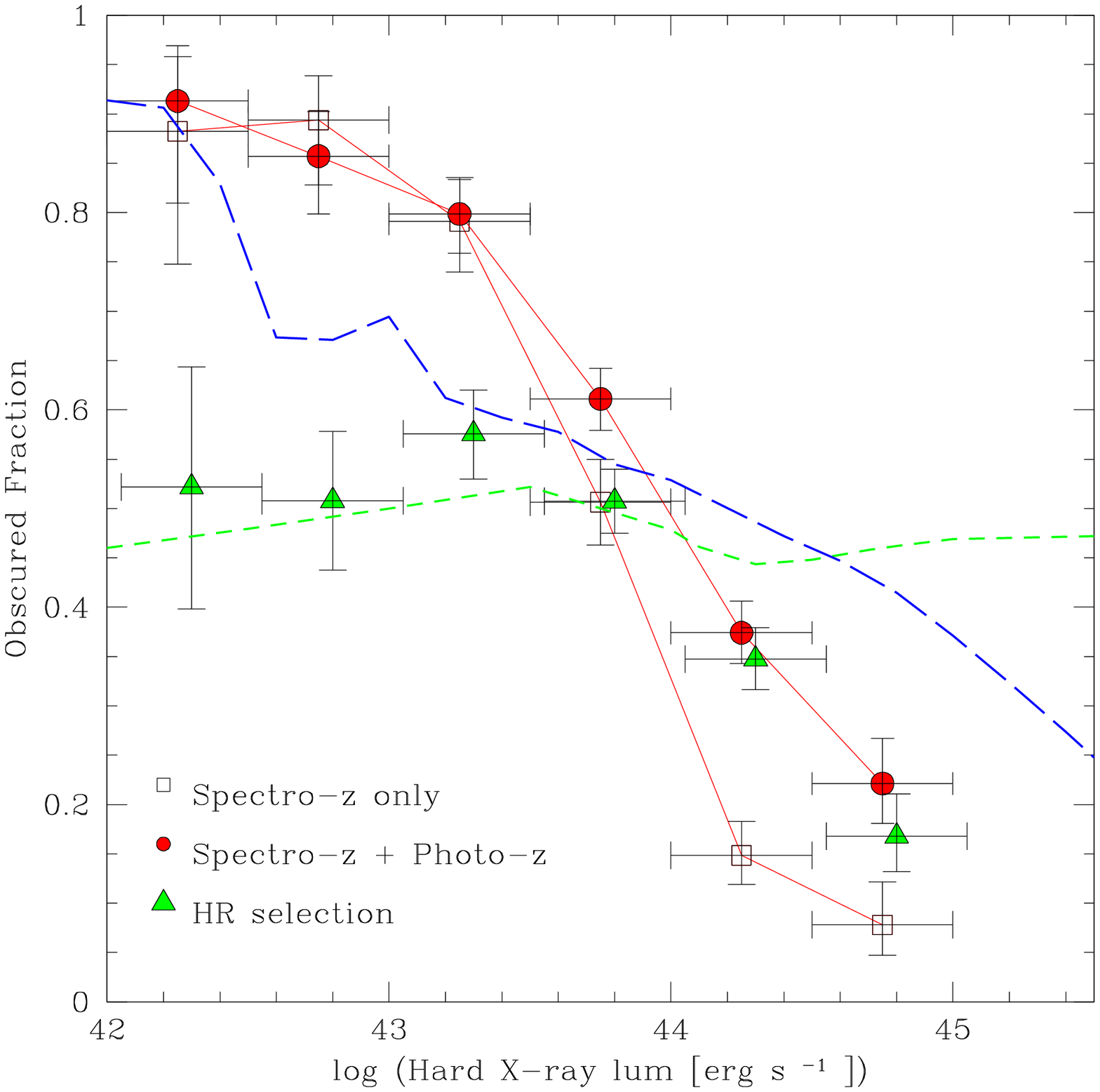}
\end{center}
\caption{
Ratio of obscured to total AGN as a function of hard 
X--ray luminosity. The open squares with error bars show the obscured 
fraction computed from the sources in the 2-10 keV sample with spectroscopic information
($\sim 550$ objects). The red filled circles represent the fraction
obtained for the full 2-10 keV sample, where the classification
in obscured and unobscured sources is made on the combination of 
spectroscopic information and the best fit SED template from sources
without spectroscopic redshifts ($\sim 900$ objects). The green triangles 
mark the fraction 
computed when a purely X--ray selection (based on the HR) is adopted. 
The green short-dashed and  blue long-dashed lines represent the predictions 
on the fraction of AGN with N$_{\rm H}>10^{22}$ cm$^{-2}$ from the XRB synthesis 
models by Gilli et al. (2007) and Treister \& Urry (2006), respectively. 
Errors are 1 $\sigma$, calculated following Gehrels (1986). 
} 
 \label{fracl}
 \end{figure*}
  \end{center}
%%%%%%%%%%%%%%%%%%%%%%%%%%%%%%%%%%%%%%%%%%%%%%%%

It is interesting to note that a non negligible fraction ($\sim18$\%) of the high-z 
obscured AGN candidates lie at [3.6]-[4.5]$>0$ ($\sim0.25$) and 
[5.8]-[8.0]$<0$, and would be missed by IRAC selections based on AGN SED 
templates, like the one proposed by Stern et al. (2005).  
As a comparison, the fraction of objects satisfying the same selection for 
the spectroscopically identified population is only $\sim5$\%. 
This locus is where both star-forming and normal galaxies colors at z$\sim 1-3$ are 
expected (see Barmby et al. 2006). Therefore, 
in our X--ray selected AGN sample, also this locus is populated by objects 
where the host galaxy dominates and/or the AGN is obscured 
up to 8 micron ($\sim 2.5$ micron rest-frame).  
Given the high inferred X-ray luminosity (L$_{\rm X}\gsimeq10^{43.5}$ erg s$^{-1}$), 
this also means that these objects should have a 
luminous (and massive) host, 
in agreement with the results of Mignoli et al. (2004), Severgnini et
al. (2005), and others. 
  
The IRAC broad band color ([3.6]-[8.0]) can be used to disentangle 
sources with an infrared-rising (red) SED, e.g. AGN,
power-law dominated sources, and sources with an inverted (blue)
infrared SED, with more flux at shorter wavelengths than at longer
wavelengths, e.g. normal galaxies at low-z 
(see also discussion in Barmby et al. 2006).   
Sources with a flat luminosity per given wavelength range 
($\lambda f_{\lambda}$=constant) have [3.6]-[8.0]$\sim 0.86$. 
Figure~\ref{iraccm} shows the distribution of the [3.6]-[8.0] color 
for the entire XMM-COSMOS population (black histogram), the AGN 1 sample
(blue shaded histogram), the AGN 2 sample (red shaded histogram) and 
the high-z obscured AGN candidates sample (yellow shaded histogram). 
AGN 1 populate the red power-law part of the distribution, as expected, with
a median value of $\langle [3.6]-[8.0]\rangle=1.46$ (0.34 dispersion). 
On the contrary, 
the XMM-COSMOS sample of spectroscopically identified AGN 2 (limited 
mainly at z$<1$ and to moderate luminosity objects) preferentially show 
IRAC colors consistent with an inverted power-law, i.e. in the host
galaxy dominated regime (the median value of the distribution is 
$\langle[3.6]-[8.0]\rangle=0.65$ and a wider dispersion, 0.55)
The probability that the 2 distributions are drawn from
the same parent population is $<10^{-7}$ according to a KS test. 
The high-z obscured AGN candidates set in between the 2 spectroscopically
selected samples, around the median value $\langle[3.6]-[8.0]\rangle=1.28$ 
(0.55 dispersion), with only $\sim25$\% of the sources having 
[3.6]-[8.0]$<0.865$. 
This is a further confirmation that the devised strategy is efficiently 
in selecting luminous AGN with red power-law (i.e. AGN dominated) 
SED but elusive in the optical bands.

\section{Luminosity dependence of the obscured AGN fraction}  

The large sample of X--ray sources in XMM-COSMOS can be used to 
explore the luminosity dependence of the obscured AGN fraction. 
In Figure~\ref{fracl} we plot the fraction of obscured AGN (defined
as the fraction of objects classified as NL AGN over the total number of 
AGN sources) in different luminosity bins for the spectroscopically
identified hard X--ray selected XMM-COSMOS sample (open squares). 
This fraction is a strong function
of the X--ray luminosity, being almost 90\% at L$_{\rm X}\sim10^{42}$ erg s$^{-1}$
and $<10$\% at L$_{\rm X}\sim10^{45}$ erg s$^{-1}$. 

It is important to note, however, that, even if the spectroscopic sample
is representative of $\sim 60$\% of the XMM-COSMOS source population, 
high-redshift (i.e. high-luminosity) obscured AGN are optically faint 
(typically I$\sim$23-24 or fainter) and have not been targeted yet with 
dedicated spectroscopic campaigns. The resulting fraction at high luminosity 
is therefore expected to be biased, in particular {\it against} the 
obscured objects (see discussion in previous subsection). 
In order to extend the study of the AGN fraction as a function of the
luminosity to the full XMM-COSMOS sample, and reduce the bias against 
high luminosity obscured AGN, we calculated the fraction of obscured AGN 
fully exploiting the information contained in the multiwavelength catalog. 
For sources with spectroscopic redshifts,
we used the optical/X--ray classification as defined in Section~5.1;
for sources without spectroscopic redshifts, we used the 
classification from Salvato et al. (2009) based on the best-fitting
SED procedure (see also Section 7.1), which turned out to be well 
matched with the optical classification and X--ray hardness ratios from X--ray 
color diagrams (see discussion in Salvato et al. 2009 and C09): 
sources with best-fit SED templates of 
unobscured quasars (model numbers SED from 19 to 30 in Salvato et al. 2009;
see also Lusso et al. 2010) are classified as Type 1 (unobscured) AGN, while 
sources best-fit with a galaxy-dominated template (model numbers 
SED $<19$ in Salvato et al. 2009) are classified as Type 2 (obscured) AGN. 
We then computed a new estimate of the fraction of obscured AGN as a function
of X--ray luminosity (red circles in Fig~\ref{fracl}).
Not surprisingly, the fraction of obscured AGN derived in this way at 
L$_{\rm X}>10^{44}$ erg s$^{-1}$  is considerably higher (a factor of $\sim2-3$) 
than that computed for the spectroscopically identified sample. 

Finally, we computed the obscured fraction derived
by dividing sources in obscured and unobscured on the basis of the X--ray 
HR, which can be used as a {\it crude} indicator of the X--ray (nuclear) 
obscuration, although suffering from large uncertainties (see, e.g. 
Tozzi et al. 2006 and Section 7.2). We chose as threshold for classifying an
object as obscured AGN the value HR$=-0.2$ (corresponding to N$_{\rm H}
\sim10^{22}$ cm$^{-2}$ for a $\Gamma=1.8$ spectrum), 
as motivated in Section 7.2. 
The results are plotted as green triangles in Figure~\ref{fracl}. 
All the obscured AGN fraction estimates refer to {\it observed} quantities and 
have not been corrected for selection effects. 
%

%%%%%%%%%%%%%%%%%%%%%%%%%%%%%%%%%%%%%%%%%%%%%%%%
\begin{center}
\begin{figure*}[!t]
\begin{center}
\includegraphics[width=8cm]{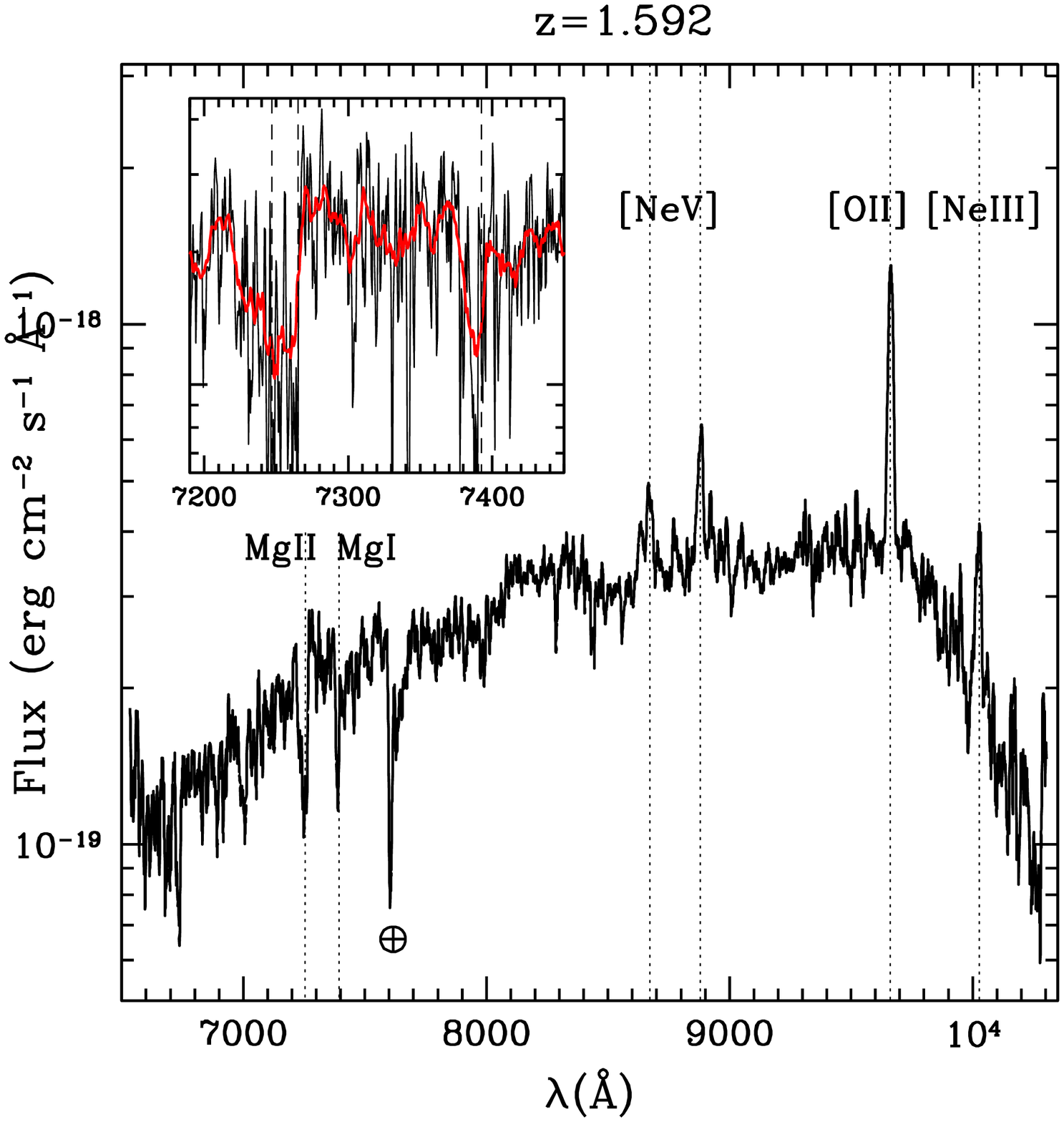}
\includegraphics[width=9cm]{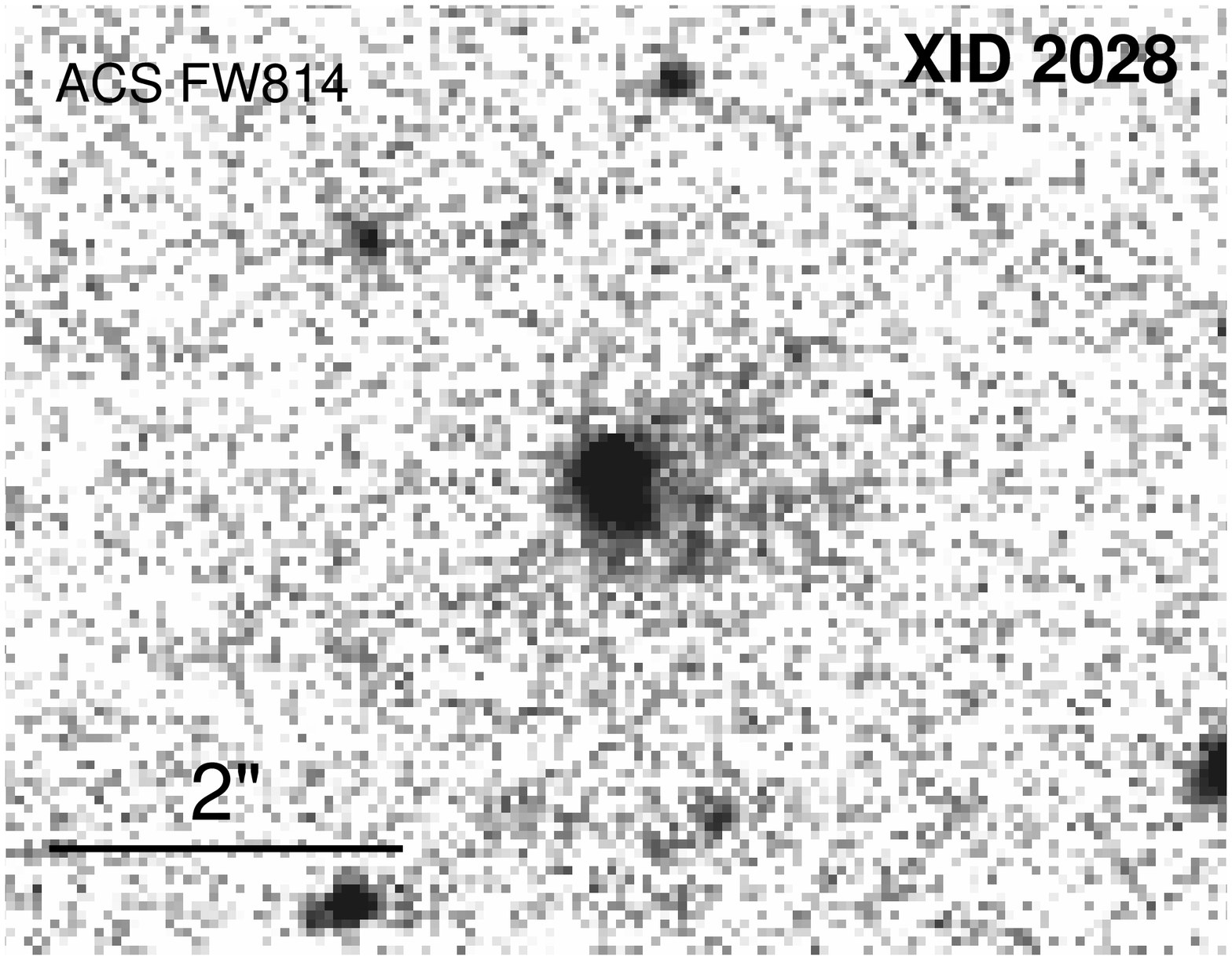}\\
\end{center}
\caption{{\it Left panel}: observed frame Keck spectrum of XID 2028, 
with superimposed the various emission and absorption lines (dotted vertical lines,
as labeled); the inset shows the zoom on the MgII and MgI lines. The black
curve shows the unbinned data, while the red curve shows the smoothed spectrum
(boxcar of 15). 
The vertical dashed lines show the expected positions of the absorption lines 
at the redshift derived from the O[II] emission line. {\it Right panel}: HST/ACS image centered
around XID~2028. }
 \label{spe2028}
 \end{figure*}
  \end{center}
%%%%%%%%%%%%%%%%%%%%%%%%%%%%%%%%%%%%%%%%%%%%%%%%

In Figure~\ref{fracl} we also compare our different  estimates of the obscured AGN fraction 
as a function of the luminosity derived as outlined above with those predicted by the XRB synthesis 
models of Gilli et al. (2007; from an X--ray based classification, 
long-dashed green curve) and Treister \& Urry  (2006; from a combined optical and X--ray  
classification, short-dashed blue curve), for the N$_{\rm H}>22$ cm$^{-2}$ population, 
folded with the survey sky coverage. 
In the luminosity range logL$_{\rm X}=43.5-44.5$ erg s$^{-1}$, i.e. around L$_{*}$ 
luminosity for the XMM-COSMOS sample sources, the two predictions are 
broadly consistent with each other and with the observed estimates derived 
from the HR and from the combined (spectroscopic plus SED) optical classification. 
At low luminosities (logL$_{\rm X}<43$ erg s$^{-1}$), 
optically-based classifications seem to yield higher obscured AGN fractions.
Similarly, XRB models predictions largely differ: the Gilli et al. (2007) 
curve well trace the X--ray based estimate of the obscured AGN
fraction, while the Treister \& Urry (2006) model predictions are
closer to the estimate from optical classification\footnote{At variance with Treister et al.
2009a, for the open squares we use only a spectroscopic 
classification. However, we note that the hybrid classification in 
Treister et al. (2009a) is used only for sources at z$<0.5$, which translates
to L$_{\rm X}<10^{42.5}$ erg s$^{-1}$ at the limiting flux
of XMM-COSMOS sources, and  L$_{\rm X}<10^{43}$ erg s$^{-1}$ at the
limiting flux of the ECDFS sample, i.e. the first two bins in Fig.~\ref{fracl}. 
The results at higher X--ray luminosities (L$_{\rm X}>10^{43}$ erg s$^{-1}$)
can therefore be considered comparable.}.
This reflects the paucity of BL AGN in the XMM-COSMOS sample 
at low-L (logL$_{\rm X}<43$ erg s$^{-1}$, see Figure~\ref{xolx} and \ref{rklx}) and might 
be due, at least in part, to the effect of the host galaxy contribution in 
the observed spectral range. Indeed,  broad lines 
signatures from low-luminosity AGN may get diluted in the host galaxy emission
(see discussion in Caccianiga et al. 2007, Civano et al. 2007), and 
this effect is likely to be important at low Eddington ratios 
and at low luminosities (Hopkins et al., 2009). 
At higher X--ray luminosities (logL$_{\rm X}>44.5$ erg s$^{-1}$) both models 
overpredict the observed obscured fraction by a factor of $\sim2-3$. 
 
The luminosity dependence of the obscured AGN fraction can also explain
why low-luminosity AGN lie preferentially outside the IRAC selection
wedges populated by spectroscopically selected unobscured quasars: 
the lower the luminosity, the higher is the percentage of obscured sources 
and therefore on average the higher is the contribution from host galaxy 
on the observed SED and colors, which eventually 
drives the galaxies outside the selection wedges (see, e.g., Sajina et al. 
2005, Barmby et al. 2006, Georgantopulos et al. 2008).

\section{A prototypical Obscured QSO at $z=1.59$}

An enhancement in the bolometric luminosity and column density, coupled with a 
low B-band luminosity (and therefore red optical to near infrared colors, 
and high X/O and MIPS/O flux ratios), as well as strong outflows from AGN 
and/or stellar winds  are predicted for objects 
that are experiencing a transition from being starburst dominated to AGN dominated 
by  most recent models of AGN galaxy coevolution (Menci et al. 2008, Hopkins et al. 2008, see 
also Narayanan et al. 2009). 

Almost by definition, most of the high-z obscured AGN candidates isolated 
Section 7
are expected to be undergoing this peculiar phase of their evolution. 
The most extreme object 
in the spectroscopically identified sample which satisfies the selection
criteria described in Section 7 is XID~2028 (z=1.592),
marked with a magenta star in the previous diagnostics
(Fig.~\ref{xolx}, Fig.~\ref{rklx} and Fig.~\ref{xork}).
This is the object with the reddest optical color ($R-K$=6.46)
and the second highest X--ray to optical ($logX/O$=1.79) flux ratio
in the spectroscopically  identified sample.
The high X--ray luminosity (L$_{\rm X})\sim10^{45}$ erg s$^{-1}$, 2-10 keV) and the high column
density derived from the spectral analysis (logN$_{\rm H}=22.0\pm0.1$ cm$^{-2}$,
Mainieri et al. in preparation\footnote{In
Mainieri et al. (2007) this object has been assigned a lower redshift,
z=0.784, on the basis of a lower-quality optical spectrum available at that time. The higher
redshift translates in a higher column density than the one reported in Mainieri et
al. 2007, Log N$_{\rm H}=21.83$}) classify this object as an obscured Type 2 Quasar.
The high mid--infrared to optical flux ratio ($\nu_{24}F_{24}/\nu_{R}F_{R}>50$) further
classifies XID~2028 as a Dust Obscured Galaxy (DOG, see Dey et al. 2008; Fiore et al. 2008).

The optical spectrum for this source was obtained   with
DEIMOS (in MOS mode) at the Keck-II telescope (Faber et al. 2003)
on Jan 8, 2008 under mostly clear  conditions with seeing of $\sim1"$.
The data were collected with the  830l/mm
grating tilted to 8300\AA\ and the OG550 blocker, with a resolution of 2.5 \AA. 
The spectra were dithered $\pm±2"$ along the slit to remove ghosting
(internal reflections) from the 830l/mm grating.
Each exposure was 20 minutes for a total of 2.6 hours.
The spectrum (5500-10000 \AA) is shown in the left panel of Fig.~\ref{spe2028}.
It presents a red, host-galaxy dominated continuum superimposed with
high-ionization narrow emission lines typical of obscured AGN (e.g.
[NeV]~$\lambda\lambda$3426,3346 doublet), starburst indicators (e.g. [OII]~$\lambda$3727), 
and MgII~$\lambda$2798 and MgI~$\lambda$2852 lines in absorption.
The redshift z=1.592 has been derived from the [OII] emission line
and it is consistent (within 0.1\%) with that measured from the other 3 emission lines 
visible in the spectrum.
The inset in the left panel of Figure~\ref{spe2028} shows a zoom of the spectrum 
(black = unbinned data, red =smoothed data, with a 15 pixel boxcar) around the 
Mg absorption complex. 
The MgII and MgI absorption lines are slightly blueshifted 
with respect to the systemic velocity defined by [OII] (dashed lines in the inset) 
indicating the possible presence of outflows in the system (see, e.g., Weiner et al. 2009).
The measured offset ($\sim 6 \AA$) implies a velocity of $\sim$300 km/s when  
the uncertainties in the wavelength scale and in the redshift measurements ($\sim 3.\AA$)
are taken into account. 
Although we cannot uniquely assess which is the process 
which drives the outflowing material, due to the concomitant presence of 
high star formation and AGN activity in this system, it is interesting 
to note that the value observed for XID~2028 is slightly larger than 
what is generally observed for star formation driven outflows ($\sim150$ km s$^{-1}$, 
Shapley et al. 2003). 

The SFR derived from the [OII] emission applying the low redshift calibration 
(Kennicutt et al. 1998) is $\sim$10$^3$M$_{\odot}$/year. 
XID~2028 is the highest-redshift X--ray selected Type
2 Quasar detected in the 70 micron survey of the COSMOS field (Kartaltepe et
al. 2010).
The high 70 micron flux implies an infrared luminosity greater than $10^{13}$
L$_\odot$, further classifying this source as a HyLIRG (Hyper Luminous
Infrared Galaxy, Sanders et al. 1988), with a SFR of few$\times$10$^3$ M$_\odot$/year,
in agreement, within a factor of 2-3, with the one estimated from the optical
spectrum.  
However, this value is likely to be an upper limit given that the [OII] emission
and the 70 $\mu$m emission (which at the redshift of the source corresponds to 
the rest-frame $\sim 25 \mu$m emission) can be severely contaminated by AGN activity
(see discussions in Silverman et al. 2009a, Daddi et al. 2007b). 

\begin{center}
\begin{figure*}[!t]
\begin{center}
\includegraphics[width=8cm]{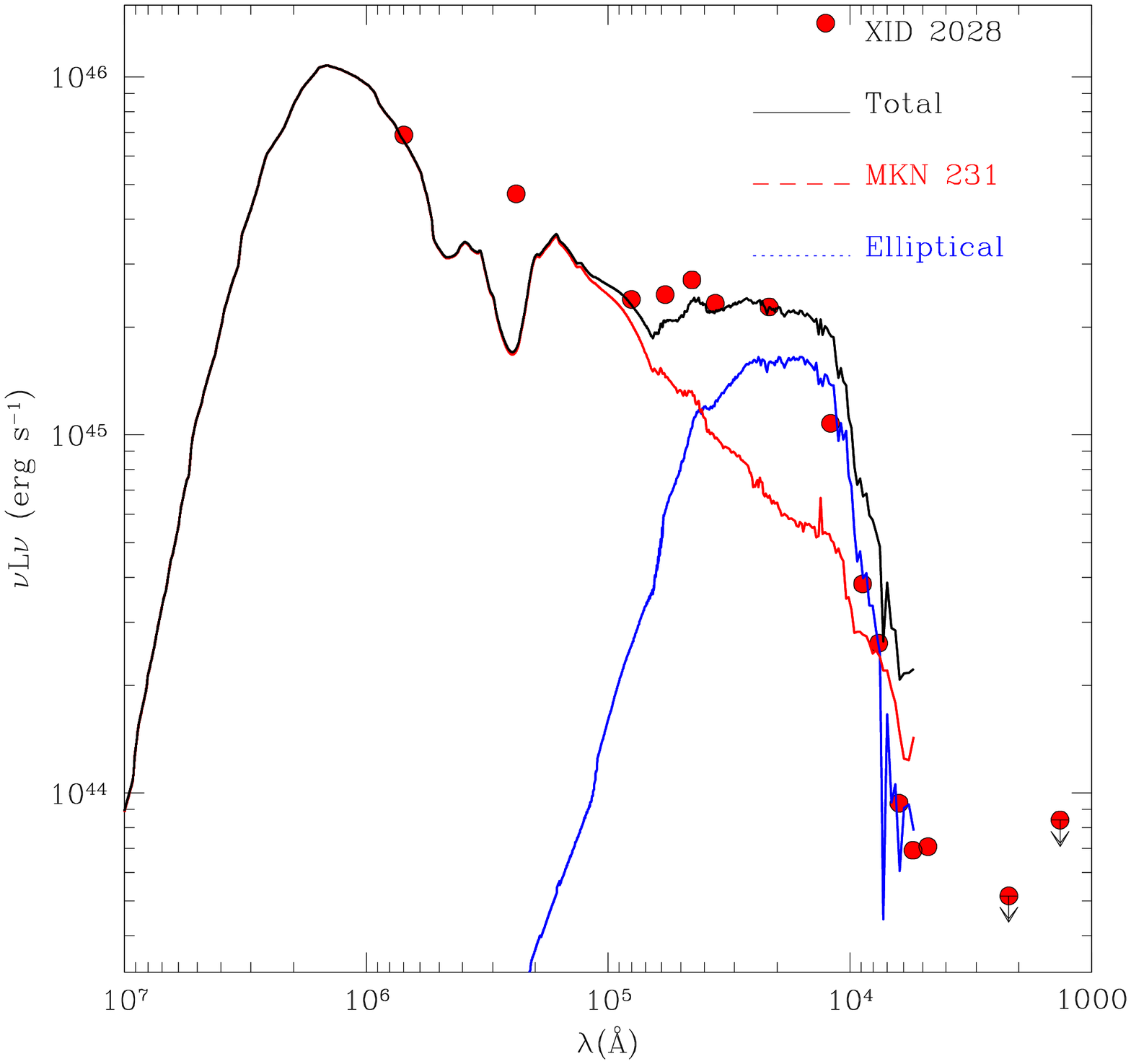}
\includegraphics[width=8cm]{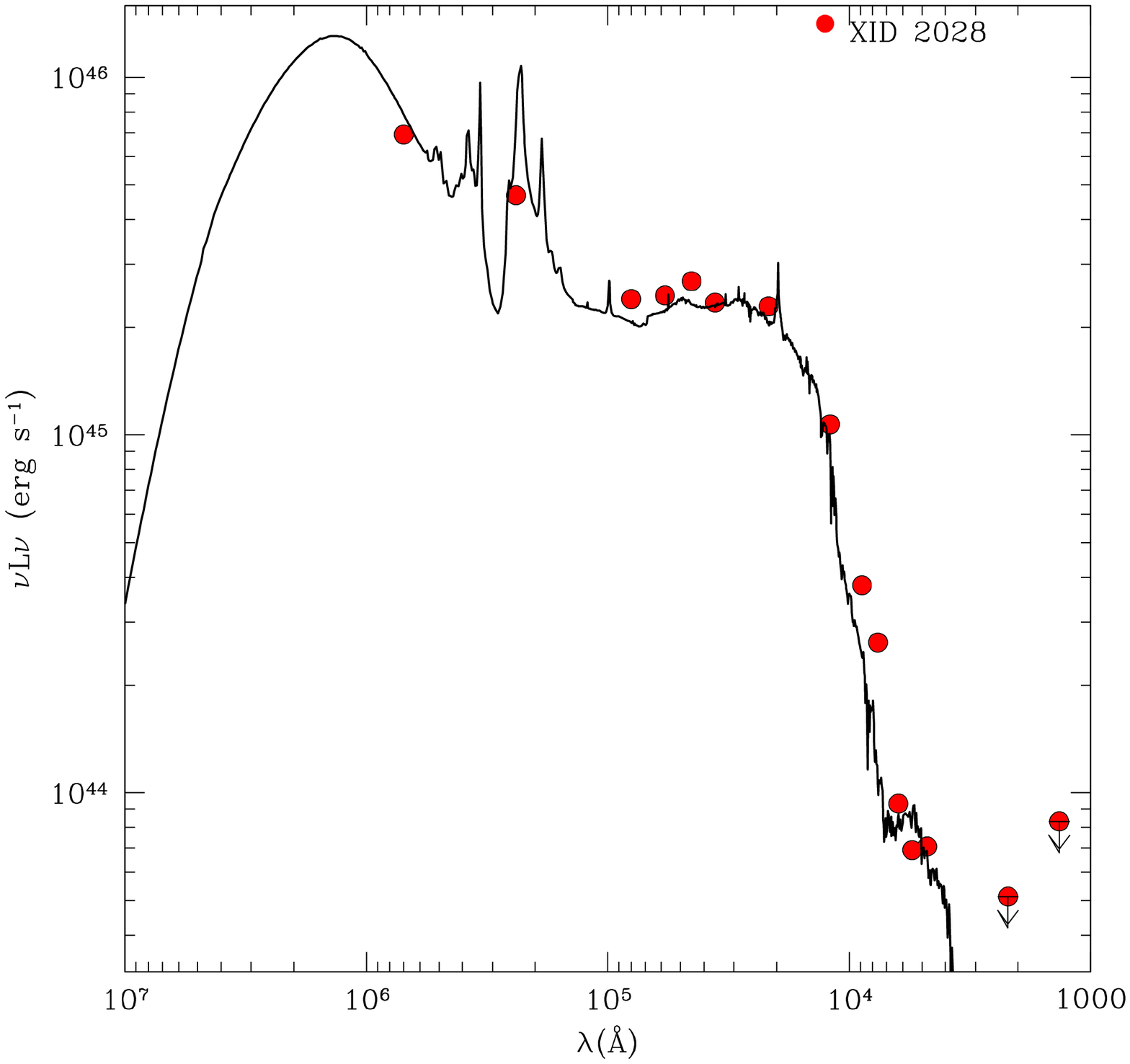}
\end{center}
\caption{SED from the radio to the UV (GALEX) of XID~2028. 
In the left panel, the red line is the SED of Mrk 231 while the blue line is the SED of an elliptical
galaxy 2 Gyr old (both SEDs are from Polletta et al. 2007). The black line 
is the sum of the two SEDs combined in order to provide a good representation of the
SED of XID 2028. In the right panel, the black template is taken from Narayanan et al. (2009). See text for details.} 
 \label{sed2028}
 \end{figure*}
  \end{center}

The broad-band SED (70 $\mu$m Spitzer to UV, including the GALEX points retrieved 
from Zamojski et al. 2007) of XID~2028 is reported in Figure~\ref{sed2028}.
The left panel shows  the XID 2028 SED with superimposed the template of Mkn 231,
a well known Compton Thick AGN-ULIRG system (red curve)
and the template of an Elliptical galaxy (2 Gyr old, blue curve)
both taken from Polletta et al. (2007).
The sum of the above templates (black) is shown in order to illustrate that the SED of
XID 2028  is likely to be the result of the combination of a massive, luminous, host 
galaxy which dominates the optical light (M$_*\sim5\times10^{11}$ M$_\odot$, 
assuming M/L=0.22 from Ilbert et al. 2010), with a starburst/AGN component, seen unblocked 
in the IR bands. The composite AGN/starburst nature of this source is also revealed in optical
morphology. The right panel of Figure~\ref{spe2028} shows the high resolution 
HST/ACS image cutout centered on XID~2028. 
From the image, a point-like nucleus is clearly visibile,
likely responsible for the X--ray emission, as well as a residual diffuse 
component (i.e. the host galaxy or a merger remnant). 

The SED of XID~2028 is shown again in the right panel of Figure~\ref{sed2028},
this time superimposed with a numerical (representative) SED resulting from 
the models for the z$\sim$2 DOG population published by Narayanan et al. (2009).
Apart for a scaling of the overall normalization of the system (0.66 dex higher), 
this template reproduces almost  perfectly the observed datapoints. 
The  stellar mass used as input parameter
of the numerical models ranges from M$_*\sim4.7-5.3\times10^{11}$ M$_\odot$, almost coincident 
with the value derived from the SED fitting. 
The input M$_{\rm BH}$ is also available from the simulations,
and it ranges from $7\times10^{8}$ to $10^9$ M$_\odot$.

The bolometric luminosity can be estimated from the X-ray emission 
assuming an appropriate bolometric correction (Elvis et al. 1994 
or more recently Lusso et al. 2010), and it is of the order of 
$\sim 2\times 10^{46}$ erg s$^{-1}$, consistent with the bolometric 
luminosity obtained from the overall observed SED (dominated by
the IR emission). 
Assuming that for XID~2028 the energy is released in a 
radiatively efficient way (L/L$_{\rm Edd}\sim0.1-1$), as expected for
objects in the transitioning phase described above, 
it is therefore possible to derive the BH mass under the 
Eddington limited accretion scenario. 
This translates into a BH mass of the order of $\sim 10^{8}-10^{9}$ M$_\odot$,
in agreement with what can be inferred from the host galaxy mass
assuming the local M$_{\rm BH}-$M$_{\rm Bulge}$ relation and its evolution
(Gultekin et al . 2009, Merloni et al. 2010, Jahnke et al. 2009),
and within the range of the input BH mass assumed in the numerical 
SED simulations. 

A comparison between the broad band properties of XID~2028 with those of
other luminous, obscured quasars reported in the literature shows
some similarities, but also remarkable differences.
The SED of other X--ray selected luminous obscured quasars (e.g. Mainieri et al. 2005, 
Severgnini et al. 2006, Le Floc'h et al. 2007, Pozzi et al. 2007, Vignali et al. 2009) 
shows evidences favoring the presence of a massive starburst component and obscured 
accretion, with similar optical and X-ray spectra. 
However, we also note that the optical spectrum of XID~2028 is quite different from that
of the high-z, Type 2 QSO prototype CXO~202 discovered in the CDFS (Norman et al.
2002, at z=3.7). In the Norman et al. (2002) object the optical spectrum is 
dominated by narrow emission line with almost no underlying continuum, 
while the host galaxy contribution is clearly visible and dominant in XID~2028.
However, CXO~202 is at much higher redshift than XID~2028, and it is therefore
well possible that XID~2028 would show some of the same high-ionization lines 
and less continuum were it at higher redshift (and viceversa). 
Also, the different level of obscuration ($\sim10^{22}$ cm$^{-2}$ for XID~2028 
vs. $>10^{24}$ cm$^{-2}$ for CXO~202) may cause the differences we see in the
optical spectra. 
In any case, the observations above highlight the power of a full multiwavelength
approach to fully characterize a key phase in AGN evolution.

%%%%%%%%%%%%%%%%%%%%%%%%%%%%%%%%%%%%%%%%%%%%%%%%
%%%%%%%%%%%%%%%%%%%%%%%%%%%%%%%%%%%%%%%%%%%%%%%%

\section{Conclusions}

We presented the catalog of optical and infrared identifications for 
$\sim$1800 X--ray sources detected in the XMM-COSMOS survey. 
This catalog comprises the totality of the XMM sources presented
in C09, with the exception of 65 faint sources
($\sim$3\%) detected in two additional fields obtained in AO6. 
The optical/IR matches with the X--ray positions were driven 
by the {\it Chandra} subarcsec centroids, when available, 
or estimated through statistical methods already successfully
tested in different XMM and {\it Chandra} fields (see B07, Cardamomone
et al. 2008). 
For each source, a flag on the reliability of the X--ray to optical/infrared
association is provided, classifying the counterparts as ``reliable" (87.7\%), 
``ambiguous" (11.3\%) and ``not identified" (1\%; see Table 1). 
For sources with two possible associations both entries are reported in the 
associated catalog (Table 2). The availability of \chandra\ data allowed us
also to quantify the reliability of the likelihood ratio technique when
applied to XMM sources matched to deep optical and infrared images,
which turned out to be very high ($\gs98.5$\%, see Section 3.1).  

Along with the positions of the optical counterparts we publish 
a wealth of multiwavelength information, from 24$\mu$m to X--ray
(see Section 4 and Table 2), 
most notably  a collection of all the available spectroscopic redshifts 
($\sim 900$) obtained through different programs ongoing in the
COSMOS field (see Section 5 for details). 
To maximize the completeness over a well defined large area and, at the same
time, keep selection effects under control, for the multiwavelength analysis 
presented in this paper we further considered 
the subsample of 1640 identified sources detected above the limiting fluxes corresponding 
to a coverage of more than 1 deg$^{2}$ in at least one  of the three X--ray 
energy ranges considered (see Section 3.1). The multiwavelength properties
of the fainter X--ray sources are investigated in a much greater
detail in the framework of the C-COSMOS survey (Elvis et al. 2009,
Civano et al. in prep.). 

The spectroscopic completeness in the soft (0.5-2 keV) and hard 
(2-10 keV) subsamples is $\sim$50\% and $\sim$60\%, respectively, 
remarkably 
high for identification campaigns of X--ray sources, to be  compared 
with the spectroscopic completeness in the AEGIS survey (40\% 
in the 2-10 keV band, see Aird et al. 2010); in the X-Bootes 
survey ($\sim$50\%, Ryan Hickox, private communication; 
see also Murray et al. 2005; Kenter et al. 2005); and in the CDFS 2Ms sample 
($<50$\%, Luo et al. 2008, Luo et al. 2010).
By far, the highest spectroscopic completeness is for the sources detected 
in the ultra-hard (5-10 keV) band, where it reaches $\sim80$\%: 
for the first time, XMM-COSMOS provides a statistically 
significant sample (221) of sources detected at $>5$ keV 
with a very high spectroscopic breakdown at fluxes.
%%%%%%%%%%%%%%%%%%%%%%%%%%%%
As a comparison, the Beppo-SAX HELLAS survey had a 50\% completeness 
(71/147) at higher X-ray fluxes ($>3\times10^{-14}$ \cgs, Vignali PhD
thesis), while the XMM-HBSS has an almost complete (97\%) spectroscopic
information, but it is limited to 62 sources at fluxes $>7\times10^{-14}$ 
\cgs\ (Della Ceca et al. 2008).

For the sources without spectroscopic redshifts, accurate photometric 
redshifts (down to $\sigma(\Delta z/(1+z))\sim0.01$ at I$<22.5$) are available 
(Salvato et al. 2009), 
allowing an almost 100\% complete sample of X--ray sources with redshift 
information. This, coupled with the large number of objects, allowed us 
to sample with an unprecedented statistic
the high luminosity tail  of the X--ray luminosity
function. 
The present data allow us to derive the AGN space density
as obtained by the COSMOS survey over a relatively broad redshift range
(z$\sim$ 1--3) and luminosity interval ($Log L_{\rm X}=44.5-47$).
We compared the COSMOS results with those discussed in recent papers
(Ebrero et al. 2009; Yencho et al. 2009; Aird et al. 2010) combining data
from different surveys and covering in a fairly homogeneous way the
luminosity redshift plane. 
Our results suggest that at high luminosities ($log L_{\rm X}>44.5$)
the AGN number density peaks at z$\sim 2-2.5$, in good agreement 
with LDDE parameterization of the above mentioned XLF. 

We investigated the dependence of the X--ray to optical flux ratio versus 
the X--ray luminosity for BL AGN and NL AGN, and sources with photometric redshifts only.
The normalization of the correlation between X/O and X--ray luminosity for the obscured AGN  
increases going from the spectroscopically identified sample to the objects 
with only photometric redshifts available (i.e., 
towards fainter magnitudes; see Figure~\ref{xolx}). 
The well-known relation between the X/O and the X--ray luminosity 
(F03; Barger et al. 2005) for obscured AGN is mostly driven by the fact that, 
while the nuclear AGN X-ray luminosity can span several decades (depending on the 
BH mass {\it and} on the accretion rate), the host galaxy R-band luminosity (which dominates 
the optical emission in obscured sources) has a moderate scatter, around
the M$_{\rm BH}-M_{\rm Bulge}$ relation, i.e., depending only on the BH mass.

We then studied  the dependence of the $R-K$ color on the X--ray 
luminosity and we found an opposite behavior for BL AGN and NL AGN 
in our sample.  
For NL AGN there is evidence for a trend where high-luminosities 
sources have redder colors than low-luminosity sources (see Figure~\ref{rklx}). 
This correlation is most likely due to redshift effects, given that 
the $R-K$ color of normal and star forming galaxies increases up to redshift
z$\sim 2$.
On the contrary, while high-luminosity BL AGN show a median
$R-K$ color consistent with that observed in optically selected
samples (Barkhouse \& Hall 2001), lower luminosity BL AGN 
show a median $R-K$ color statistically different (redder) from the one
observed for the higher-luminosity ones (see Figure~\ref{rklx}). 
This difference in the $R-K$ colors for high and low luminosity sources is likely due to the
contribution of the host galaxies in the latter sample.  
We conclude that X--ray surveys are effective in isolating also 
{\it unobscured} objects 
otherwise missed by optical multicolor surveys, with preselection
on the basis of optical colors and pointlike emission (see also similar conclusions 
obtained on BL AGN samples from pure spectroscopic selection, Bongiorno et al. 2007). 
However, the effect of host galaxy contribution  may play
an important role in getting the correct classification of the nuclear source. 

The large body of COSMOS  multiwavelength data allowed us 
to devise a robust method to build a sizable sample of
luminous obscured AGN candidates at $z>1$ among the still spectroscopically
unidentified population, and control
the selection effects (see Section 7.1.1 and 7.1.2).
In total, we ended up with a robust sample of $\sim 150$ objects 
with photometric redshifts information, and multicolor properties consistent
with those of luminous, obscured quasars at z$>1$.
This sample should not be considered complete with respect to the
X--ray selected, high-z obscured AGN population,  but can be considered 
as representative of the average properties of this class of sources. 
We compared the average X--ray hardness ratio of the high-z obscured AGN 
candidate sample with the ones derived for the spectroscopically identified samples 
(Section 7.2). 
We found that the average HR for this sample, HR$=-0.17$, is considerably
higher than that derived for spectroscopically classified NL AGN
(HR$=-0.29$), further confirming that a red observed  $R-K$ color ($R-K\gsimeq5$) 
coupled with high X/O or $\nu_{24}F_{24}/\nu_{R}F_{R}$ ratios, 
is a good indicator for both optical 
and moderate X--ray obscuration in high-luminosity sources (see also 
Maiolino et al. 2006). 

For the vast majority (75-80\%) of the high-z obscured AGN candidates 
the AGN component is visible in the near-IR (roughly corresponding to the rest 
frame $\sim 1-3\mu$m range), where the dust reprocessed emission
is expected to dominate the observed SED. 
These sources would have been selected
on the basis of IRAC colors only (see Fig.~\ref{sternflux} and Fig.~\ref{iraccm}, and Section 7.3).
However, at the limiting fluxes of the IRAC COSMOS survey the contamination from
non-AGN objects within the selection wedge starts
to become important, and the selection
criterion would become efficient only when combined with an additional 
criterion (in this case, the X--ray emission). 

We studied the fraction of obscured AGN 
in the XMM-COSMOS sample as a function of the 2-10 keV X--ray luminosity, 
and discussed the differences between optical (spectroscopic
and photometric) and X--ray classifications. 
We confirmed that the fraction decreases with increasing luminosity (see Section~8
and Fig.~\ref{fracl}), for both the spectroscopically identified and
the full sample. At L$_{\rm X}\gs10^{44}$ erg s$^{-1}$, 
the obscured AGN fraction is of the order of 25-30\% in the logL$_{\rm X}=44-44.5$
bin and $\sim 15-20$\% in the  logL$_{\rm X}=44.5-45$ bin, about a
factor of 2-3 lower than the predictions 
from the Gilli et al. (2007) and Treister \& Urry (2006) models. 
At lower-luminosities, the optical and X--ray classifications
largely differ (see green triangles and squares at logL$_{\rm X}<43$ erg s$^{-1}$ 
in Figure~\ref{fracl}). 
This may be related, at least in part, to host galaxy dilution 
of the intrinsic AGN spectrum, which is expected to be important at low 
Eddington ratios and at low luminosities (Hopkins et al., 2009).

We studied in detail the SED and the spectral properties of XID 2028, that 
we consider the prototype of an obscured QSO at high-z (z=1.592) caught in a 
transition stage from being starburst dominated to AGN dominated, 
which was possible to isolate only thanks to the combination of X-ray and infrared 
observations. 
XID 2028 is one of the brightest 
XMM-COSMOS sources, with an unabsorbed X--ray luminosity 
L$_{\rm X}\sim10^{45}$ erg s$^{-1}$ and significant X--ray 
absorption (N$_{H}\sim10^{22}$), i.e. a Type 2 QSO. 
The optical spectrum of this source presents strong AGN emission lines ([NeV] doublet) 
consistent with our selection method. 
Moreover,  evidence of outflowing material  
at a velocity of $\sim 300$ km s$^{-1}$ (see Figure~\ref{spe2028})
is also clearly detected from the blueshifted MgI and MgII complex absorption.
The X--ray, optical and infrared properties and band ratios of XID 2028 are
very similar to those predicted by the above mentioned models,
and the overall SED is best represented by numerical templates 
resulting from the models published by Narayanan et al. (2009) to
describe z$\sim2$ DOG (see Fig.~\ref{sed2028}, right panel). 

XID 2028 is the brightest object for which such a detailed
analysis can be conducted in the XMM-COSMOS catalog. 
However, as demonstrated in Section 7, it is reasonable to assume that a large fraction of the sources
among the high-z obscured AGN candidates share the same properties,
although at lower X--ray (L$_{\rm X}\gs10^{44}$ erg s$^{-1}$) and bolometric
luminosities. Most of these sources are well within the reach of the {\it Herschel} (Pilbratt 2005)
instruments between 75 and 500 $\mu$m, which in the GT project in the COSMOS
field will reach limiting fluxes of few mJy. Such longer wavelength observations
can greatly help in separating nuclear activity and star formation, and
assessing the real bolometric output of obscured AGN. 
Further investigations, as for example, direct infrared spectroscopy with 
X-Shooter@VLT (D'Odorico et al. 2004) or 
LUCIFER@LBT (Mandel et al. 2007)
may be crucial in confirming the nature of these candidates.

\acknowledgments
This work is based on observations obtained with XMM-{\it Newton}, 
an ESA Science Mission with instruments
and contributions directly funded by ESA Member States and the
USA (NASA). In Germany, the XMM-{\it Newton} project is supported by the
Bundesministerium f\"ur Wirtschaft und Technologie/Deutsches Zentrum
f\"ur Luft- und Raumfahrt (BMWI/DLR, FKZ 50 OX 0001), the Max-Planck
Society and the Heidenhain-Stiftung. 
Part of this work was supported by the Deutsches Zentrum f\"ur Luft-- und
Raumfahrt, DLR project numbers 50 OR 0207 and 50 OR 0405. In Italy, the
XMM-COSMOS project is supported 
by PRIN/MIUR under grant 2006-02-5203, ASI-INAF grants I/023/05/00, I/088/06 
and ASI/COFIS/WP3110,I/026/07/0. 
This work was supported in part by NASA Chandra grant number GO7-8136A 
(FC, ME, AF, HH).
TM acknowledges support from CONACyT 83564 DGAPA/PAPIIT IN10209 to IA-UNAM 
as well as the NASA ADP (NNX07AT02G) grant to UCSD. 
GH and MS ackwnoledge a contribution from the Leibniz Prize of 
the Deutsche Forschungsgemeinschaft under the grant HA 1850/28-1.
NC and AF were partially supported from a NASA grant NNX07AV03G to UMBC.
KJ is supported by the Emmy Noether-Programme of the German Science 
Foundation DFG.
ET is supported by the National Aeronautics and Space Administration through
{\it Chandra} Postdoctoral Fellowship Award Number PF8-90055.
We thank James Aird and Jacobo Ebrero for sending us 
machine-readable tables of their X-ray luminosity functions,
Desika Narayanan for help with his model SEDs, and Ryan Hickox for
providing unpublished information about the XBootes survey.
We gratefully thank Nick Wright for a carefully reading of the
manuscript.  
This work is based in part on observations obtained with 
MegaPrime/MegaCam, a
joint project of CFHT and CEA/DAPNIA, at the Canada-France-Hawaii Telescope
(CFHT), and on data products produced at TERAPIX data centre located at the
Institut d'Astophysique de Paris.
This research has made use of the Keck Observatory Archive (KOA), which is 
operated by the W. M. Keck Observatory and the NASA Exoplanet Science Institute 
(NExScI), under contract with the National Aeronautics and Space Administration.
We gratefully acknowledge the contribution of the entire COSMOS
collaboration; more information on the COSMOS survey is available at 
\verb+http://www.astro.caltech.edu/~cosmos+. This research has made  
use of the NASA/IPAC Extragalactic Database (NED) and the SDSS spectral 
archive. Finally, we thank the anonympus referee for detailed and 
constructive comments to the first version of this paper.

%%%%%%%%%%%%%%% TABLE 2 %%%%%%%%%%%%%%%%%%%%%%%%%%%%%%%%%%%%%%
%
\clearpage
\begin{landscape}
\begin{deluxetable}{lrccccccccccccccccccccccccc}
\tabletypesize{\tiny}
\tablecaption{Basic properties of the XMM-COSMOS counterparts$^a$} 
\tablewidth{0pt}
\tablehead{
\colhead{IAU Name} & 
\colhead{XID} & \colhead{RA} & \colhead{DEC} & 
\colhead{S$_{0.5-2}$} & \colhead{S$_{2-10}$} & \colhead{S$_{5-10}$} & flag & HR & CHID &
\colhead{flagID} & \colhead{ID(Capak)} & \colhead{ID(Ilbert)} &
\colhead{r} & \colhead{I} & \colhead{K} &
\colhead{3.6$\mu$m} & \colhead{4.5$\mu$m} & \colhead{5.8$\mu$m} & \colhead{8.0$\mu$m} & \colhead{24$\mu$m} &
\colhead{zspec} & \colhead{class} & \colhead{NOTES} & \colhead{zphot} \\
%%%%%%
    &  
    & deg & deg & 
cgs & cgs & cgs &  &  &  &
    &     &     & 
AB & AB & AB & 
AB & AB & AB & AB & AB & 
    &     &   & \\}
\scriptsize
\startdata
XMMC\_150.10515+1.98082 &   1 &   150.10521 &    1.981183 &      139.00 &      228.00 &      122.00 &     1 &    -0.50 &      358 &          1 &   1268521 &   786683 &    19.24 &    19.12 &    18.21 &    17.49 &  17.20 &    16.89 &    16.67 &    15.10 &     0.373 &       1 &      1 &     0.37 \\  
XMMC\_149.73919+2.22053 &  2 &   149.73896 &    2.220675 &      105.00 &      211.00 &      107.00 &     1 &    -0.42 &      329 &          1 &   1695261 &  1054439 &    20.34 &    19.96 &    19.00 &    18.23 &  17.93 &    17.47 &    17.08 &    15.84 &     1.024 &       1 &      1 &     1.05 \\ 
XMMC\_149.76154+2.31849 &   3 &   149.76148 &    2.318458 &      150.00 &      258.00 &      142.00 &     1 &    -0.49 &      440 &          1 &   2072529 &  1290981 &    19.39 &    18.67 &    17.39 &    17.08 &  16.88 &    16.61 &    16.26 &    14.48 &     0.345 &       1 &      1 &     0.36 \\ 
XMMC\_149.74418+2.24948 &   4 &   149.74389 &    2.249753 &       78.40 &      159.00 &       95.60 &     1 &    -0.41 &      418 &          1 &   1688156 &  1048950 &    18.51 &    16.90 &    17.02 &    17.31 &  17.49 &    17.55 &    16.88 &    15.33 &     0.132 &       2 &      1 &     0.13 \\ 
XMMC\_149.82819+2.16421 &   5 &   149.82793 &    2.164360 &       72.80 &      148.00 &       75.40 &     1 &    -0.42 &      320 &          1 &   1709129 &  1063264 &    19.90 &    19.47 &    18.75 &    18.15 &  17.78 &    17.35 &    17.00 &    15.93 &     1.157 &       1 &      1 &     1.18 \\ 
XMMC\_150.17978+2.11015 &   6 &   150.17978 &    2.110380 &       38.20 &       60.80 &       28.70 &     1 &    -0.53 &       42 &          1 &   1236435 &   767213 &    19.31 &    18.34 &    17.95 &    18.03 &  17.91 &    17.85 &    17.35 &    15.63 &     0.360 &       1 &      1 &     0.32 \\ 
XMMC\_150.52108+2.62525 &   7 &   150.52096 &    2.625412 &       96.60 &      189.00 &      101.00 &     1 &    -0.44 &      -99 &          1 &   2277781 &  1418792 &    20.35 &    18.94 &    18.07 &    17.73 &  17.62 &    17.43 &    17.36 &    15.73 &     0.519 &       1 &      5 &     1.36 \\ 
XMMC\_150.05383+2.58967 &   8 &   150.05378 &    2.589671 &       57.80 &       54.00 &       19.40 &     1 &    -0.69 &      142 &          1 &   2362650 &  1472056 &    19.29 &    18.79 &    18.20 &    17.22 &  16.89 &    16.53 &    16.24 &    15.07 &     0.699 &       1 &      1 &     0.71 \\ 
XMMC\_149.91983+2.32747 &   9 &   149.91976 &    2.327460 &       28.30 &       47.00 &       22.20 &     1 &    -0.51 &      499 &          1 &   2030952 &  1265494 &    20.50 &    20.07 &    19.47 &    18.89 &  18.36 &    17.88 &    17.41 &    16.05 &     1.459 &       1 &      2 &     1.46 \\ 
XMMC\_149.91261+2.20032 &  10 &   149.91244 &    2.200366 &       27.20 &       22.10 &       10.90 &     1 &    -0.72 &      446 &          1 &   1660949 &  1032058 &    20.88 &    20.39 &    19.37 &    18.72 &  18.50 &    18.22 &    17.93 &    15.74 &     0.689 &       1 &      2 &     0.68 \\ 
\enddata
\pn
\tablenotetext{a}{Only a portion of the table is shown here for guidance regarding the
form and content of the catalog. \\ 
The entire table is available in electronic form 
at http://www.mpe.mpg.de/XMMCosmos/xmm53\_release/xmm53fields\_table\_all.dat. \\
The full table contains 25 columns of information on 1797 sources. The information on the possible second counterpart for the sources 
in the ``ambiguous" ID class are listed at the end of the table. \\ 
The catalog README file is available at
http://www.mpe.mpg.de/XMMCosmos/xmm53\_release/} \\

\par
\tablenotetext{NOTES}{-- 
Col. [1]: XMM-COSMOS IAU designation;
Col. [2]: XMM-COSMOS identifier number (from Cappelluti et al. 2009);
Col. [3-4]: coordinates of the optical/IR counterpart (J2000); 
Col. [5-7]: X--ray fluxes in the soft, hard and ultra-hard bands (from Cappelluti et al. 2009);
Col. [8]: flag identifying the sources included in the flux limited sample (1) or not (0); 
Col. [9]: X--ray hardness ratio, HR; 
Col. [10]: Chandra-COSMOS identifier number (from Elvis et al. 2009); 
Col. [11]: flag for the optical identification, according to the classes described in Table 1: 
Sources flagged with ``1'' are the ``reliable'' 
counterparts; sources flagged with ``2'' are the ambiguous counterparts (in this case 
the positions and basic properties of both candidate counterparts are listed); 
sources flagged with ``0''are statistically not identified. See Section~3 for  details.
Col. [12]: identifier number from Capak et al. (2007) catalog; 
Col. [13]: identifier number from the Ilbert et al. (2009) catalog; 
Col. [14-15]: the r-band and I-band magnitudes (AB system, from Capak et al. 2007);
Col. [16]: K-band magnitude (AB system, from Mc Cracken et al. 2010);
Col. [17-20]: magnitudes in the four IRAC channels (AB; from Ilbert et al. 2009);
Col. [21]: MIPS 24$\mu$m magnitude (AB, from Le Floc'h et al. 2009);
Col. [22]: Spectroscopic redshift (see Section 5);  
Col. [23]: Spectroscopic classification according to the classes described in Section 5.1: 1 = BL AGN; 2 = NL AGN; 
3 = normal/starforming galaxy; 
Col. [24]: Origin of the spectroscopic redshifts, with relevant notes when needed. The code for the 
source of the spectroscopic redshift is the following: 1 = SDSS; 2 = MMT (Prescott et al. 2006);
3, 4 = IMACS runs (Trump et al. 2007; 2009); 5 = zCOSMOS 20k catalog (Lilly et al. 2007); 
6 = zCOSMOS faint 4.5k catalog; 7 = Keck runs; 
Col. [25]: photometric redshift (from Salvato et al. 2009)}

\end{deluxetable}
\clearpage
\end{landscape}

%%%%%%%%%%%%%%% TABLE 2 %%%%%%%%%%%%%%%%%%%%%%%%%%%%%%%%%%%%%%

\end{document}